%% file: root.tex
\documentclass[lettersize,journal]{IEEEtran}
\usepackage[font=normalsize]{caption}
\usepackage{printlen}
\usepackage{amsmath,amssymb,amsfonts}
\usepackage{amsthm}
\newtheorem{theorem}{Theorem}

\theoremstyle{remark}
\newtheorem{remark}{Remark}
\usepackage{graphicx}
\usepackage{algorithm,algorithmic}

\usepackage{enumitem}
\usepackage{caption}
\usepackage{subcaption}
\usepackage{textcomp}
\usepackage{mathtools}
\usepackage{cancel}

\usepackage{xcolor}
\usepackage{nicefrac}
\usepackage{bbm}
\usepackage{booktabs}
\usepackage{multirow}
\usepackage{nameref}
\usepackage{float}
\usepackage{placeins}
\usepackage{romannum}
\usepackage{pgfplots}
\usepackage[final]{changes}
\definechangesauthor[name=Ricardo, color=orange]{RA}
\setauthormarkup{} 

\newcommand{\myadd}[1]{\added[id=RA]{#1}}

\setaddedmarkup{\textcolor{blue}{#1}}
\usepackage[backend=biber,style=ieee]{biblatex}
\usepackage{hyperref}
\hypersetup{
    colorlinks=true,
    linkcolor=blue,
    citecolor=blue,
    urlcolor=blue
}
\usepackage{cleveref}

\def\BibTeX{{\rm B\kern-.05em{\sc i\kern-.025em b}\kern-.08em
    T\kern-.1667em\lower.7ex\hbox{E}\kern-.125emX}}
    
\begin{document}
\pagenumbering{arabic}
\title{A Convex Obstacle Avoidance Formulation}

\author{Ricardo Tapia, Iman Soltani
\thanks{Ricardo Tapia is with the Laboratory for AI, Robotics and Automation, University of California at Davis, Davis, CA 95616 USA. Email: ricardo.tapia.m@proton.me.}%
\thanks{Iman Soltani (Corresponding Author, Lab PI) is with the Laboratory for AI, Robotics and Automation, University of California at Davis, Davis, CA 95616 USA. Email: isoltani@ucdavis.edu }%
\thanks{Author contributions: Ricardo Tapia conceived the project, designed and carried out the simulations, performed the analysis, and wrote the manuscript. Iman Soltani provided editorial revisions improving clarity and structure.}
}

\maketitle
\begin{abstract}
Autonomous driving requires reliable collision avoidance in dynamic environments. Nonlinear Model Predictive Controllers (NMPCs) are suitable for this task, but struggle in time-critical scenarios requiring high frequency. To meet this demand, optimization problems are often simplified via linearization, narrowing the horizon window, or reduced temporal nodes, each compromising accuracy or reliability. This work presents the first general convex obstacle avoidance formulation, enabled by a novel approach to integrating logic. This facilitates the incorporation of an obstacle avoidance formulation into convex MPC schemes, enabling a convex optimization framework with substantially improved computational efficiency relative to conventional nonconvex methods. A key property of the formulation is that obstacle avoidance remains effective even when obstacles lie outside the prediction horizon, allowing shorter horizons for real-time deployment. In scenarios where nonconvex formulations are unavoidable, the proposed method meets or exceeds the performance of representative nonconvex alternatives.  The method is evaluated in autonomous vehicle applications, where system dynamics are highly nonlinear.
\end{abstract}


\section{Introduction}
\label{sec:introduction}
\IEEEPARstart{I}{n} \added{the domain of autonomous vehicles, optimization-based obstacle avoidance (OA) controllers offer significant advantages over heuristic approaches, such as Potential Field Methods (PFM) \cite{Khatib1985Real-TimeRobots} or path-planning algorithms like Rapidly Exploring Random Tree (RRT) \cite{S.M.LaValle2020Rapidly-ExploringProspects}. While heuristic methods often rely on probabilistic completeness or offline boundary value solvers, optimization-based methods provide two key benefits: (\romannum{1}) they generate obstacle-free trajectories that explicitly account for complex system dynamics and operational constraints, and (\romannum{2}) they can readily determine whether a collision is imminent by assessing the feasibility of the optimal control problem (OCP).}

Optimization-based OA admits several distinct mathematical formulations, each with different structural properties and computational implications. Notable approaches include Mixed Integer Convex Programming (MICP) \cite{Marcucci2022MotionOptimization}, barrier function formulations \cite{Zeng2021Safety-CriticalFunction}, and shortest-path methods in Graphs of Convex Sets (GCS) \cite{Marcucci2022MotionOptimization}, \cite{Marcucci2024ShortestSets}. The choice of formulation directly influences key characteristics of the resulting OCP, including convergence, convergence rate, and overall computational efficiency. The first challenge addressed in this work is therefore to identify an OA formulation whose inherent structure provides advantages in these properties.

The second challenge arises from the dependence of most OA formulations, particularly those with hard constraints, on the obstacle lying within the prediction horizon of the OCP. When the obstacle is outside the horizon, the OA constraint will be trivially satisfied and may not influence the resulting trajectory. This motivates the use of a longer prediction horizon. However, extending the horizon introduces a fundamental compromise: while it increases the likelihood of detecting and reacting to obstacles in time, it also raises computational demands and can reduce the controller's responsiveness. Balancing this robust convergence against the inevitable computational burden of longer horizons remains a critical design challenge.

Finally, real-world problems are often nonlinear and nonconvex, which necessitates the solution of NMPCs. To achieve high frequency rates, real-time performance schemes necessitate linearization or convexification of any nonconvex constraints \cite{DiehlMandBock2006, Stella2017}. Linearization of the OA formulation introduces additional approximation error. This creates a third challenge: designing an OA formulation that remains computationally tractable under real-time NMPC while minimizing the degradation introduced by such approximations.

The principal contribution of this work is the introduction of a convex OA formulation, termed Relaxed Convex Obstacle Avoidance (RCOA), which directly addresses the limitations of existing approaches. By exploiting convexity, RCOA enhances convergence toward obstacle-free trajectories and reduces computational effort relative to nonconvex formulations. While the incorporation of a continuous penalty function trades the strict algebraic certificate of infeasibility for computational speed, feasibility is readily re-established in practice due to the formulation's structural equivalence to mixed-integer counterparts that encode logical constraints explicitly. A distinctive property of the formulation is its effectiveness even when obstacles lie outside the prediction horizon, thereby permitting shorter horizons and improving computational efficiency.

To assess the capability of RCOA, a series of simulations are conducted and benchmarked against established OA formulations, including a general nonconvex approach and a mixed integer formulation. The OCPs are solved using several modern state-of-the-art solvers and algorithms, namely FATROP \cite{Vanroye2023FATROPControl}, HSL code MA57 \cite{Duff2004}, Gurobi \cite{GurobiOptimization2016GurobiOptimizer}, and Successive Convexification \cite{Mao2016SuccessiveProperties}. Utilizing these solvers allows the formulations to be evaluated in both fully nonconvex environments and Successive Convexification schemes, explicitly demonstrating the computational advantages of convex OCPs. 

The formulation is evaluated on an autonomous vehicle application utilizing a three degree of freedom bicycle model with advanced tire models. These nonlinearities are particularly pronounced involving aggressive maneuvers, such as severe braking and sharp turning. \added{Capturing these limits is essential for collision-imminent safety. Thus,} both nonlinear and linearized dynamic models are examined, exposing the influence of nonconvex functions on computational tractability. Videos of the NMPC simulations are provided as supplementary material for reference; included are videos with a two-second prediction horizon for each formulation. The reference material does not include the speed penalty in the cost function for RCOA, which is not necessary for longer prediction horizons.

Simulation results highlight two central findings. First, the RCOA formulation exhibits consistent performance even in configurations approaching infeasibility, a property notably absent in the nonconvex and mixed integer formulations used for comparison. Second, RCOA achieves computational efficiency that is on par with, and in several cases superior to, the benchmark formulations, thereby reinforcing its suitability for real-time applications.

\added{This paper presents the first stage of this work, in which RCOA is defined and rigorously assessed in a two-dimensional optimal control setting. To maintain the computational efficiency required for real-time deployment, the physical dimensions of the ego vehicle are assumed to be accounted for by geometrically inflating the obstacle boundaries. This standard reduction guarantees collision-free navigation while allowing the optimal control problem to efficiently treat the vehicle as a kinematic point-mass.}

This paper is organized as follows: section~\ref{sec: Preliminaries} introduces the preliminaries and summarizes the key technical properties of different optimization classes. Section~\ref{sec: Related Work} reviews selected works on optimization-based OA. Section~\ref{sec:RCOA} presents RCOA, together with the relevant system dynamics, path tracking equations, and any notable modifications to algorithms used to solve the resulting OCPs. Section~\ref{sec: Experiments} details the experimental simulations along with results and discussion on the evaluation of RCOA. Finally, section~\ref{sec: Conclusion} offers concluding remarks.

\section{Preliminaries} \label{sec: Preliminaries}

\subsection{Class of Optimization and Their Properties}\label{sec:Algorithms}
Optimization problems can generally be classified into three categories: convex, nonconvex, and mixed integer. Although mixed integer problems are nonconvex, their discrete structure requires fundamentally different algorithmic strategies. This section summarizes the guarantees, convergence rates, and iteration bounds associated with each class. These properties directly influence the performance of OA formulations, and they highlight the inherent computational advantages of convex OCP. 

\subsubsection{Convex Programming Algorithms}
Convex programming (CP) offers the strongest theoretical guarantees. Any local optimum is globally optimal \cite{Boyd2014ConvexOptimization}. Under standard assumptions, these methods are proven to converge, often achieve superlinear or even quadratic convergence rates \cite{Ralph1996SuperlinearInequalities}.  In addition, theoretical results establish explicit upper bounds on the number of iterations required \cite{Boyd2014ConvexOptimization}. This applies to widely used algorithms, such as interior point methods  (barrier and primal-dual variants) \cite{Gondzio2012InteriorLater}.

These properties make convex OA formulations especially attractive: they provide predictable convergence behavior, fast rates, and bounded iteration counts, key requirements for real‑time applications.

\subsubsection{Nonconvex Programming Algorithms}
Nonconvex programming introduces significant challenges \cite{Nocedal2006NumericalOptimization}, but established algorithms such as interior point methods and Sequential Quadratic Programming (SQP), remain effective. Under suitable assumptions, both achieve local convergence \cite{Wachter2005LineConvergence, Ulbrich2004AProgramming}, with the potential for superlinear convergence \cite{Byrd1997OnProgramming}. Robustness enhancements, such as filter methods and the feasibility restoration phases, improve reliability but increase computational cost. Additional assumptions on problem structure (e.g., Section 3.1 of \cite{Wachter2005LineConvergence}) apply. 

SQP is particularly appealing because it repeatedly solves convex subproblems, leveraging efficient solvers while incorporating modifications needed for nonconvexity \cite{Gill2012SequentialMethods}. A related subclass, Successive Convex Programming (SSCP), addresses nonconvexity by solving a sequence of convex approximations. In this work, a specific SSCP variant is adopted, Successive Convexification (SCvx) \cite{Mao2016SuccessiveProperties}, which is later extended to mixed integer programming. This connection between nonconvex algorithms and convex subproblems further underscores the value of convex OA formulations.

\subsubsection{Mixed Integer Programming Algorithms} \label{subsec: Algorithms,MIP}

Mixed Integer Programming (MIP) algorithms includes both the MILP and MINLP variants. Modern MILP solvers rely on \textit{branch-and-cut} methods, combining branch-and-bound and cutting-plane techniques, supplemented by heuristics to accelerate convergence \cite{Conforti2014IntegerProgramming}. Global convergence is guaranteed \cite{Conforti2014IntegerProgramming}, but worst-case complexity remains NP-hard. For binary variables, the worst-case bound number of branch-and-bound nodes is \( 2^n \) \cite{Balakrishnan1991BranchSystems}, illustrating the exponential scaling.

For MINLPs, \cite{Lee2011MixedProgramming} provides a comprehensive review of algorithms developed for both convex and nonconvex problems. Convex MINLPs inherit many of the favorable properties of MILPs. Lemma 1.3 in \cite{Conforti2014IntegerProgramming} can be extended to convex MINLPs \cite{DelPia2012OnProgramming} and the finite convergence of branch-and-bound holds under certain assumptions \cite{Balakrishnan1991BranchSystems, Belotti2013Mixed-integerOptimization}. Algorithms for MILP are adapted for MINLP, for instance, NLP-based branch-and-bound (NLP-BB) algorithms \cite{Gupta1985BranchProgramming, Quesada1992AnProblems}. 

Nonconvex MINLPs pose significantly greater challenges. Global optimality is rarely guaranteed, and solvers typically return only locally optimal or $\varepsilon$-optimal solutions, assuming they converge at all \cite{Liberti2008IntroductionOptimization, Bonami2008AnPrograms, Byrd2006Knitro:Optimization}. 

Hybrid methods that combine MILP techniques with SQP or SSCP strategies \cite{Leyffer2001IntegratingProgramming, Exler2012AOptimization, Pratt1984OptimisationSystems, GonzalezRueda2019ANetworks} have been proposed to mitigate MINLP challenges.  In Section \ref{subsubsec:sequentialAlgorithms}, the SCvx algorithm is extended to address nonconvex MINLPs, resulting in a \textit{Successive Mixed Integer Linear Programming (SMILP)} approach. Despite these advances, MIP algorithms generally remain at a disadvantage relative to convex and nonconvex approaches due to their nondeterministic convergence rates and potentially large iteration counts, further reinforcing the computational benefits of convex OA formulations.

\section{Related Work} \label{sec: Related Work}
\subsection{Optimization-Based Obstacle Avoidance} \label{subsec: Literature_OB ObsAvoid}
The study of OA within optimization can be traced back to the reach-avoid games of the 1960s \cite{RufusIsaacs1965DifferentialWiley}, in which a pursuer sought to intercept an evader whose objective was to reach a designated target while avoiding capture. Research in OA advanced gradually, and by the 1980s, one of the first optimization-based formulations was introduced \cite{Gilbert1985}. This early approach, grounded in distance measures, was inherently nonconvex and highlighted both the potential and the computational challenges of optimization-driven OA. As such, the most fundamental OA formulations are based on minimum distance constraints \cite{Gilbert1985, Szmuk2019Real-TimeConstraints, TanAnVehicles}. 

Modern approaches define obstacles as spatial regions with specific geometric shapes, such as rectangles or ellipses \cite{Schouwenaars2001MixedPlanning, Chen2017ConstrainedPlanning, Ioan2020Mixed-integerPlanning}. Others characterize obstacle-free regions instead \cite{Tordesillas2022FASTER:Environments, DeitsRobin2015ComputingProgramming}. Current research trends integrate both strategies, defining obstacles as spatial regions while simultaneously enforcing a minimum distance constraint \cite{Zhang2021Optimization-BasedAvoidance, Thirugnanam2022}. Here, the focus is placed on fundamental formulations that exemplify the core methodologies most relevant to RCOA. These representative formulations are deliberately simple, yet they often yield the best computational efficiency within their respective optimization classes.

\deleted{ A common starting point is to reduce the controlled system to a point mass and represent each obstacle as a bounded region in space \cite{Schouwenaars2001MixedPlanning}.} \added{The methods are strictly point-to-obstacle, although robust body-to-obstacle OA formulation are preferred, they often require bi-level optimization \cite{Tracy2023} or result in significant computational expense \cite{Zhang2021Optimization-BasedAvoidance}}. Two of the reviewed formulations are included as benchmarks for RCOA. 

\added{In \cite{Schouwenaars2001MixedPlanning}, the obstacle is enclosed by a rectangular region defined by its lower-left and upper-right vertices \(\boldsymbol{v}_{\text{ll}}=(x^o_{\text{min}},y^o_{\text{min}})\) and \(\boldsymbol{v}_{\text{ur}}=(x^o_{\text{max}},y^o_{\text{max}})\). The obstacle set is expressed as $\mathcal{O}=\{(x,y) : \boldsymbol{v}_{\text{ll}}\preccurlyeq (x,y) \preccurlyeq \boldsymbol{v}_{\text{ur}}\}$. A valid obstacle-free trajectory must satisfy the logical disjunction that the vehicle is positioned strictly to the left, right, below, or above these boundaries at all times. To integrate these logical conditions into an optimization framework, the well-established big-M method \cite{Bemporad1999ControlConstraints} is employed, yielding the following inequalities:}
\begin{subequations}\label{grpobs1}
\begin{align}
    &x\leqslant x^o_{\text{min}} + M_1\gamma_{I,1} \label{grpobsa}\\
    -&x\leqslant -x^o_{\text{max}} + M_2\gamma_{I,2} \label{grpobsb}\\
    &y\leqslant y^o_{\text{min}} + M_3\gamma_{I,3} \label{grpobsc}\\
    -&y\leqslant -y^o_{\text{max}} + M_4\gamma_{I,4} \label{grpobsd}\\
    &\sum_{i=1}^{4}\gamma_{I,i}\leqslant{3},\quad\gamma_{I,i}\in\{0,1\} \label{grpobse}
\end{align}
\end{subequations}

The big-M parameters \(M_1\) through \(M_4\) ensure that inequality constraints \eqref{grpobsa} through \eqref{grpobsd} remain feasible throughout the prediction horizon window. The binary variables \(\gamma_I \in \mathbb{Z}\) act as logical switches, taking values of either ``on'' (1) or ``off'' (0). \added{The summation constraint ensures that at least one logical condition is active ($\gamma_i = 0$).} \added{Directly enforcing all four boundaries in a continuous solver creates an infeasible intersection, as the vehicle cannot simultaneously exist outside all sides of a finite volume. Therefore, exact obstacle avoidance strictly requires a disjunctive formulation, necessitating auxiliary integer variables to establish valid feasible spaces without introducing severe continuous nonconvexities.}

The resulting OCP is inherently nonconvex and belongs to the class of MILP, provided that all remaining functions are linear. This formulation can be easily extended to polyhedra obstacles \cite{Alrifaee2014CentralizedVehicles}. Hence, this formulation extends the constraints defined in \eqref{grpobs1} to accommodate polyhedral regions. In general, while OA constraints are enforced at each temporal node $t_k$, obstacle intersections can potentially occur in-between successive nodes $t_k$ and $t_{k+1}$. A solution is provided in \cite{Maia2009OnConstraints} and \cite{Richards2015Inter-sampleProgramming}, which propose refinements that address the discrete-time nature of the resulting OCP. Their work provides an extension to the formulations of equations \eqref{grpobs1}, guaranteeing inter-sample OA.

Over the past two decades, various nonconvex OA formulations have been developed without binary variables. \added{For example, a prominent formulation \cite{Chen2017ConstrainedPlanning} bounds obstacles within ellipsoids. Given an obstacle set $\mathcal{O}=\{y\in \mathbb{R}^2:\:(y-\text{c})^T \mathrm{P} (y-\text{c}) < 1\}$, where $\text{c}$ is the center and $\mathrm{P}$ is a positive definite shape matrix, the following continuous hard constraint is imposed:}
\begin{equation}
    1-(y-\text{c})^T \mathrm{P} (y-\text{c}) \leqslant\:0 \label{obsellipc}
\end{equation}
\noindent This constraint ensures that $y$ remains outside the obstacle. Constraints \eqref{grpobs1} and \eqref{obsellipc} are commonly referred to as \textit{hard constraints}, which means that the domain defined by these constraints is strictly enforced. 

In contrast, some formulations introduce \textit{soft constraints}, where constraint violations are penalized by augmenting the original cost function with additional penalty terms. For example, in \cite{Sathya2018EmbeddedPANOC}, an obstacle defined by continuous boundary functions $\mathcal{O}=\{y\in \mathbb{R}^2 : h_i(y)>0,\:i=1,\dots,m\}$ is transformed completely into a penalty formulation:
    \begin{equation}
        \psi(y) = \prod_{i=1}^m[h_i(y)]_+=0 \label{obsProdEq}
    \end{equation}    
where the operator $[*]_+$ is defined as $\max\{*,0\}$, ensuring that violations of obstacle constraints contribute positively. This generalized formulation can accommodate various obstacle definitions, including the rectangular and ellipsoidal regions above. Since \eqref{obsProdEq} is a nonconvex equality constraint, the authors, motivated by the \textit{quadratic penalty method} \cite{Bertsekas1999NonlinearProgramming}, replaced the constraint by augmenting the cost function with the following penalty function, $\Tilde{\psi}=\frac{1}{2}\,\psi(y)^2$. By incorporating $\Tilde{\psi}$ into the cost function, obstacle violations are minimized rather than strictly enforced, resulting in a smooth optimization framework where the lower bound of the penalty term is zero by definition. Hence, theoretically equivalent to the hard constrained problem.

\section{Methodology} \label{sec:RCOA}
\subsection{Relaxed Convex Obstacle Avoidance Formulation}
This section derives the relaxed convex OA formulation. This approach is inspired by the incorporation of logic into optimization used in MIP \cite{Bemporad1999ControlConstraints}, particularly that of \cite{Schouwenaars2001MixedPlanning}. 
\added{These MIP formulations are attractive because they are generally linear, with the exception of integral variables.} 
In traditional MI\added{L}P formulations, logical conditional statements are enforced via binary variables and solved using branch-and-bound, which entails solving many relaxed Linear Program (LP) subproblems to explore a combinatorial search tree. The goal of RCOA is to bypass this combinatorial search by embedding relaxed binary behavior directly into a convex optimization framework. 

\added{First, a specific MILP OA formulation is presented, specifically developed to function under relaxation or equivalence. To improve numerical performance,} rather than logically enforcing the entire domain of the obstacle, the domain of the obstacle is split into two independent regions (e.g., "above" vs. "below"). In each region, the obstacle domain is enforced using a conditional statement that is formulated as a convex constraint. This replaces an exponential branch-and-bound tree with one or two convex \added{or nonconvex} problems per obstacle. By doing so, RCOA retains the essential logical structure of MILP-based OA while benefiting from efficiency, scalability, and robustness inherent to convex optimization.  

\added{It's standard in robotics to assign a coordinate system to every object, and since the object of concern is rectangular, the obstacle is modeled as an axis-aligned bounding box}\deleted{\eqref{Obs1}}. The core logic underpinning this formulation is:
\begin{equation}\label{RCOAifstatement1}
if(x^o_{\text{min}} \leqslant X \leqslant x^o_{\text{max}}) \rightarrow \,Y \geqslant y^o_{\text{max}}
\end{equation}
\begin{center}
\textit{or}
\end{center}
\begin{equation}\label{RCOAifstatement2}
    if(x^o_{\text{min}} \leqslant X \leqslant x^o_{\text{max}}) \rightarrow \,Y \leqslant y^o_{\text{min}} 
\end{equation}
\noindent where, $(X,Y)$ denote position of the vehicle.

These logical conditional statements imply that if the vehicle lies between the vertical boundaries of an obstacle, defined by $x^o_{\min}$ and $x^o_{\max}$, it must either be above the top boundary ($Y \geqslant y^o_{\max}$) or below the bottom boundary ($Y \leqslant y^o_{\min}$). Only one of these conditions needs to be enforced per problem instance, which enables the decomposition of the overall problem into two independent subproblems, which can be solved in parallel.

By applying the big-M method \cite{Bemporad1999ControlConstraints}, where \(M\in R_+\), the logical condition statement can be encoded using binary variables \added{$\gamma_{I_i}\in \mathbb{Z}$} as follows:
\begin{subequations} \label{convexCons1}
\begin{align}
-X &\leq -x^o_{\text{min}} + M_1 \gamma_{I_1} \label{convexCons1a} \\
X &\leq x^o_{\text{max}} + M_2 \gamma_{I_2} \label{convexCons1b} \\
Y &\geq y^o_{\text{max}} - M_3(\gamma_{I_1} + \gamma_{I_2}) \label{convexCons1c} \\
\gamma_{I_1} + \gamma_{I_2} &\leq 1,\quad \gamma_{I_i} \in \{0,1\} \label{convexCons1d}
\end{align}
\end{subequations}
When $\gamma_{I_1} = \gamma_{I_2} = 0$, the conditional statement is satisfied, the vehicle must satisfy $Y \geqslant y^o_{\text{max}}$ in this case. If either $\gamma_{I_1}$ or $\gamma_{I_2}$ is set to one, the constraints are trivially satisfied due to the large terms $M$, indicating the conditional statement is not satisfied. The inequality in \eqref{convexCons1d} ensures that at most one of $\gamma_{I_{1,2}}$ is nonzero. \added{Importantly, note that integral variables are necessary to strictly enforce the domain of the obstacle. Rational values of $\gamma$ values open the domain of the obstacle. As presented, the OA falls into the class of MILP.}

To encode the complementary scenario of \eqref{RCOAifstatement2}, the constraint \eqref{convexCons1c} is replaced with:
\begin{equation}
    Y \leqslant y^o_{\text{min}} + M_3 (\gamma_{I,1} + \gamma_{I,2}) \label{convexCons2c} 
\end{equation}
\added{Unlike the OA formulation in \eqref{grpobs1}, \eqref{convexCons1} can be transformed into a continuous nonconvex formulation without introducing spurious local optima via an exact penalty function \cite{Zhang1999,Ge1989,Murray2010}. Stated here without proof, this relaxation bounds the variables to $\gamma_i \in [0,1]$ and augments the cost function as follows:}
\begin{equation} \label{eq: RNCOA equi f0}
    \myadd{f_{\text{obs}} = w\sum_i\gamma_i(1-\gamma_i)}
\end{equation}

\myadd{To capture the entire domain surrounding the obstacle—simultaneously applying \eqref{convexCons1c} and \eqref{convexCons2c}—a third variable $\gamma_3 \in [0,1]$ is introduced as a logical switch. It is added to the $(\gamma_1+\gamma_2)$ summation in the first constraint, and as $(1-\gamma_3)$ in the second. However, dividing the domain into independent subproblems generally offers better computational efficiency.}

To obtain a convex version of this formulation, \myadd{and further improve computational efficiency, along with relaxation of integral constraints, instead of \eqref{eq: RNCOA equi f0}, the following penalty equation is applied}:
\begin{equation}
f_{\text{obs}} = w(\gamma_1 + \gamma_2) \label{penfunc}
\end{equation}
Here, $w$ is a weight parameter that penalizes violations of the conditional OA statement. For example, if the vehicle lies between the vertical boundaries, then $\gamma_i\rightarrow 0$ as $w\gg  0$. This converts hard constraints into \textit{soft constraints}, allowing the problem to remain convex while still achieving the desired response. \myadd{As a consequence of \eqref{penfunc}, the summation constraint in \eqref{convexCons1d} serves to strictly define the feasible space.}

The feasible set of a convex optimization problem is itself convex \cite{Boyd2014ConvexOptimization}. After relaxation of the integral constraint, all inequality constraints in \eqref{convexCons1} become convex, as they are linear functions of the variables \((X,Y,\gamma)\). The domain defined by the intersection of these constraints is convex. To show why the resulting feasible set is convex, constraints \eqref{convexCons1a}, \eqref{convexCons1c}, and the relaxed integral constraint \eqref{convexCons1d} can be restated as:
\begin{subequations}
    \begin{align}
        X &\geqslant x^o_{\text{min}} -M_1\gamma_1 \label{convexCons1av2} \\
        X &\leqslant x^o_{\text{max}} + M_2\gamma_2  \label{convexCons1bv2}  \\
        Y &\geqslant y^o_{\text{max}} - M_3(\gamma_1+\gamma_2)   \label{convexCons1cv2} \\
        0 &\leqslant \gamma_i \leqslant 1 \label{convexCons1dv2}
    \end{align}
\end{subequations}

\added{Constraint \eqref{convexCons1av2} establishes a lower bound on the vehicle's position along the $X$-axis, $X_{lb}=x_{\min}^o-M_1\gamma_1$, which reaches its minimum when $\gamma_1=1$. This constraint is affine because it forms a linear inequality in the variables $X$ and $\gamma_1$. Furthermore, since $\gamma_1\in [0,1]$, the term $M_1\gamma_1$ maps to the convex set $[0,M_1]$. Similarly, constraint \eqref{convexCons1bv2} establishes an upper bound on the vehicle's position along the $X$-axis, $X_{ub}=x_{\max}^o+M_2\gamma_2$, which reaches its maximum when $\gamma_2=1$. This constraint is also affine, and $M_2\gamma_2$ defines the convex set $[0,M_2]$. To verify that these two bounds form a valid closed interval, note that the physical dimensions of the obstacle necessitate $x_{\min}^o<x_{\max}^o$. Because $M_1,\,M_2\in \mathbb{R}_{+}$ and $\gamma_1,\,\gamma_2 \geqslant 0$, it strictly follows that $x_{\min}^o-M_1\gamma_1 <x_{\max}^o+ M_2\gamma_2$. Therefore, $X_{lb}\leqslant X \leqslant X_{ub}$ defines a valid convex set (specifically, a non-empty closed interval) for the feasible spatial domain of $X$.}

An analogous argument applies to the \(Y\)-axis. Constraint \eqref{convexCons1cv2} defines a lower bound \((Y_{\text{lb}})\) on the distance between the vehicle and the top boundary of the obstacle, resulting in the convex domain \(Y_{\text{lb}}\leqslant Y \leqslant\infty\). Finally, the relaxed binary variables and associated cost function are trivially convex. Therefore, the intersection of all domains in this formulation is convex. When the inequality of \eqref{convexCons2c} is applied, it introduces an upper bound. 
\begin{remark}[Convexity of RCOA] \label{Remark: RCOA is convex}
The inequalities in \eqref{convexCons1} are affine in $(X,Y,\gamma)$, and the bounds derived above show that the feasible region in both $X$ and $Y$ is a convex interval. Together with the relaxed binary variables and the affine penalty function, the feasible set is therefore convex.
\end{remark}

\added{The RCOA OA formulation is summarized by \eqref{FinalObs}. To determine whether constraint \eqref{eq: RCOA Ymax} or \eqref{eq: RCOA Ymin} should be enforced, two distinct strategies are proposed. The first strategy seeks the global optimum by systematically evaluating the spatial permutations. For an environment containing $n_{\mathrm{obs}}$ obstacles, the domain partitioning yields $2^{n_{\text{obs}}}$ candidate routing sequences. Because these subproblems are mathematically decoupled, they can be solved simultaneously via parallel computing. The subproblem that produces a dynamically feasible trajectory with the minimum deviation from the reference path is selected as the optimal solution. By leveraging parallelization, this method systematically explores multiple valid trajectories while mitigating the latency typically associated with combinatorial searches.}
\begin{subequations}\label{FinalObs}
    \begin{gather}
        f_{\text{aug}} = f_0(y_n) + f_{\text{obs}}(\gamma) \\ 
        -X \leqslant -x^o_{\text{min}} + M_1 \gamma_1  \\ 
        X \leqslant x^o_{\text{max}} + M_2 \gamma_2  \\
        \sum_{i=1}^2{\gamma_i} \leqslant 1,\quad \gamma_i \in [0,1] \\
        \text{-----------------} \nonumber  \\
        Y \geqslant y^o_{\text{max}} - M_3 (\gamma_1 + \gamma_2) \label{eq: RCOA Ymax} \\
        \textbf{ or } \nonumber \\
        Y \leqslant y^o_{\text{min}} + M_3 (\gamma_1 + \gamma_2) \label{eq: RCOA Ymin} 
    \end{gather}
\end{subequations}

\added{The second strategy, which is particularly advantageous when deploying RCOA as a real-time local trajectory generator, determines the circumvention topology a priori. In this hierarchical approach, a high-level global planner (e.g., RRT) identifies the preferred spatial corridor, which inherently prescribes the appropriate inequality constraint for each obstacle and supplies a viable initial guess to the optimizer. This structure shares conceptual similarities with the CIAO framework \cite{SCHOELS20206555}, but avoids the computational overhead of solving auxiliary optimization components. Furthermore, environmental heuristics can be easily integrated to dictate the routing logic; for example, if one side of an obstacle is densely cluttered, the formulation can be deterministically constrained to the unobstructed domain, effectively bypassing the combinatorial search entirely. }

The capability of this formulation to serve as an effective OA formulation is established in the following theorem.

\begin{theorem}[Conditional Statement Equivalence] \label{Theorem 1}
	\deleted{Applying the conditional statement of \eqref{RCOAifstatement1} or \eqref{RCOAifstatement2} to generate an obstacle-free trajectory in an optimal control problem is equivalent to \eqref{FinalObs}, in the sense that RCOA yields the same obstacle avoidance behavior as the mixed integer form of \eqref{convexCons1}, under assumptions A1-A4 below.}
	\added{Under assumptions A1–A4, the continuous RCOA formulation \eqref{FinalObs} exhibits conditional equivalence to the mixed-integer formulation \eqref{convexCons1}. Specifically, the continuous relaxation strictly enforces identical spatial avoidance boundaries (e.g., $Y \geqslant y^o_{\max}$ when $X \in [x^o_{\min}, x^o_{\max}]$) as the original discrete logical conditions at the optimal solution.}
\end{theorem}

\begin{enumerate}[label=\textup{(A\arabic*)},ref=\textup{A\arabic*}]
	\item An obstacle-free trajectory is feasible.\label{Asumption A: 1}
	\item The cost function of $f_0$ is convex and penalizes path deviation with unit weight. \label{Asumption A: 2}
	\item The system dynamics do not depend explicitly on $(X,Y)$\added{, applicable to automobile and quadrotor $(X,Y,Z)$ models.}
	\item $M_i$ are chosen sufficiently large to preserve feasibility when $\gamma_i=1$; the upper bounds on $\gamma_i$ are inactive. \label{Asumption A: 4}
\end{enumerate} 
\noindent The assumptions ensure feasibility and simplify the analysis. 

\renewcommand{\proof}{\textbf{Proof. }}
\noindent \begin{proof}
	Consider a vehicle traveling along the $X$-axis\myadd{, i.e., the reference path, }approaching a rectangular obstacle centered at the origin. The OA maneuver is obtained from the following OCP:
	\begin{equation} \tag{P1} \label{eq:P1}
		\begin{aligned}
			\min\limits_{y_n,\gamma} \,&f_0(y_n)\:+\:f_{\text{obs}}(\gamma) \\
			s.t.\quad&c_E(y_n) = 0 \\
			\text{c}_{I,1}:\quad& -X \leqslant -x^o_{\text{min}} + M_1 \gamma_1  \\ 
			\text{c}_{I,2}:\quad& X \leqslant x^o_{\text{max}} + M_2 \gamma_2  \\
			\text{c}_{I,3}:\quad& Y \geqslant y^o_{\text{max}} - M_3 (\gamma_1 + \gamma_2)  \\
			\text{c}_{I,4/5}:\quad& -\gamma_1 \leqslant 0,\;-\gamma_2 \leqslant 0 \\
			\text{c}_{I,6/7}:\quad& \gamma_1 \leqslant 1,\;\gamma_2 \leqslant 1 \\
			\myadd{\text{c}_{I,8}: }\quad& \myadd{\gamma_1 + \gamma_2-1\leqslant0 }
		\end{aligned}
	\end{equation}
	where $y_n\in \mathbb{R}^n$ includes position variables $(X,Y)$, and $c_{I,*}$ are inequality constraints with $*\in \{1,2,\dots,8\}$. Because the equality constraints $c_E(y_n)$ are affine dynamics, Assumption \ref{Asumption A: 2} and Remark~\ref{Remark: RCOA is convex} guarantee the OCP is convex. \deleted{The summation constraint in \eqref{convexCons1d} is omitted without loss of generality.}
	
	\added{Let $(\text{P}2)$ denote the optimal control problem formulated using the discrete mixed-integer constraints of \eqref{convexCons1}. Geometrically, the feasible set of $(\text{P}2)$ is restricted to the binary vertices of the continuous set defined by \eqref{eq:P1} \cite[Lemma 1.3]{Conforti2014IntegerProgramming}. If the optimal solution of \eqref{eq:P1} inherently yields $\gamma_i \in \{0,1\}$, it optimally solves the discrete problem $(\text{P}2)$ \cite{Conforti2014IntegerProgramming, Murray2010}. Because standard continuous relaxations do not generally guarantee binary convergence, conditional equivalence is proven by analyzing the optimality conditions of \eqref{eq:P1} to demonstrate that it inherently enforces this exact discrete boundary behavior.}
	
	Applying Assumption~\ref{Asumption A: 4}, the Lagrangian \added{of \eqref{eq:P1}} is:
	\begin{equation*} 
		\begin{split}
			\mathcal{L}(y_n,\lambda,v) = f_0(Y)+f_{\text{obs}}+\sum_{i \in \{1,\dots,5,\, \myadd{8}\}}\lambda_i(\text{c}_{I,i})+v^T(c_E) \nonumber 
		\end{split}
	\end{equation*}
	with dual variables $\lambda_i\geqslant 0$ and $v$. The Karush-Kuhn-Tucker (KKT) stationary conditions \cite{Boyd2014ConvexOptimization} with respect to $Y$ and $\gamma_i$ are:
	\begin{subequations} \label{eq: primal lagrange KKT}
		\begin{align}
			\mathcal{L}_Y  &= \frac{\partial f_0 }{\partial Y}-\lambda_3  =0  \label{eq: primal lagrange KKT, Y} \\
			\mathcal{L}_{\gamma_1} &= w-\lambda_1 M_1-\lambda_3 M_3 -\lambda_4\myadd{ + \lambda_8} = 0 \label{eq: primal lagrange KKT, gamma1} \\
			\mathcal{L}_{\gamma_2} &= w-\lambda_2M_2-\lambda_3 M_3 - \lambda_5\myadd{ + \lambda_8} =0 \label{eq: primal lagrange KKT, gamma2} 
		\end{align}
	\end{subequations}
	Equation \eqref{eq: primal lagrange KKT, Y} shows path deviation is driven by $\lambda_3$. By complementary slackness ($\lambda_i^* \,\text{c}_{I,i}^*=0$), for a vehicle approaching from the left, feasibility of $c_{I,1}$ requires $\gamma_1^*\neq0$, hence $\lambda_4^*=0$. Equation~\eqref{eq: primal lagrange KKT, gamma1} reduces to:
	\begin{gather} \label{eq: primal lagrange KKT gamma1 2}
		\lambda_1^* M_1+\lambda_3^* M_3\myadd{-\lambda_8^*} = w
	\end{gather}
	Since $w>0$, \eqref{eq: primal lagrange KKT gamma1 2} dictates that at most one dual variable ($\lambda_1^*$ or $\lambda_3^*$) can be zero. The respective constraints become active when:
	\begin{subequations} \label{proof: slackness cI1 and cI3}
		\begin{align}
			\text{c}_{I,1}=0\rightarrow \: \gamma_1&=(x^o_{\text{min}}-X)/M_1 \label{proof: slackness cI1 and cI3 (b)}\\
			\text{c}_{I,3}=0\rightarrow \: \gamma_1&=(y^o_{\text{max}}-Y)/M_3-\gamma_2 \label{proof: slackness cI1 and cI3 (a)}
		\end{align}
	\end{subequations}
	\vspace{-0.5cm}
	\begin{subequations} \label{proof: slackness cI2 and cI3}
		\begin{align}
			\text{c}_{I,2}=0\rightarrow \: \gamma_2&=(X-x^o_{\text{max}})/M_2 \label{proof: slackness cI2 and cI3 (b)} \\
			\text{c}_{I,3}=0\rightarrow \: \gamma_2&=(y^o_{\text{max}}-Y)/M_3-\gamma_1 \label{proof: slackness cI2 and cI3 (a)}
		\end{align}
	\end{subequations}
	
	At optimality, each $\gamma_i$ matches \added{the larger affine value required to satisfy primal feasibility}. Approaching from the left implies $\gamma_1>\gamma_2$ and \added{$\gamma_2^* \in [0,\epsilon]$}, where $\epsilon$ is an arbitrarily small positive tolerance, as the cost function minimizes $\gamma_i$. \added{Under Assumption \ref{Asumption A: 4}, the summation constraint $c_{I,8}$ remains strictly inactive ($\lambda^*_8=0$), reducing \eqref{eq: primal lagrange KKT gamma1 2} and \eqref{eq: primal lagrange KKT, gamma2} to:}
	\begin{equation} \label{proof: dual bound KKT}
		\begin{aligned}
			\myadd{\lambda_1^* M_1+\lambda_3^* M_3} &\myadd{= w}\\
			\added{\lambda_3^* M_3+\lambda_5^*} &\myadd{= w}
		\end{aligned}
	\end{equation}
	
	Comparing the two expressions for $\gamma_1$ yields three mutually exclusive cases:
	
	\textbf{Case 1 ($\gamma_1$ from $\text{c}_{I,3}$ is larger):} Here, $\lambda_1^*=0\text{ and }\lambda_3^*>0$. From \eqref{eq: primal lagrange KKT, Y}, $\partial f_0/\partial Y=\lambda_3^*>0$, driving $Y^*$ up until $Y\geqslant y^o_{max}$, $\gamma_1^*=0\text{, and }\gamma_2^*=0$, or a transition occurs. Additionally, from \eqref{proof: dual bound KKT}:
	\begin{equation} \label{proof: case 1 opt}
		\myadd{\lambda_3^* M_3=w, \quad \lambda_5^*=0}
	\end{equation}
	where $\lambda_5^*=0$ implies $\gamma_2^*=\epsilon$.
	
	\textbf{Case 2 (The expressions are equal):} Here, both $\text{c}_{I,1}$ and $\text{c}_{I,3}$ are active ($\lambda_1^*>0\text{, }\lambda_3^*>0$), \added{implying} $\partial f_0/\partial Y=\lambda_3^*>0$. \added{Substituting \eqref{eq: primal lagrange KKT, Y} into \eqref{proof: dual bound KKT} yields $\lambda_1^* M_1 + \frac{\partial f_0}{\partial Y} M_3 = w$. Because dual feasibility requires $\lambda_1^* > 0$, this establishes that the obstacle weight must satisfy $w > M_3 \frac{\partial f_0}{\partial Y}$ to successfully enforce boundary avoidance against the competing path-tracking cost.} Also, \eqref{proof: dual bound KKT} dictates that $\lambda_5^*>0 \rightarrow \gamma_2^*=0$. Equating \eqref{proof: slackness cI1 and cI3 (a)} and \eqref{proof: slackness cI1 and cI3 (b)} yields:
	\begin{gather} \label{proof: Y* optimality 1}
		Y^*=y^o_{\max}-\frac{M_3}{M_1}(x^o_{\min}-X^*)
	\end{gather}
	
	\textbf{Case 3 ($\gamma_1$ from $\text{c}_{I,1}$ is larger):} Here, $\lambda_1^*>0\text{ and }\lambda_3^*=0$, yielding $\partial f_0/\partial Y=0$, which strictly minimizes path deviation. Evaluating \eqref{proof: dual bound KKT} gives:
	\begin{equation} \label{proof: case 3}
		\myadd{\lambda_1^* M_1=w, \quad \lambda_5^*=w}
	\end{equation}
	Concurrently, $\gamma_1^*=(x_{\min}^o-X)/M_1$, which substituted into $c_{I,3}$ yields:
	\begin{equation} \label{proof: Y* optimality 2}
		Y^*> y_{\max}^o -\frac{M_3}{M_1}(x_{\min}^o-X^*)
	\end{equation}
	
    \added{Case 1 characterizes conditions where $Y^*$ falls below the spatial bound defined in \eqref{proof: Y* optimality 1}, which can arise from specific reference path geometries, initial conditions, or scenarios of near-infeasibility. Although structural or dynamic near-infeasibility is strictly eliminated by Assumption \ref{Asumption A: 1}, the tracking of demanding reference paths or initial conditions may yield Case 1 conditions. Nevertheless, minimization of the cost function ensures that a transition to Case 2 or Case 3 is achieved at the optimal solution. Because Case 2 represents an exact boundary alignment where $\partial f_0/\partial Y > 0$, the stationary conditions naturally drive a continuous transition toward Case 3.}
	
	Analyzing the maximization of the dual function $g(\lambda,v)=\mathcal{L}(y_n^*,\gamma^*,\lambda,v)$ provides additional insight. \added{Cases 1, 2, and 3 bound the active dual variables in \eqref{proof: dual bound KKT} via the obstacle penalty weight ($w$). Comparing these allocations reveals that Case 1 only maximizes $\lambda_3^*$, whereas Cases 2 and 3 allow positive values across multiple variables. Because Case 2 requires $\lambda_1^*M_1=\lambda^*_5$ from \eqref{proof: dual bound KKT}, assuming $M_i\geqslant 1$ ensures that the structure of \eqref{proof: case 3} in Case 3 yields the supreme bound that maximizes the dual objective function.}
	
	Thus, when approaching from the left, the optimal solution of \eqref{FinalObs} will generally follow \eqref{proof: Y* optimality 2}, resulting in $Y\geqslant y_{\max}^o$ when $X \in [x^o_{\min},x^o_{\max}]$, adhering to the conditional statement of \eqref{RCOAifstatement1}. By formulation symmetry, identical logic applies for right-side approaches or navigating below the obstacle \eqref{RCOAifstatement2}, completing the proof of \added{conditional} equivalence. 
\end{proof}

\added{A key attribute of the formulation, as revealed by \eqref{proof: Y* optimality 2}, is its capacity to initiate avoidance maneuvers even when the physical boundaries of the obstacle reside beyond the immediate prediction horizon. While this out-of-horizon capability improves computational tractability and facilitates real-time deployment, it is important to note that, similar to standard path tracking, overly reducing the prediction horizon can still degrade overall controller performance. Ultimately, this flexibility proves most advantageous in designing robust control architectures, as it enables the simultaneous deployment of multiple MPCs with varying prediction horizons without sacrificing early evasion capabilities.}

To extend the formulation to dynamic or irregularly shaped obstacles, RCOA can be generalized using functions  $g(\boldsymbol{z}(t))$ and $h(\boldsymbol{z}(t))$, as shown in equation~\eqref{eq: Generalized RCOA}. Here, $\boldsymbol{z}(t) \in \mathbb{R}^2$ denotes the obstacle's position as a function of time in the inertial frame. The function $g(z(t))$ specifies the obstacle's position and activates the conditional statement (i.e., $[x^o_{\text{min}},x^o_{\text{max}}]$), see section~\ref{subsec:MPC} for an example. Similarly, $h(z(t))$ defines the vertical bounding functions of the obstacle. These functions need not be convex, which preserves the generality of the formulation.
\begin{equation} \label{eq: Generalized RCOA}
    \begin{gathered}
        -X \leqslant -g(\boldsymbol{z}(t))_{\text{min}} + M_1 \gamma_1  \\ 
        X \leqslant g(\boldsymbol{z}(t))_{\text{max}} + M_2 \gamma_2  \\
            \text{-----------------}  \\
        Y \geqslant h(\boldsymbol{z}(t))_{\text{max}} - M_3 (\gamma_1 + \gamma_2)  \\
        \textbf{ or } \\
        Y \leqslant h(\boldsymbol{z}(t))_{\text{min}} + M_3 (\gamma_1 + \gamma_2)
    \end{gathered}
\end{equation}

RCOA can be simplified in scenarios where only limited obstacle data is available. In such cases, the obstacle may be approximated by a unit step representation. For example, consider the unit step function \(Au(t-a)\), where \(a\) represents a time shift and \(A\) is a scaling factor. In this context, time shift refers to the position of a vertex, either left or right, of the obstacle. The scaling factor represents the position of the top surface; \((a,A)=(x^o_{\text{min}},y^o_{\text{max}})\), under the assumption that the obstacle extends indefinitely beyond this boundary. This abstraction provides a compact representation of the obstacle as defined below.
\begin{equation*} \label{eq: unit step representation}
    \begin{gathered}
        -X \leqslant -x^o_{\text{min}} + M_1 \gamma_1  \\ 
        \quad \gamma_1 \in [0,1] \\
        \text{-----------------}  \\
        Y \geqslant y^o_{\text{max}} - M_3\gamma_1 \\
        \textbf{ or } \\
        Y \leqslant y^o_{\text{min}} + M_3\gamma_1 
    \end{gathered}
\end{equation*}
Together, these variants demonstrate that RCOA is flexible enough to handle static, dynamic, and data-limited obstacle scenarios while remaining computationally tractable for real-time applications. Moreover, the generality of this formulation makes it versatile, offering a principled framework that can be adapted to a wide range of engineering problems beyond OA.
\subsection{Limitations and Parameter Sensitivity} \label{subsec: Parameter Study}
\added{While RCOA enforces strict obstacle avoidance, its performance is sensitive to the selection of the big-$M$ parameters and the penalty weight $w$, particularly in multi-obstacle scenarios. The optimality conditions derived for Cases 1 and 3 (\eqref{proof: case 1 opt} and \eqref{proof: case 3}) explicitly highlight this parameter dependence. In Case 1, for example, the cost function's optimal condition is directly governed by $\lambda_3^*$, where a larger value induces a steeper spatial gradient. Furthermore, \eqref{proof: case 1 opt} establishes that $\lambda_3^* = w/M_3$, directly coupling the dual variable to both the penalty weight and the spatial bound. Similarly, in \eqref{proof: case 3}, a small value for $\lambda_1^*$ corresponds to a weak enforcement of case 1 optimality, as this variable dictates the minimization of the variable $\gamma_1^*$ associated with constraint $c_{I,1}$.}

\myadd{As established in \eqref{proof: Y* optimality 2}, the ratio of the big-$M$ parameters (e.g., $M_3/M_1$) defines a continuous spatial gradient that dictates when the lateral avoidance maneuver initiates. In general, it is best to maintain a relaxed bound and avoid large ratios by artificially increasing $M_1$ and $M_2$. Specifically, a smaller ratio produces a shallower gradient, forcing the optimal trajectory to initiate lateral deviation further away from the obstacle, resulting in a more gradual and conservative avoidance maneuver. Conversely, a larger ratio allows the vehicle to approach closer longitudinally before abruptly steering laterally. However, these extended, overlapping gradients can introduce artificial constraints in densely cluttered environments, artificial in the sense that they arise from the mathematical structure of the optimality conditions rather than physical dynamic infeasibility.}

\added{Consider a scenario with two consecutive static obstacles where the vehicle must navigate above Obstacle 1 and below Obstacle 2. From the proof for Theorem~\ref{Theorem 1}, the vehicle becomes subject to the extended spatial gradients of both obstacles. Assuming a uniform longitudinal parameter $M_1$ is applied across the environment, the lateral position $Y$ is bounded below by Obstacle 1 as long as $X \geqslant x_{\max,1}^o$, and bounded above by Obstacle 2 as long as $X \leqslant x_{\min,2}^o$, as shown by \eqref{eq:lim_obs1} and \eqref{eq:lim_obs2}.}

\added{When the vehicle is situated longitudinally between the obstacles (i.e., $X \in [x_{\max,1}^o, x_{\min,2}^o]$), both conditions \eqref{eq:lim_obs1} and \eqref{eq:lim_obs2} apply simultaneously. If the longitudinal gap between the obstacles is small relative to the chosen $M$ ratios, these inequalities may conflict, artificially truncating the dynamically feasible space. Consequently, parameters like $M_3$ and $M_4$ must be carefully tuned relative to the global $M_1$ to ensure a valid spatial corridor remains open when alternating constraints are applied. If these bounds conflict, or if dynamic infeasibility forces mutually exclusive variables to become simultaneously active (e.g., $\gamma_1^* > 0$ and $\gamma_2^* > 0$), a formal diagnostic mechanism is required; the subsequent section details the generation of an infeasibility certificate to restore OA safety.}
\begin{align}
    Y &\geqslant y_{\max,1}^o-\frac{M_3}{M_1}(X-x_{\max,1}^o) \label{eq:lim_obs1}\\
    Y &\leqslant y_{\min,2}^o-\frac{M_4}{M_1}(x_{\min,2}^o-X) \label{eq:lim_obs2}
\end{align}
\subsection{Restoring Certificate of Infeasibility} \label{subsec:correctionstep}
The RCOA formulation of \eqref{FinalObs} is inherently a soft constraint, which makes infeasibility difficult to verify. To address this limitation, a secondary problem can be formulated that restores infeasibility assessment, as with hard-constrained formulations.

This correction leverages the formulation's mixed integer origins, since the relaxed integral constraints allow each variable to be restricted to either zero or one. Specifically, with reference to \eqref{FinalObs}, when \(X \in [x^o_{\min}, x^o_{\max}]\) and an obstacle-free trajectory requires \(Y \geq y^o_{\max}\), it follows that \(\gamma_1 = 0\) and \(\gamma_2 = 0\). Enforcing these conditions as equality constraints yields a new convex problem that can determine infeasibility.   

To formalize this correction mechanism, consider the near-infeasible three-obstacle environment illustrated in Fig.~\ref{fig: Correction Example}, where the initial soft-constrained trajectory violates the spatial boundaries. Let $I$, $G$, and $K$ denote the sets of temporal nodes $k$ where the vehicle is longitudinally parallel to Obstacles 1, 2, and 3, respectively (i.e., where $X \in [x^o_{\min}, x^o_{\max}]$). To definitively verify infeasibility, the optimal control problem \eqref{FinalObs} is re-solved using the initial trajectory as a warm-start, converting the soft penalty into strict hard constraints by appending \eqref{ConstCorrection}.

\begin{equation}\label{ConstCorrection}
\begin{aligned}
    \gamma_1^{(k)} &= 0,\quad \gamma_2^{(k)} = 0, \quad \forall k \in I, \\
    \gamma_3^{(k)} &= 0,\quad \gamma_4^{(k)} = 0, \quad \forall k \in G, \\
    \gamma_5^{(k)} &= 0,\quad \gamma_6^{(k)} = 0, \quad \forall k \in K.
\end{aligned}
\end{equation}

Here, \((\gamma_1,\gamma_2)\) correspond to obstacle 1, \((\gamma_3,\gamma_4)\) correspond to obstacle 2, and so forth.  See section~\ref{subsubsec: Correction Setup} for a continuation \added{and configuration details} of this example.

\added{Furthermore, Figure~\ref{fig: Correction Example} explicitly demonstrates the formulation's sensitivity to the big-$M$ parameters. The initial RCOA trajectory was generated using the baseline parameters provided in Table~\ref{tbl:NLP OCP Para} for $E\Romannum{1}$, which were originally tuned for smaller obstacles. When applied to this scenario, the increased obstacle dimensions induced the conflicting spatial gradients detailed in Section~\ref{subsec: Parameter Study}. To resolve this, the parameters were updated to $M=[52,52,11,13,15]$ and $w_2=400$. While this adjustment significantly improved the trajectory by opening a valid spatial corridor, the relaxed variables associated with the first obstacle still converged to strictly positive, mutually exclusive values ($\gamma_1^*>0$ and $\gamma_2^*>0$). As previously established, this condition signals marginal dynamic infeasibility, thereby necessitating the formal correction step.}

\begin{figure}[h]
\centerline{\input{Correction_example_1_R4.pgf}}
\caption{\added{Baseline trajectory illustrating near-infeasibility in a densely cluttered environment. The trajectory is discretized into 31 temporal nodes, with even nodes explicitly labeled. The initial RCOA formulation (orange) experiences conflicting spatial gradients that artificially truncate the feasible space. Although a secondary adjustment to the $M$-parameter ratios (blue) visually opens a spatial corridor, the formulation remains mathematically near-infeasible ($\gamma_1^*>0, \gamma_2^*>0$). This directly motivates the formal correction step, which enforces strict binary constraints at the locally active nodes (e.g., $k \in \{9, 17, 27, 28\}$).}}
\label{fig: Correction Example}
\end{figure}

\added{On final note, the penalty function of \eqref{penfunc} can be replaced by \eqref{eq: RNCOA equi f0} at any point in this process (initial or correction), theoretically providing stronger conditions of optimality. }
\subsection{Vehicle Dynamics}\label{Dynamics}

\begin{figure}[!h] 
\centerline{\includegraphics[width=\columnwidth]{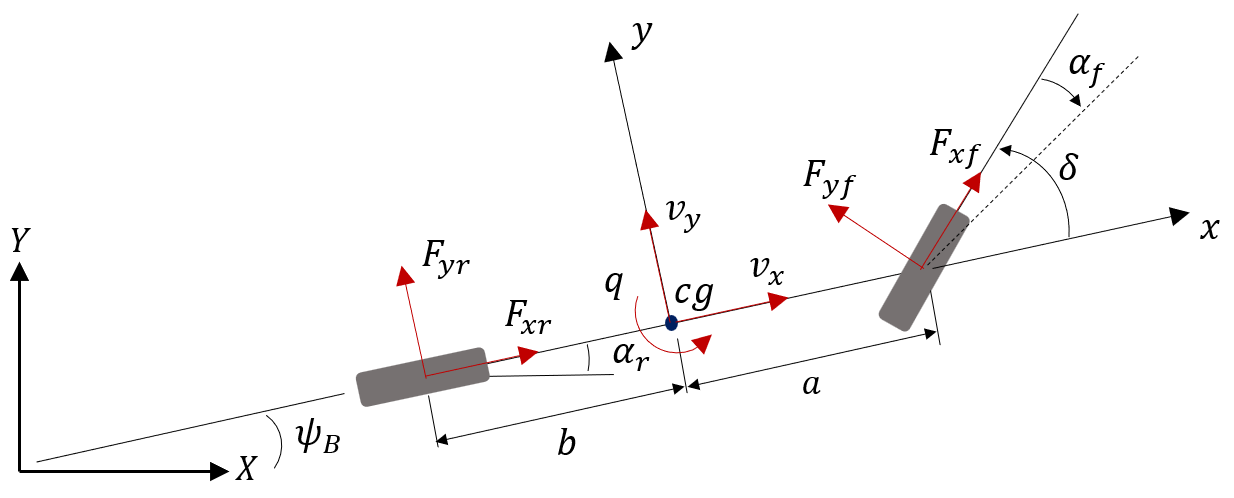}}
\caption{Free-body diagram of single-track bicycle model}
\label{fig1}
\end{figure}

\added{The rigid-body nonlinear dynamics of the vehicle are modeled using the standard single-track (bicycle) model with three degrees of freedom \cite{Jazar2018VehicleApplications, Bobier2013StayingSurface}. A free-body diagram of this model is shown in Figure~\ref{fig1}, where $v_x$ and $v_y$ are the body-frame velocities, $q$ is the yaw rate, and $\delta$ is the steering input. The front steering input is \( \delta \), \( \mathrm{F}_{ij} \) denotes the forces applied to the front (\(f\)) and rear (\(r\)) tires, and \( a \) and \( b \) are the distances from the center of mass of to the front and rear axles, respectively. The physical parameters: mass ($m$), yaw moment of inertia ($\mathrm{I}_z$), and distances from the center of mass to the front and rear axles ($a$ and $b$).} 

To accurately capture dynamic behavior during aggressive evasive maneuvers, the longitudinal ($\mathrm{F}_{x}$) and lateral ($\mathrm{F}_{y}$) tire forces are computed utilizing the nonlinear brush model \cite{Pacejka2012TireDynamics}, which accounts for both pure lateral and combined longitudinal/lateral slip. The brush model for pure lateral slip is given below.
\begin{align}
    |\alpha|\leqslant\alpha_{\text{sl}},\;\mathrm{F}_y &= 3\mu \mathrm{F}_z\theta_y\sigma_y\{1 - |\theta_y\sigma_y| + \frac{1}{3}(\theta_y\sigma_y)^2\} \label{LslipForce1} \\
    |\alpha|>\alpha_{\text{sl}},\; \mathrm{F}_y &= \mu \mathrm{F}_z\text{sign}(\alpha) 
\end{align}
\noindent where \( \alpha \) is the side slip angle, \( \sigma_y = \tan{\alpha} \), \( \mathrm{F}_z \) is the normal force, \( \mu \) is the friction coefficient, and \( \theta_y = \nicefrac{C_{\text{F}\alpha}}{3\mu \mathrm{F}_z} = \sec{\alpha_{\text{sl}}} \). \( C_{\text{F}\alpha} \) is the lateral stiffness at zero side slip, and \( \alpha_{\text{sl}} \) is the side slip angle limit where pure sliding begins.

For combined longitudinal and lateral slip, assuming isotropic stiffness \( (\theta_x = \theta_y = \theta) \), the brush model becomes:
\begin{equation}\label{combined_slipEq}
    \begin{aligned}
        \mathbf{F} &= \mathrm{F} \frac{\boldsymbol{\sigma}}{\sigma} \\
        \sigma \leq \sigma_{\mathrm{sl}}:\quad \mathrm{F} &= \mu \mathrm{F}_z\left(3\theta \sigma - 3(\theta \sigma)^2 + (\theta \sigma)^3\right) \\
        \sigma > \sigma_{\mathrm{sl}}:\quad \mathrm{F} &= \mu \mathrm{F}_z  \\
        \boldsymbol{\sigma} &= (\sigma_x, \sigma_y) = \left(\frac{1}{1+\kappa}, \frac{\tan{\alpha}}{1+\kappa}\right) 
    \end{aligned}
\end{equation}
where \( \kappa \) is the longitudinal slip ratio, \deleted{ see \eqref{longslip},} $\boldsymbol{\sigma}$ is the theoretical slip vector, $ \sigma = \|\boldsymbol{\sigma}\| $, and $ \sigma_{\text{sl}} = 1/\theta$ is the sliding threshold.

Local side slip angles for a front-steered, front-driven vehicle are:
\begin{equation} \label{sideslipangs}
    \begin{aligned}
        \alpha_f &= \tan^{-1}\left(\frac{v_y + a q}{v_x}\right) - \delta \\
        \alpha_r &= \tan^{-1}\left(\frac{v_y - b q}{v_x}\right) 
    \end{aligned}
\end{equation}


To evaluate solver performance across differing degrees of nonconvexity, a corresponding linearized state-space model, $\boldsymbol{\dot{x}}_L = \text{A}\boldsymbol{x}_L + \text{B}\delta$, where, \( \boldsymbol{x}_L = (v_y, q, \psi_B) \), is also utilized. This linear variant is derived via standard small-angle approximations and constant longitudinal velocity assumptions, where $\text{A}$ and $\text{B}$ represent the Jacobian state and input matrices evaluated with linear tire stiffness $C_{\mathrm{F}\alpha, i}$ \cite{Jazar2018VehicleApplications}.

\subsection{Path Tracking} \label{subsec: LR, Contouring Control}
\begin{figure}[h!]
\centerline{\includegraphics[height=0.45\columnwidth]{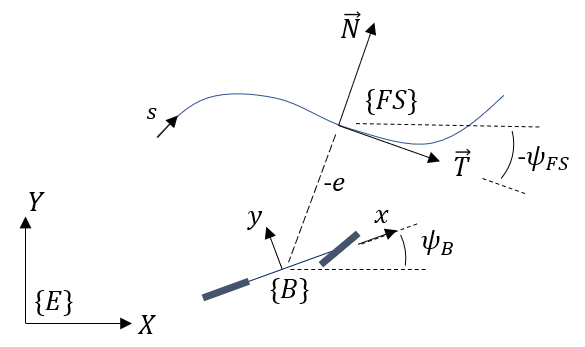}}
\caption{Path Tracking Kinematics: \{\textbf{E}\} is the inertial frame, \{\textbf{FS}\} is the FrenetSerret frame along the path \( s(t) \), and \{\textbf{B}\} is the vehicle body-fixed frame.}
\label{fig: Frenet-Serret}
\end{figure}

A path tracking (PT) formulation based on the Frenet-Serret (TNB) frame \cite{Micaelli1994TrajectoryRobots, Skjetne2001NonlinearShips}, is adopted and illustrated in Figure~\ref{fig: Frenet-Serret}. The Frenet-Serret PT formulation tracks three quantities: arc length or position \( s(t) \) along the path, lateral deviation \( e(t) \), and the heading error \( \Bar{\psi}(t) \) that is defined as the angle between the vehicle's body frame and the path tangent (\(\psi_B - \psi_{\text{FS}}\)).

The PT dynamics relate the curvature and kinematics properties of the path, expressed in the Frenet frame \{FS\}, to the vehicle kinematics in its local frame \{B\}. This method is formulated such that the vehicle is always located along the Normal vector \( (\Vec{N} \)), so the path deviation is lateral to the Tangential vector (\( \Vec{T} \)).

The PT dynamics are defined by \eqref{eq: error dynamics}, these first-order equations describe how the vehicle progresses along the path \( s(t) \), deviates laterally \( e(t) \), and heading error \( \Bar{\psi}(t) \) in terms of \( (v_x, v_y, \dot{\psi}_B) \). The PT state vector is defined as \( \boldsymbol{e} = (s, e, \Bar{\psi})\), $\boldsymbol{p}=(v_x,v_y,\dot{\psi}_B)$, and the path curvature $\kappa_{\text{FS}}(s)$ defines the reference path.

\begin{equation} 
    \begin{aligned}
        \mathrm{FS}(e, \kappa_{\text{FS}}) \dot{\boldsymbol{e}} &= \mathrm{R}(\Bar{\psi}) \boldsymbol{p} \\
        \mathrm{FS}(e, \kappa_{\text{FS}}) &=
        \begin{bmatrix}
            1 - e \kappa_{\text{FS}} & 0 & 0 \\
            0 & 1 & 0 \\
            \kappa_{\text{FS}} & 0 & 1
        \end{bmatrix} \\ 
                \mathrm{R}(\Bar{\psi}) &= 
        \begin{bmatrix}
            \cos{\Bar{\psi}} & -\sin{\Bar{\psi}} & 0\\
             \sin{\Bar{\psi}} & \cos{\Bar{\psi}}  & 0\\
             0 & 0 & 1
        \end{bmatrix} \label{eq: error dynamics}
    \end{aligned}
\end{equation}

\subsection{Solvers and Algorithms} \label{subsection:solvers}
\added{The continuous nonconvex programming problems in Table~\ref{tbl:case1formualtions} are solved using FATROP~\cite{Vanroye2023FATROPControl} and the SCvx algorithm (utilizing GUROBI~\cite{GurobiOptimization2016GurobiOptimizer}). The MINLP cases are handled using a two-phase hybrid scheme, detailed below.}

\subsubsection{Successive Algorithms} \label{subsubsec:sequentialAlgorithms}
\added{The SCvx algorithm solves a sequence of convexified elastic subproblems subject to a trust region \cite{Gill2012SequentialMethods}. To prevent premature stalling and mitigate phantom solutions \cite{Gill2012SequentialMethods} caused by slack variables \cite{Fletcher2002NonlinearFunction}, the standard algorithm was modified to use absolute values for predicted cost reductions and to dynamically expand the trust region upon repeated rejected steps.}

\added{The algorithm terminates using the state and control difference criterion from \cite{Malyuta2022ConvexEfficiently}:}
\begin{equation}
    \left\lVert p^* - p \right\rVert_{\hat{q}} + \max_{k \in \{1, \dots, N\}} \left\lVert y^*_k - \bar{y}_k \right\rVert_{\hat{q}} \leqslant \varepsilon \label{eq: SCvx exit criteria 2}
\end{equation}
\added{with $\varepsilon=0.02$. This formulation was utilized because traditional cost-based exit criteria proved numerically unstable under large penalty weights.}

\subsubsection{\added{SMILP and Hybrid MINLP Algorithms}}\label{subsubsec: Hybrid Scheme}
\added{Sequential Mixed-Integer Linear Programming (SMILP) operates as a standalone solver that preserves integrality constraints while linearizing continuous non-convexities identically to SCvx. Because direct MINLP solvers failed to converge within practical timeframes (72 hours), a two-phase hybrid algorithm is employed for MINLP problems. Phase 1 utilizes SMILP to extract a feasible binary sequence. Phase 2 fixes these binary variables, reducing the problem to a continuous NLP that is rapidly refined to optimality using FATROP, efficiently overcoming the relaxed tolerance limitations inherent to the standalone SMILP.}

\section{Experimental Simulations} \label{sec: Experiments}
This section evaluates the performance of RCOA via the selected simulations. The first evaluation involves implementing an open-loop OCP to generate a trajectory in which the vehicle navigates through a cluttered environment of static obstacles. RCOA is benchmarked against a nonconvex and a mixed integer OA formulation. Second, one of the key limitations of this formulation is addressed: the capability to determine the infeasibility of an obstacle-free trajectory. The third evaluation considers a single dynamic obstacle, while implementing an NMPC controller with the RCOA formulation embedded. 

\subsection{Cluttered Environments: Open-Loop RCOA}\label{subsec: Cluttered Environment}
\subsubsection{Definition of Environment and Evaluation }
The RCOA formulation is compared against the OA formulations of \eqref{grpobs1} and \eqref{obsellipc}, which are referred to as the Mixed Integer Obstacle Avoidance (MIOA) and Elliptical Obstacle Avoidance (EOA), respectively. The evaluation includes two distinct environments designed to induce zigzag maneuvers. This pattern occurs because the reference path is purely horizontal, and the minimization of the PT dynamics of \eqref{eq: error dynamics} enforces the vehicle to navigate alternately above and below obstacles. This setup specifically challenges "soft" constraints, as the obstacle penalty competes directly with tracking error minimization.

Environment I ($E_{\Romannum{1}}$) features tall, narrow obstacles, while Environment II ($E_{\Romannum{2}}$) uses short, wide geometries. Rectangular representations serve \eqref{grpobs1} and \eqref{convexCons1}, with the vertices defined in Table~\ref{tbl:obsvert}. While elliptical representations \eqref{obsellipc} use major and minor radii equal to half the rectangular dimensions. \deleted{Note that this configuration results in a conservative assessment for RCOA, especially for $E_{\Romannum{2}}$, where the resulting trajectory is far more conservative.} \added{Because the rectangular bounding box strictly contains the ellipse, this configuration inherently forces RCOA to solve a more conservative and spatially restrictive avoidance problem, particularly in $E_{\Romannum{2}}$.}

To reduce the complexity of the OCPs, the following assumptions are made:
\begin{enumerate}[label=\textup{(B\arabic*)},ref=\textup{B\arabic*}] 
    \item For the nonlinear equations of motion (EOM): \label{Assumptions B: 1} 
    \begin{enumerate}[label=\roman*., ref=\roman*]
        \item The vehicle is always free rolling: $\kappa_{f}=\kappa_{r}=0$
        \item Small angle approximation is applied to the local side slip angles of equations \eqref{sideslipangs}: $\tan{(*)}\approx (*)$
    \end{enumerate}
    \item Weight distribution is assumed constant over the prediction horizon (no dynamic load transfer). \label{Assumptions B: 2} 
    \item Linear EOM: Longitudinal speed is constant, $v_x=U$ 
\end{enumerate}
\begin{table}[h]
    \centering
    \caption{Obstacle vertices in inertial frame, units of (m)}
    \renewcommand{\arraystretch}{1.25}
    \begin{tabular}{|c|c|c|c|}
        \hline
         & Obstacle& $(x^o_{\text{min}},y^o_{\text{min}})$& $(x^o_{\text{max}},y^o_{\text{max}}) $\\
        \hline
         \multirow{3}{*}{\textit{E\Romannum{1}}} & 1& (-1,-4)& (1,1.25) \\
          & 2& (11,0)& (13,8) \\
          & 3& (25,-4)& (27,1.75) \\
          \hline
         \multirow{2}{*}{\textit{E\Romannum{2}}} & 1& (-5,-2)& (5,1.5) \\
          & 2& (20,-0.5)& (27,3)  \\
          \hline
          \multirow{3}{*}{\textit{E\Romannum{3}}$^*$} & 1 & $(-1, -4)$ & $(1, 1.75)$ \\
            & 2 & $(11, 0)$ & $(13, 8)$ \\
            & 3 & $(25, -4)$ & $(27, 1.75)$ \\
        \hline
        \multicolumn{4}{l}{*Applied in section~\ref{subsubsec: Correction Setup}}
    \end{tabular}
    \label{tbl:obsvert}
\end{table}

\paragraph{Control Problem formulation} \label{parag: cluttered env OCP formulation}
The OCP with the nonlinear dynamics ($P_\Romannum{1}$) is defined in \eqref{sim1_con}, with the full state vector defined as $\boldsymbol{x} = (v_x, v_y, q, X, Y, \psi_B)$. Lateral tire forces $(\mathrm{F}_{yf}, \mathrm{F}_{yr})$ are governed by \eqref{LslipForce1}, and applicable vehicle parameters are listed in Table~\ref{tbl: Vehicle Parameters}. Safety constraints $|\alpha| \leq \alpha_{\text{sl}}$ are enforced to prevent drifting.

The objective function of \eqref{sim1_obj} consists of the $L_1$ norm of the path error, which simplifies to $e=Y$, and the RCOA penalty function of \eqref{penfunc}. The OA constraints of \eqref{sim1_con:f} through \eqref{sim1_con:l} represent the constraints of \eqref{FinalObs} applied to three obstacles defined by the vertices of Table \ref{tbl:obsvert}. \added{The NLP OCP parameters $(w_i,M_i)$ utilized are provided in Table~\ref{tbl:NLP OCP Para}.} The avoidance direction (above or below) is determined by the minimum of independent subproblems; here, the most intuitive reaction for minimum deviation from the reference is selected for simplicity.
\begin{subequations} \label{sim1_con}
    \begin{align}
        \min\limits_{x,\gamma}\; &w_1\lVert Y \rVert_1  + w_2\lVert \gamma_i \rVert_1 \label{sim1_obj}\\ 
        \textbf{s.t. } &\dot{\textit{\textbf{x}}} = f(\textit{\textbf{x}},\delta)  \label{sim1_con:b} \\
        & \lvert\delta\rvert \leqslant 35\deg \\
        &\lvert \alpha_f \rvert \leqslant \alpha_{\mathrm{sl},f} \label{sim1_con:d} \\
        &\lvert \alpha_r \rvert \leqslant \alpha_{\mathrm{sl},r} \label{sim1_con:e} \\
        &x^o_{\text{min},j} - X \leqslant M_1\gamma_{2j-1} \label{sim1_con:f} \\ 
        &X -x^o_{\text{max},j} \leqslant M_2\gamma_{2j} \label{sim1_con:g} \\
        &\gamma_{2j-1} +\gamma_{2j} \leqslant 1,\, ... \text{ for } j=1,2,3 \label{sim1_con:h} \\
        & 0 \leqslant \gamma_i \leqslant 1,\quad \text{for } i=1,..,6 \label{sim1_con:i}\\
        &Y \geqslant y^o_{\text{max},1} - M_3(\gamma_1+\gamma_2) \label{sim1_con:j} \\
        &Y \leqslant y^o_{\text{min},2} + M_4(\gamma_3+\gamma_4) \label{sim1_con:k} \\
        &Y \geqslant y^o_{\text{max},3} - M_5(\gamma_5+\gamma_6) \label{sim1_con:l}
        \end{align}
\end{subequations}
\begin{equation} \label{nonlineardynamics}
    \small f(\textit{\textbf{x}},\delta) = \left\{
    \begin{array}{l}
         \nicefrac{1}{m}(-\mathrm{F}_{yf}\sin{\delta}) + qv_y \\
         \nicefrac{1}{m}(\mathrm{F}_{yf}\cos{\delta} + \mathrm{F}_{yr}) - qv_x \\
         \nicefrac{1}{\mathrm{I}_z}\{a\mathrm{F}_{yf}\cos{\delta} - b\mathrm{F}_{yr}\} \\
         v_x\cos{\psi_B} - v_y\sin{\psi_B} \\
         v_x\sin{\psi_B} + v_y\cos{\psi_B} \\
         q 
    \end{array} \right.
\end{equation}
\begin{table}[h]
    \centering
    \caption{Vehicle Parameters}
    \renewcommand{\arraystretch}{1.5}
    \begin{tabular}{|c|c||c|c|}
          \hline
          $m$ & $1636.364\,\mathrm{kg}$ & b & $1.153\,\mathrm{m}$\\
          \hline
          $\mathrm{I}_z$ & $925.02\,\mathrm{kg}\,\mathrm{m}^2$ & $C_{\mathrm{F}\alpha,f} $ & $59649 \,\mathrm{N/rad}$ \\
          \hline
          $a$ & $0.9803\,\mathrm{m}$ & $C_{\mathrm{F}\alpha,r}$ & $61138\,\mathrm{N/rad}$ \\
          \hline
    \end{tabular}
    \label{tbl: Vehicle Parameters}
\end{table}

For the linear dynamics constraints \eqref{sim1_con:b}, \eqref{sim1_con:d}, and \eqref{sim1_con:e} are replaced by the following. 
\begin{equation} \label{sim1_optlin}
    \begin{aligned}
        &\boldsymbol{\dot{x}}_L = \text{A}\boldsymbol{x}_L+\text{B}\delta \\
        &\dot{\bar{\boldsymbol{x}}} = f_2(\boldsymbol{x}_L) \\
        &\lvert C_{\mathrm{F}\alpha,f}\alpha_f \rvert \leqslant \mu \mathrm{F}_z/2  \\
        &\lvert C_{\mathrm{F}\alpha,r}\alpha_r \rvert \leqslant \mu \mathrm{F}_z/2 \\
        \text{where}\;&f_2(\boldsymbol{x}_L) = \mathrm{R}_2(\psi_B)(U,v_y)^T 
        \end{aligned}
\end{equation}
\noindent where the matrix $\mathrm{R}_2(\psi_B) \in SO(2)$ corresponds to the transformation of the body fixed velocities $v_x, v_y$ (upper left $2 \times 2$ of the full $SO(3)$ rotation matrix in \eqref{eq: error dynamics}), and $\bar{\boldsymbol{x}}= (X, Y)$ is the vehicle position in the inertial frame. The OCP with linear dynamics is referred to as $P_\Romannum{2}$.

The performance of the three OA formulations is benchmarked using the problem matrix of Table~\ref{tbl:case1formualtions}. Simulations for $P_\Romannum{1}$ and $P_\Romannum{2}$ utilize the Runge-Kutta (RK41) integration scheme, with four additional intermediate nodes for $P_\Romannum{1}$. For SCvx implementations, all nonlinearities, excluding integral constraints where applicable, are linearized via a first-order Taylor approximation. The prediction horizons are $T=3.5\,\text{s}$ for $E_{\Romannum{1}}$ and $T=4.0\,\text{s}$ for $E_{\Romannum{2}}$; selected to ensure the entire obstacle field is captured within the prediction horizon.
\begin{table}[h]
    \centering
    \caption{RCOA NLP OCP Parameters}
    \renewcommand{\arraystretch}{1.25}
    \begin{tabular}{|c|c|c|c|}
        \hline
         Env. & $w_1$ & $w_2$ & $M_1,M_2,M_3,M_4,M_5$\\
        \hline
        \textit{E\Romannum{1}} & 3 & 350 & 52, 52, 7, 9, 15\\
        \hline
        \textit{E\Romannum{2}} & 3 & 50 & 52, 52, 7, 7, NA \\
        \hline
        \textit{E\Romannum{3}$^*$} & 3 & 1300/20 & 52, 52, 7, 9, 15 \\
        \hline
        \multicolumn{4}{l}{*Applied in section~\ref{subsubsec: Correction Setup}}
    \end{tabular}
    \label{tbl:NLP OCP Para}
\end{table}
Temporal node counts were adjusted to ensure adequate obstacle resolution across formulations. In particular, the formulation of the EOA in the environment $E_{\Romannum{1}}$ requires a significantly larger set of nodes ($N=75$) to resolve the narrow minor axis of the obstacles, see Table~\ref{tbl:temporalNodes}. 
\begin{table}[h] 
    \centering
    \caption{Cluttered environment problem matrix}
    \renewcommand{\arraystretch}{1.25} 
    \begin{tabular}[h]{|l|c|c|c|c|}  
        \hline
        &
        \multicolumn{2}{c|}{$P_\Romannum{1}$} &
        \multicolumn{2}{c|}{$P_\Romannum{2}$} \\
        \cline{2-5}
        & \multicolumn{4}{c|}{Solver / Algorithm *} \\
        \hline
        $\mathrm{RCOA}$   & FATROP     & SCvx     & FATROP  & SCvx \\
        $\mathrm{EOA}$   & FATROP     & SCvx    & FATROP     & SCvx  \\
        $\mathrm{MIOA}$ & SMILP/FATROP & SMILP    & SMILP/FATROP & SMILP  \\
        \hline
        \multicolumn{5}{l}{*see section~\ref{subsection:solvers}}
    \end{tabular}
    \label{tbl:case1formualtions}
\end{table}
\begin{table}[h]
    \centering
    \caption{Number of temporal nodes assigned to each problem.}
    \renewcommand{\arraystretch}{1.25}
    \begin{tabular}{|c| c|c|}
     \hline
         & $E_\Romannum{1}$ & $E_\Romannum{2}$ \\
        \cline{2-3}
        Formulation& NLP/SSCP& NLP/SSCP\\
    \hline
         RCOA& 30 & 30/34\\
         EOA& 75 & 30/34\\
         MIOA& 30 & 34/34\\
         \hline
    \end{tabular}
    \label{tbl:temporalNodes}
\end{table}

\subsubsection{Results and Discussion} \label{subsubsec: results}

The OCPs in Table \ref{tbl:case1formualtions} were modeled using CVXPY \cite{Diamond2016CVXPY:Optimization} and CasADi \cite{Andersson2018CasADiControl}, for SCvx algorithms and NLP problems, respectively. The initial conditions were set to $ \boldsymbol{x}_0 = [15,0,0,-15,0] $ for $E_\Romannum{1}$ and $\boldsymbol{x}_0 = [15,0,0,-20,0]$ for $E_\Romannum{2}$. Computations were performed on an HP OMEN 35L desktop with an Intel i7-14700F processor. 

To assess the performance of the three OAs, \deleted{three}\added{two} key metrics are evaluated:
\begin{enumerate}[label=\arabic*., ref=\Roman*, leftmargin=1.5cm]
    \item Computational Efficiency: Solver execution time.
    \item Trajectory Quality: Minimization of obstacle penetration at both nodal and inter-sample points.
\end{enumerate}
\begin{figure}[t]
   \begin{minipage}{0.5\textwidth}
       \centerline{\input{P1_Cluttered_Environment_R4.pgf}}
       \vspace{0.1cm}
   \end{minipage}
   \begin{minipage}{0.5\textwidth}
        \centerline{\input{P2_Cluttered_Environment_R4.pgf}}
   \end{minipage}
    \caption{Simulation results for scenario $E_\Romannum{1}$ (static cluttered environment), showing the resulting trajectories for nonlinear dynamics $P_\Romannum{1}$ (top) and linear dynamics $P_\Romannum{2}$ (bottom). The reference trajectory is the dashed horizontal line in light blue.} 
   \label{fig: Environment I, Trajectories}
\end{figure}
\begin{figure}[!h]
   \begin{minipage}{0.5\textwidth}
       \centerline{\input{P1_EnvironmentII_R4.pgf}}
       \vspace{0.1cm}
   \end{minipage}
   \begin{minipage}{0.5\textwidth}
        \centerline{\input{P2_EnvironmentII_R4.pgf}}
   \end{minipage}
    \caption{\added{Simulation results for scenario} $E_\Romannum{2}$ \added{(static cluttered environment), showing the} resulting trajectories for \added{nonlinear dynamics} $P_\Romannum{1}$ (top) and \added{linear dynamics} $P_\Romannum{2}$ (bottom). The reference trajectory is the dashed horizontal line in light blue.}
   \label{fig: Environment II, Trajectories}
\end{figure}

\paragraph{Dynamics and Trajectories}
The results shown in Figures \ref{fig: Environment I, Trajectories}--\ref{fig: Environment II, Trajectories} demonstrate that SCvx trajectories align closely with NLP counterparts. However, $P_\Romannum{2}$ problems consistently exhibit under-actuation, as the linear tire model overestimates lateral force capacity. This highlights the necessity of NL dynamics for generating safe and feasible trajectories for aggressive maneuvers. Due to the difference in obstacle dimensions, $E_\Romannum{2}$ exhibits the most pronounced divergence among the formulations, with RCOA and MIOA generating strictly conservative trajectories compared to EOA.

\deleted{Under reasonable assumptions regarding the OCP, which are generally satisfied in this study, convergence is closely tied to the problem structure and the selection of the solver. RCOA is convex, and assuming CP, then it's guaranteed to converge using convex solvers. In a nonconvex setting or when using a nonconvex OA formulation, convergence is still guaranteed using robust nonlinear solvers like IPOPT's \cite{Wachter2006OnProgramming} interior point method. This applies under similar assumptions to CPs (e.g. smoothness, boundedness), but additionally, integrator stability, and a sufficient initial guess. Alternative solvers such as FATROP \cite{Vanroye2023FATROPControl} exploit structural sparsity and generally see significant performance improvements, although it was found to be less robust in some problems. 

Considering convergence alone, among the three formulations,  RCOA is the most robust for both convex and nonconvex scenarios. Conversely, MIOA tends to be more challenging in nonconvex settings. 
}

\paragraph{Computational Efficiency} \label{para: speed}

To evaluate computational efficiency, each configuration in the problem matrix of Table~\ref{tbl:case1formualtions} was executed 100 times with the solver/algorithm listed. Table~\ref{tbl:E1, total runtime} reports the mean total run times for each problem, while Table~\ref{tbl:E1, runtime per SCP iteration} details the mean run time per SCvx iteration. These values exclude pre-processing and capture only solver execution time.  \myadd{These trials consisted of sequential executions of identical problems to capture natural timing variance, which is notably wider for the MILP baseline due to its heuristic combinatorial search.}
\begin{table}[H]
    \centering
    \caption{Total solver mean runtime in (sec) for formulations using SCvx algorithm and NLP solver.}
    \renewcommand{\arraystretch}{1.4}
    \setlength{\tabcolsep}{5pt}
    \begin{tabular}{|c|c|c|c|c|c|c|}
        \hline
         & \multicolumn{3}{c|}{SCvx} & \multicolumn{3}{c|}{NLP} \\
         \cline{2-7}
        & RCOA & EOA & MIOA & RCOA & EOA & MIOA \\
        \hline
         \textit{E\Romannum{1}-P1} & \textbf{0.0589} & 0.1987 & 1.3428 & 0.3139 & 0.3047 & 1.4558 \\
         \textit{E\Romannum{1}-P2} & 0.0106 & 0.0255 & 0.2785 & \textbf{0.0104} & 0.0213 & 0.2852 \\
         \textit{E\Romannum{2}-P1} & 0.0754 & \textbf{0.0339} & 1.8940 & 0.0991 & 0.1034 & 1.9408 \\
         \textit{E\Romannum{2}-P2} & 0.0117 & 0.0078 & 0.4307 & 0.0086 & \textbf{0.0072} & 0.4379 \\
         \textit{E\Romannum{3}-P1}$^*$ & NA & NA & NA & \textbf{0.4534}* & 12.2158 & NA \\
         \hline
         \multicolumn{7}{l}{*see section~\ref{subsubsec:correction step results}}
    \end{tabular}
    \label{tbl:E1, total runtime}
\end{table}
\begin{table}[H]
    \centering
    \caption{Mean solver run time in (sec) per SCvx iteration.}
    \renewcommand{\arraystretch}{1.4}
    \begin{tabular}{|c|c|c|c|}
    \hline
        & RCOA & EOA & MIOA \\
        \hline
         \textit{E\Romannum{1}-P1} & \textbf{0.0045} & 0.0083 & 0.0959  \\
         \textit{E\Romannum{1}-P2} & \textbf{0.0035} & 0.0064 & 0.0928 \\
         \textit{E\Romannum{2}-P1} & 0.0044 & \textbf{0.0038} & 0.1052  \\
         \textit{E\Romannum{2}-P2} & 0.0039 & \textbf{0.0026} & 0.1436  \\
         \hline
    \end{tabular}
    \label{tbl:E1, runtime per SCP iteration}
\end{table}
The most significant time differences appear between $P_\Romannum{1}$ and $P_\Romannum{2}$ problems. In all cases, $P_\Romannum{2}$ problems are at a minimum 4.8X faster than their $P_\Romannum{1}$ counterparts. Furthermore, the gap between SCvx and NLP run times is smaller in $P_\Romannum{2}$, due to reduced sources of nonconvexity. RCOA demonstrates the potential of a convex OCP in SCvx implementations where a single iteration is solved in under $4.5\,\text{ms}$. \deleted{Despite the difference in obstacle area for $E_\Romannum{2}$, RCOA remains competitive against EOA.} 

\added{When evaluating these run times, particularly for $E_\Romannum{2}$, it is critical to recall the geometric discrepancy established during the environment definition. Because the rectangular bounding box strictly contains the ellipse used by EOA, RCOA is evaluated against an inherently more conservative and restrictive spatial domain. However, rather than invalidating the comparison, this geometric handicap actually underscores the computational efficacy of the proposed continuous relaxation. The data demonstrates that RCOA achieves highly competitive or lower solve times compared to EOA, even when artificially penalized by a larger geometric footprint that reduces the available free space. Thus, the computational advantages of RCOA hold robustly despite the strict conservatism of its spatial bounds.}

From a structural standpoint, EOA introduces the fewest constraints and variables; no new variables are introduced, and one constraint per obstacle is required. RCOA introduces two variables and four constraints per obstacle, yet remains computationally efficient. Ignoring integral constraints, the MIOA formulation requires four variables and five constraints, making it structurally the most expensive of the three.

\paragraph{Trajectory Quality} \label{subsubsection: Quality of trajectory}
Obstacle penetration is defined as the depth of penetration in the $Y$-axis, as it aligns with the obstacle's major or minor axis. The maximum penetration for both environments is reported in Table~\ref{tbl: obstacle penetration}. \deleted{RCOA and MIOA inter-sample penetration is evaluated at boundary points, while EOA considers all intervals between nodes.} \added{Obstacle penetration is defined as the maximum depth of penetration along the $Y$-axis. Because RCOA and MIOA define safe domains using linear half-planes, evaluating the trajectory at the discrete geometric boundaries inherently captures the maximum spatial violation. In contrast, EOA uses a curved elliptical boundary. Due to this curvature, the straight-line trajectory connecting two feasible discrete nodes may still intersect the elliptical arc (corner-cutting). Therefore, accurately capturing the true maximum penetration depth for EOA necessitates evaluating the continuous inter-sample intervals, ensuring a mathematically fair comparison across differing geometric formulations.}  For this assessment, only the $P_\Romannum{1}$ problems are considered, as $P_\Romannum{2}$ simulations all exhibit violations.

\begin{table}[h]
    \centering
    \caption{$E_{\Romannum{1}/\Romannum{2}}$ trajectory obstacle penetration (m) for $P_\Romannum{1}$ simulations.}
    \renewcommand{\arraystretch}{1.5}
    \setlength{\tabcolsep}{5pt}
    \begin{tabular}{|c|c|c|c|c|c|c|c|}
        \hline
         \multicolumn{2}{|c|}{} & \multicolumn{3}{c |}{SCvx} & \multicolumn{3}{c|}{NLP} \\
         \cline{3-8}
        \multicolumn{2}{|c|}{} & RCOA & EOA & MIOA & RCOA & EOA & MIOA \\
        \hline
         \multirow{2}{*}{\centering \textit{E\Romannum{1}}}  & Node & \textbf{0} & 0.143 & 0.017 & \textbf{0} & \textbf{0} & \textbf{0} \\
         \cline{2-8}
         & Intersample & \textbf{0.024} & 0.128 & 0.069 & 0.033 & 0.116 & 0.068 \\
         \hline 
         \multirow{2}{*}{\centering \textit{E\Romannum{2}}}  & Node & 0.016 & 0.065 & 0.140 & 0.008 & \textbf{0} & \textbf{0} \\
         \cline{2-8}
         & Intersample & \textbf{0.044} & 0.089 & 0.283 & 0.058 & 0.051 & 0.147 \\
         \hline
    \end{tabular}
    \label{tbl: obstacle penetration}
\end{table}

$E_\Romannum{1}$ shows the least amount of penetration at both the nodal and inter-sample \added{levels}. Surprisingly, the RCOA problem solved with SCvx also results in an obstacle-free trajectory. In terms of node violations, the hard-constrained OCPs solved with an NLP solver result in no violations across both environments. However, all approaches experience inter-sample penetration at comparable levels. The high node count required by the EOA formulation in $E_\Romannum{1}$ is justified compared to $E_\Romannum{2}$ for $P_\Romannum{1}$ problems. Despite the increased node count, inter-sample penetration remains compatible across both environments.

This demonstrates RCOA's ability to generate suitable trajectories,  with the worst-case node infraction of $\mathrm{0.008\,m}$. \myadd{In practice, the minor nodal penetrations inherent to soft-constraint relaxations are readily mitigated by applying a slight geometric inflation to the obstacle boundaries.}

\subsection{Recovery of Infeasibility Certificate in Cluttered Environment} \label{subsubsec: Correction Setup}
To demonstrate the feasibility correction method from section~\ref{subsec:correctionstep}, a simulation similar to $E_\Romannum{1}$ but with taller obstacles is assessed. This setup is denoted as $E_\Romannum{3}$, with the modified obstacle vertices detailed in Table~\ref{tbl:obsvert}. Under the RCOA formulation, this environment is near-infeasible. Only the $P_\Romannum{1}$ configuration is considered, corresponding to the trajectory in Figure~\ref{fig: Correction Example}.

Since EOA and RCOA have similar performance, EOA is used for comparison. The number of temporal nodes is the same as $E_\Romannum{1}$ (Table~\ref{tbl:temporalNodes}). Simulations are solved using IPOPT with HSL code MA57 \cite{Duff2004}. 

\subsubsection{Infeasibility Results} \label{subsubsec:correction step results}
Using the same prediction horizon, initial condition, and modeling as $E_\Romannum{1}$, in Figure~\ref{fig: Correction example 2 of 2}, the RCOA (initial and corrected) and EOA trajectories are shown. Initially, several temporal nodes penetrate obstacles 1 and 2. After applying the constraints in \eqref{ConstCorrection}, nodal penetrations are eliminated, \added{even when applying poor M-parameters for the correction step as discussed in section~\ref{subsubsec: Correction Setup}}. Although an obstacle-free trajectory at node level is feasible, inter-sample violations persist. Further refinement could constrain adjacent nodes to improve continuous-time collision avoidance.

Note that it was confirmed that increasing the obstacles' sizing does eventually lead to infeasibility, specifically the zigzag path, of both RCOA and EOA.

\begin{figure}[h]
    \centerline{\input{Correction_example_2_R5.pgf}}
    \caption{Simulation results for scenario $E_\Romannum{3}$. The feasibility restoration step successfully corrects the spatial violations of the Initial RCOA trajectory, allowing it to safely navigate the constrained environment.}
    \label{fig: Correction example 2 of 2}
\end{figure}

Table~\ref{tbl:E1, total runtime} reports the mean solver runtime for $E_\Romannum{3}$ problems over 100 trials. With RCOA, both problems were solved in under 0.5 seconds on average, while EOA, despite its less conservative obstacle area, requires over 10 seconds. This highlights RCOA's efficiency near infeasible regimes.

\added{A direct runtime for MIOA in scenario $E_\Romannum{3}$ is not applicable (NA); however, the computational advantage of the proposed framework is evident when comparing across problem types. Even when RCOA is forced to execute the secondary hard-constrained restoration step (combined requiring $0.4534\,\text{s}$), its total solve time remains strictly lower than MIOA's baseline performance on standard scenarios ($1.4558\,\text{s}$ and $1.9408\,\text{s}$ for $E_\Romannum{1}$ and $E_\Romannum{2}$ respectively).}
\subsection{Autonomous Driving Involving Dynamic Obstacles: Closed-Loop RCOA} \label{subsec:MPC}
\subsubsection{Environment Setup}

To evaluate the RCOA formulation in a closed-loop setting, it is embedded within a NMPC framework. The illustration in Figure~\ref{fig: intersection scenario} represents a left‑turn maneuver at a four-way intersection. Shown is the vehicle’s reference path, the oncoming obstacle, and its heading. This scenario represents the leading cause of fatal motorcycle accidents \cite{NHTSA2023Motorcycles}. \added{This configuration is specifically designed to demonstrate the formulation's performance under a critically reduced prediction horizon.}

The obstacle’s initial position and velocity are selected such that a collision would occur without evasive action. The RCOA NMPC switches between two OCPs: an OA OCP when the vehicle and obstacle are moving towards each other, and a PT OCP once they deviate away from each other. \added{The EOA formulation is also evaluated as a baseline comparison.}

Assumptions~\ref{Assumptions B: 1} and \ref{Assumptions B: 2} are applicable except for the following: \ref{Assumptions B: 1}$\mathrm{.\romannum{1}}$ is relaxed by modeling braking and acceleration through the longitudinal slip \deleted{angles, } \added{ratios,} $(\kappa_f,\kappa_r)$. In addition to the steering, the longitudinal slip \deleted{angles} \added{ratios} will also serve as inputs.

\begin{figure}[!h]
\centerline{\input{Intersection_Scenario6.pgf}}
\caption{\added{Simulation configuration for a left-turn maneuver at a four-way intersection with oncoming traffic. This setup tests the NMPC formulation's ability to plan a safe trajectory while accounting for a dynamic obstacle. The dashed reference path defines a constant-radius quarter circle, with the labeled spatial coordinates marking the entry and exit points of the circular path.}}
\label{fig: intersection scenario}
\end{figure}

\paragraph{Control Problem Formulation}
\added{The complete OCP is detailed in \eqref{eq:def_MPC_opt_problem_1}, where the objective minimizes the $L_1$-norm of the penalty vector $\boldsymbol{y}_0=(e,\kappa_f,\kappa_r,\delta,\gamma_i)$ and the final term mitigates the tendency for the vehicle to aggressively accelerate toward the obstacle—a direct consequence of minimizing $\gamma_1$.} \added{The state vector is defined as $\boldsymbol{x}=(v_x,v_y,q,X_{\text{v}},Y_{\text{v}},\psi_B,s,e,\bar{\psi},X_{\text{o}},Y_{\text{o}})$, where $(X_{\text{v}},Y_{\text{v}})$ and $(X_{\text{o}},Y_{\text{o}})$ represent the inertial positions of the vehicle and obstacle, respectively.}

\added{$U_{\text{avg}}=6\, \mathrm{m/s}$ represents the average longitudinal velocity ($v_x$) along the circular portion, acting as a safe curve-navigation speed analogous to the admissible velocity in the Dynamic Window Approach (DWA) \cite{Fox1997}. This baseline value is extracted from an isolated PT OCP simulation, where the PT OCP is defined by \eqref{eq:def_MPC_opt_problem_2}}.

Compared to the OCP of \eqref{sim1_con}, additional terms appear in both the cost and constraints due to the combined‑slip tire model. These additions mitigate issues arising from the increased stiffness of the system dynamics, due to the inputs $\kappa_i$. \deleted{as well as the velocity‑induced stiffness that emerges as the longitudinal speed approaches zero \cite{Kim2019AdvancedRatio}.} The longitudinal velocity is bounded by a lower limit away from zero and an upper limit. Sensor feedback includes obstacle position and velocity, and the instantaneous PT error.

The magnitudes of the slip vectors $(\boldsymbol{\sigma}_f,\boldsymbol{\sigma}_r)$ are constrained via the first constraint in \eqref{eq:def_MPC_opt_problem_1 d} to avoid entering the full sliding regime defined by $(\sigma_{\text{sl},f},\sigma_{\text{sl},r})$. The rear wheel longitudinal slip ratio ($\kappa_r \leqslant 0$) limits the rear axle to strictly braking forces, which is consistent with a front-wheel drive configuration. Finally, constraints \eqref{eq: def_MPC_opt_problem_1 i}--\eqref{eq: def_MPC_opt_problem_1 M} define the OA formulation, see \eqref{eq: Generalized RCOA}. Initial conditions for states and inputs complete the NMPC specification. \added{The EOA obstacle is geometrically sized as the minimum bounding ellipse of the rectangular vertices defined below:}
\begin{align}   
    x^o_{\text{min}} &= -(a+0.5)\,\mathrm{m},\quad x^o_{\text{max}} = (b+0.5) \,\mathrm{m} \nonumber \\
    y^o_{\text{min}} &= -1\,\mathrm{m},\quad y^o_{\text{max}} = 1\,\mathrm{m} \nonumber
\end{align}
\added{In contrast to the final cost term in \eqref{eq: def_MPC_opt_problem_1 a}, the PT OCP purpose is to prevent the vehicle from decelerating as it departs the obstacle—a direct numerical consequence of minimizing $\gamma_2$.}
\begin{subequations} \label{eq:def_MPC_opt_problem_1}
    \begin{align}
        \min\limits_{\gamma,\delta,\kappa_f,\kappa_r}\; & \sum_{j=1}^5 {w_j \lVert \boldsymbol{y}_0(j) \rVert} + 0.3 (U_{\text{avg}}-v_x^N)^2 \label{eq: def_MPC_opt_problem_1 a} \\
        \text{s.t. } \quad &\dot{\boldsymbol{x}} = f(\boldsymbol{x},\delta,\kappa_r,\kappa_f), \\
        &\lvert\delta\rvert \leqslant 35\deg,\;\lvert\dot{\delta}\rvert \leqslant \dot{\delta}_{\text{max}},\;\lvert\ddot{\delta}\rvert \leqslant \ddot{\delta}_{\text{max}} \label{eq:def_MPC_opt_problem_1 c} \\
        & \sigma_i \leqslant \sigma_{\text{sl},i},\;\lvert \kappa_i \rvert \leqslant \dot{\kappa}_{\max},\; \kappa_r \leqslant 0, \label{eq:def_MPC_opt_problem_1 d} \\
        & 3 \leqslant v_x \leqslant 10 \; \mathrm{m/s}, \\
        & - X_{\text{v}} \leqslant -(X_{\text{o}}+x^o_{\text{min}}) + M_1\gamma_{1}, \label{eq: def_MPC_opt_problem_1 i} \\
        & X_{\text{v}}  \leqslant (X_{\text{o}}+x^o_{\text{max}}) + M_2\gamma_{2}, \label{eq: def_MPC_opt_problem_1 j} \\
        &\gamma_{1} + \gamma_{2} \leqslant 1, \label{eq:h} \\
        &0 \leqslant \gamma_j \leqslant 1,\quad \text{for } j=0,1 \\
        &Y_{\text{v}} \leqslant (Y_{\text{o}}+y^o_{\text{min}}) - M_3(\gamma_1+\gamma_2), \label{eq: def_MPC_opt_problem_1 M} \\
        &\boldsymbol{x}(0) = \boldsymbol{x}_0,\quad \delta(0) = \delta_0, \quad \\
        &\kappa_i(0) = \kappa_{i,0}, \quad \text{for } i\in \{f,r\}.
    \end{align}
\end{subequations}
where \added{the continuous-time dynamics $\dot{\boldsymbol{x}} = f(\boldsymbol{x},\delta,\kappa_f,\kappa_r)$ include the nonlinear vehicle equations of motion, the Frenet-Serret PT dynamics from \eqref{eq: error dynamics}, and the constant horizontal obstacle kinematics ($\dot{X}_{\text{o}} = -U_{\text{o}}$) as follows:}
\begin{equation} \label{eq:def_MPC_opt_problem_1 dynamics}
\small
f(\boldsymbol{x},\delta,\kappa_f,\kappa_r) = 
\begin{cases}
\nicefrac{1}{m}\Bigl(\mathrm{F}_{xf}\cos{\delta} + \mathrm{F}_{xr} - \mathrm{F}_{yf}\sin{\delta}\Bigr) + qv_y\\
\nicefrac{1}{m}\Bigl(\mathrm{F}_{yf}\cos{\delta} + \mathrm{F}_{yr} + \mathrm{F}_{xf}\sin{\delta}\Bigr) - qv_x\\
\nicefrac{1}{\text{I}_z}\Bigl\{a\Bigl(\mathrm{F}_{yf}\cos{\delta} + \mathrm{F}_{xf}\sin{\delta}\Bigr) - b\mathrm{F}_{yr}\Bigr\}\\
v_x\cos{\psi_B} - v_y\sin{\psi_B}\\
v_x\sin{\psi_B} + v_y\cos{\psi_B}\\
q\\
\mathrm{FS}^{-1}\mathrm{R}(\bar{\psi})\boldsymbol{p}\\
-U_{\text{o}}\\
0
\end{cases}
\end{equation}
  
\begin{equation} \label{eq:def_MPC_opt_problem_2}
    \begin{aligned}
        \min\limits_{\delta,\kappa_f,\kappa_r}\; &  5\lVert e \rVert_1 + 0.5\lVert \kappa_f\rVert +0.5\lVert \kappa_r\rVert +0.2 \lVert \delta \rVert \\
        \text{s.t. } \quad &\dot{\boldsymbol{x}} = f(\boldsymbol{x},\delta,\kappa_r,\kappa_f) \\
        &\lvert\delta\rvert \leqslant 35\deg,\quad \sigma_i \leqslant \sigma_{\text{sl},i}  \\
        &\kappa_r \leqslant 0,\quad 0.1 \leqslant v_x  \\
        &\boldsymbol{x}(0) = \boldsymbol{x}_0,\quad \delta(0)=\delta_0, \quad \\
        &\kappa_i(0) = \kappa_{i,0}, \quad \text{for } i\in \{f,r\} \\
    \end{aligned}
\end{equation}

The OCPs are modeled in CasADi and solved using IPOPT with MA57. The NMPC operates at a frequency of $T/N$, where $T=0.5$ is the prediction horizon and $N=7$ is the number of temporal nodes. After each solve, the first optimal control inputs $\delta^k(1)$, $\kappa_r^k(1)$, and $\kappa_f^k(1)$ are applied. At each iteration, the initial state is re-evaluated to correct any deviations (e.g., path error). The inputs applied become the initial conditions for the next iteration of NMPC. 

The dynamics are solved using a Runge-Kutta integration scheme featuring four intermediate nodes. Steering-rate constraints are discretized using first-order and second-order finite difference methods, respectively.

The initial conditions are listed in Table~\ref{tbl:MPC ICs}, all initial inputs are zero, and Table~\ref{tbl:NMPC OCP Para} contains the OA OCP parameters. At the start of the RCOA simulation, the OA OCP is active. Both RCOA and EOA simulate a total of 4.2857 seconds.

\begin{table}[h]
    \centering
    \caption{Simulation Initial Conditions}
    \renewcommand{\arraystretch}{1.5} 
    \begin{tabular}{|c|}
    \hline
         $v_x,v_y,q,X_{\text{v}},Y_{\text{v}},\psi_{B},s,e,\bar{\psi},X_o,Y_o$ \\
         \hline
         $10\,\mathrm{m/s},0,0,0,0,0,0,0,0,50\,\mathrm{m},3.6576\,\mathrm{m}$ \\
         \hline
    \end{tabular}
    \label{tbl:MPC ICs}
\end{table}

\begin{table}[h]
    \centering
    \caption{NMPC OA Parameters}
    \renewcommand{\arraystretch}{1.35}
    \begin{tabular}{|c|c|c|c|c|}
        \hline
         OA & $w_1,w_2,...,w_5$ & $M_1,M_2,M_3$ & $\dot{\delta}_{\max}, \ddot{\delta}$ & $\dot{\kappa}_{\max}$\\ 
         \hline
         RCOA & 0.5, 0.75, 0.75, 0.3, 100 & 100, 100, 15 & $\pi$/2, 32/3 & 3\\
         \hline
         EOA & 2, 0.75, 0.75, 0.3, NA & NA & $\pi$/2, 32/3 & 3\\
        \hline
    \end{tabular}
    \label{tbl:NMPC OCP Para}
\end{table}

\paragraph{NMPC Results} \label{subsec: MPC example results}
\added{Figure \ref{fig: MPC Sim1 Trajectories} illustrates the resulting vehicle and obstacle trajectories at their closest proximity for both RCOA (top) and EOA (bottom). NMPC iterations 25 through 30 are highlighted to explicitly showcase the OA. RCOA successfully enforces the spatial boundaries with no collision or inter-sample penetration. Conversely, under the EOA formulation, the vehicle rides the exact obstacle boundary at iteration 28, resulting in distinct inter-sample penetration between iterations 28 and 29. While this violation can practically be mitigated by artificially inflating the obstacle's safety margin, attempts to correct it by decreasing the PT weight ($w_1$) severely compromise the turning maneuver, causing the vehicle to drift into opposing traffic. Together, these observations highlight the inherent spatial robustness of the RCOA constraints.}

\added{In addition to maintaining strict spatial boundaries, RCOA also yields favorable longitudinal dynamics. Despite the RCOA OA OCP including an active speed penalty, Figure \ref{fig: MPC Velocities} demonstrates that RCOA still maintains a higher average longitudinal velocity through the maneuver compared to the benchmark.}

\begin{figure}[!h]
    \begin{subfigure}[!h]{0.5\textwidth}
        \centerline{\input{MPC_RCOA_Trajectory_R0.pgf}}
    \end{subfigure}
    \begin{subfigure}[h]{0.5\textwidth}
        \centerline{\input{MPC_EOA_Trajectory_R0.pgf}}
    \end{subfigure} 
    \caption{\added{Closed-loop vehicle and obstacle trajectories for (top) RCOA and (bottom) EOA. Highlighted temporal NMPC nodes (25-30) correspond to $t\in[1.7857,2.1429]$ s, detailing the precise boundary interaction without inter-sample penetration for RCOA.}}
    \label{fig: MPC Sim1 Trajectories}
\end{figure}

\begin{figure}[!h]
    \centerline{\input{MPC_Vx_R0.pgf}}
    \caption{\added{Longitudinal velocity profile. Despite the inclusion of a speed penalty term in the OA objective, RCOA generally maintains a higher velocity than the EOA benchmark.}}
    \label{fig: MPC Velocities}
\end{figure}

\deleted{The numerical performance of the OA OCPs are shown in the table below. On average, they are on par, with RCOA clearly having a lower standard deviation based on the max and min. }
\added{Solver execution statistics for the OA OCPs are detailed in Table \ref{tbl: Ex2_solver_stats}. While both formulations exhibit comparable mean execution times, RCOA demonstrates tighter computational variance, evidenced by tighter bounds.}

\begin{table}[h]
    \centering
    \caption{NMPC OA OCP numerical performance (sec.)}
    \vspace{-.2cm}
    \label{tbl: Ex2_solver_stats}
    \renewcommand{\arraystretch}{1.3}
    \begin{tabular}{c c  c  c} 
        \toprule[1pt] 
        OA & mean & max & min \\ \hline
        RCOA & \textbf{0.0240} & \textbf{0.0307} & 0.0190 \\ 
        EOA   & 0.0242 & 0.0368 & \textbf{0.0096} \\ \bottomrule[1pt]
    \end{tabular}
\end{table}

\section{Conclusion}\label{sec: Conclusion}
The novel OA formulation has demonstrated performance on par with or exceeding other notable formulations in a nonconvex setting, even when at a disadvantage due to conservative obstacle definitions relative to EOA. One could have defined the rectangular obstacle so that it lies inside the ellipse. Since the formulation is convex, SCvx iterations demonstrate a clear performance advantage in a convex setting. Although structurally it should theoretically trail in performance due to the introduction of new variables and several constraints, it remains a competitive option. Furthermore, as demonstrated via experiments, in challenging environments, such as the near-infeasible cluttered setting, the RCOA computationally outperformed the EOA formulation by a large margin. 

\added{Finally, as indicated in Section~\ref{subsec: Parameter Study}, the primary penalty formulation \eqref{penfunc} may exhibit scaling limitations in highly complex, large-scale environments (e.g., dense mazes). These challenges can be mitigated by employing the alternative NLP formulation \eqref{eq: RNCOA equi f0} or by warm-starting the optimal control problem with a sampling-based planner like RRT. Ultimately, these characteristics suggest that RCOA is most effectively deployed as a rapid, local trajectory generator for real-time applications, operating within a broader hierarchical planning architecture.}

{\printbibliography
}
\vspace{-1.cm}
\begin{IEEEbiography}[{\includegraphics[width=1in,height=1.in,clip,keepaspectratio]{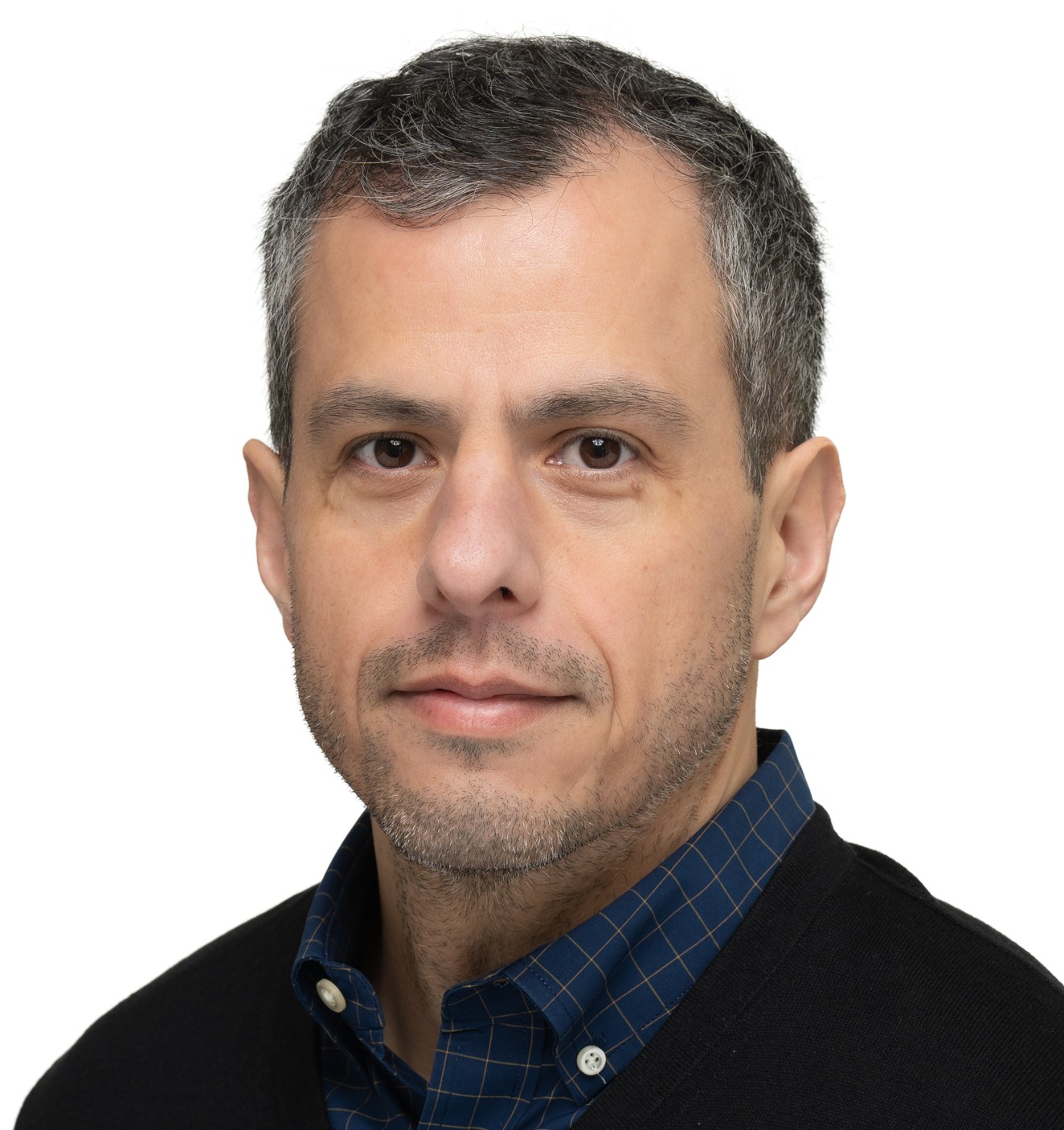}}]{Tapia, Ricardo} (member, IEEE) received a Bachelor's and Master's Degree in mechanical and aeronautical engineering at the University of California, at Davis in 2008 and 2012 respectively.

He is a doctoral candidate in the department of Mechanical and Aerospace Engineering, University of California at Davis, USA. Before pursuing his doctorate, he worked in commercial aviation as a Structural Analyst, where he led the structural analysis on the Nacelle Inlet for NASAs Quiet Technology Demonstrator 3 (QTD3).
\end{IEEEbiography}
\vspace{-1.cm}
\begin{IEEEbiography}[{\includegraphics[width=1in,height=1.in,clip,keepaspectratio]{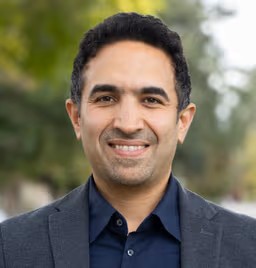}}]{Soltani, Iman} (member, IEEE) received the bachelors degree in mechanical engineering from Tehran Polytechnic, Tehran, Iran, in 2003, the masters degree in mechanical engineering from the University of Ottawa, Ottawa, ON, Canada, in 2007, and the Ph.D. degree in mechanical engineering from the Massachusetts Institute of Technology (MIT), Cambridge, MA, USA, in 2015. 

He is a Faculty Member with the Department of Mechanical and Aerospace Engineering as well as a Graduate Faculty Member with the Departments of Computer Science and Electrical and Computer Engineering, University of California at Davis (UC Davis), USA. Before joining UC Davis, he worked at the Ford Greenfield Labs, Palo Alto, CA, USA, where he founded and led the Advanced Automation Laboratory. He holds more than 17 patents and has authored over 40 journal and conference publications on topics ranging from medical imaging to autonomous driving, nanorobotics, dexterous bimanual robotics, machinery health monitoring, and precision positioning systems. His research has been featured in prominent outlets such as The Boston Globe, Elsevier Materials Today, ScienceDaily, and MIT News. His research spans the interface of artificial intelligence, instrumentation, controls, and design, with a focus on developing advanced machine learning tools and robotic and automation systems. Dr. Soltani received numerous awards, including the MIT Carl G. Sontheimer Award and the National Instruments Engineering Impact
Award.
\end{IEEEbiography}

\end{document}

%% file: Correction_example_1_R4.pgf
\begingroup%
\makeatletter%
\begin{pgfpicture}%
\pgfpathrectangle{\pgfpointorigin}{\pgfqpoint{3.367088in}{1.877947in}}%
\pgfusepath{use as bounding box, clip}%
\begin{pgfscope}%
\pgfsetbuttcap%
\pgfsetmiterjoin%
\definecolor{currentfill}{rgb}{1.000000,1.000000,1.000000}%
\pgfsetfillcolor{currentfill}%
\pgfsetlinewidth{0.000000pt}%
\definecolor{currentstroke}{rgb}{1.000000,1.000000,1.000000}%
\pgfsetstrokecolor{currentstroke}%
\pgfsetdash{}{0pt}%
\pgfpathmoveto{\pgfqpoint{0.000000in}{-0.000000in}}%
\pgfpathlineto{\pgfqpoint{3.367088in}{-0.000000in}}%
\pgfpathlineto{\pgfqpoint{3.367088in}{1.877947in}}%
\pgfpathlineto{\pgfqpoint{0.000000in}{1.877947in}}%
\pgfpathlineto{\pgfqpoint{0.000000in}{-0.000000in}}%
\pgfpathclose%
\pgfusepath{fill}%
\end{pgfscope}%
\begin{pgfscope}%
\pgfsetbuttcap%
\pgfsetmiterjoin%
\definecolor{currentfill}{rgb}{1.000000,1.000000,1.000000}%
\pgfsetfillcolor{currentfill}%
\pgfsetlinewidth{0.000000pt}%
\definecolor{currentstroke}{rgb}{0.000000,0.000000,0.000000}%
\pgfsetstrokecolor{currentstroke}%
\pgfsetstrokeopacity{0.000000}%
\pgfsetdash{}{0pt}%
\pgfpathmoveto{\pgfqpoint{0.322917in}{0.362654in}}%
\pgfpathlineto{\pgfqpoint{3.295520in}{0.362654in}}%
\pgfpathlineto{\pgfqpoint{3.295520in}{1.839367in}}%
\pgfpathlineto{\pgfqpoint{0.322917in}{1.839367in}}%
\pgfpathlineto{\pgfqpoint{0.322917in}{0.362654in}}%
\pgfpathclose%
\pgfusepath{fill}%
\end{pgfscope}%
\begin{pgfscope}%
\pgfsetbuttcap%
\pgfsetroundjoin%
\definecolor{currentfill}{rgb}{0.000000,0.000000,0.000000}%
\pgfsetfillcolor{currentfill}%
\pgfsetlinewidth{0.803000pt}%
\definecolor{currentstroke}{rgb}{0.000000,0.000000,0.000000}%
\pgfsetstrokecolor{currentstroke}%
\pgfsetdash{}{0pt}%
\pgfsys@defobject{currentmarker}{\pgfqpoint{0.000000in}{-0.048611in}}{\pgfqpoint{0.000000in}{0.000000in}}{%
\pgfpathmoveto{\pgfqpoint{0.000000in}{0.000000in}}%
\pgfpathlineto{\pgfqpoint{0.000000in}{-0.048611in}}%
\pgfusepath{stroke,fill}%
}%
\begin{pgfscope}%
\pgfsys@transformshift{0.653207in}{0.362654in}%
\pgfsys@useobject{currentmarker}{}%
\end{pgfscope}%
\end{pgfscope}%
\begin{pgfscope}%
\definecolor{textcolor}{rgb}{0.000000,0.000000,0.000000}%
\pgfsetstrokecolor{textcolor}%
\pgfsetfillcolor{textcolor}%
\pgftext[x=0.653207in,y=0.265432in,,top]{\color{textcolor}{\sffamily\fontsize{8.000000}{9.600000}\selectfont\catcode`\^=\active\def^{\ifmmode\sp\else\^{}\fi}\catcode`\%=\active\def
\end{pgfscope}%
\begin{pgfscope}%
\pgfsetbuttcap%
\pgfsetroundjoin%
\definecolor{currentfill}{rgb}{0.000000,0.000000,0.000000}%
\pgfsetfillcolor{currentfill}%
\pgfsetlinewidth{0.803000pt}%
\definecolor{currentstroke}{rgb}{0.000000,0.000000,0.000000}%
\pgfsetstrokecolor{currentstroke}%
\pgfsetdash{}{0pt}%
\pgfsys@defobject{currentmarker}{\pgfqpoint{0.000000in}{-0.048611in}}{\pgfqpoint{0.000000in}{0.000000in}}{%
\pgfpathmoveto{\pgfqpoint{0.000000in}{0.000000in}}%
\pgfpathlineto{\pgfqpoint{0.000000in}{-0.048611in}}%
\pgfusepath{stroke,fill}%
}%
\begin{pgfscope}%
\pgfsys@transformshift{1.313785in}{0.362654in}%
\pgfsys@useobject{currentmarker}{}%
\end{pgfscope}%
\end{pgfscope}%
\begin{pgfscope}%
\definecolor{textcolor}{rgb}{0.000000,0.000000,0.000000}%
\pgfsetstrokecolor{textcolor}%
\pgfsetfillcolor{textcolor}%
\pgftext[x=1.313785in,y=0.265432in,,top]{\color{textcolor}{\sffamily\fontsize{8.000000}{9.600000}\selectfont\catcode`\^=\active\def^{\ifmmode\sp\else\^{}\fi}\catcode`\%=\active\def
\end{pgfscope}%
\begin{pgfscope}%
\pgfsetbuttcap%
\pgfsetroundjoin%
\definecolor{currentfill}{rgb}{0.000000,0.000000,0.000000}%
\pgfsetfillcolor{currentfill}%
\pgfsetlinewidth{0.803000pt}%
\definecolor{currentstroke}{rgb}{0.000000,0.000000,0.000000}%
\pgfsetstrokecolor{currentstroke}%
\pgfsetdash{}{0pt}%
\pgfsys@defobject{currentmarker}{\pgfqpoint{0.000000in}{-0.048611in}}{\pgfqpoint{0.000000in}{0.000000in}}{%
\pgfpathmoveto{\pgfqpoint{0.000000in}{0.000000in}}%
\pgfpathlineto{\pgfqpoint{0.000000in}{-0.048611in}}%
\pgfusepath{stroke,fill}%
}%
\begin{pgfscope}%
\pgfsys@transformshift{1.974363in}{0.362654in}%
\pgfsys@useobject{currentmarker}{}%
\end{pgfscope}%
\end{pgfscope}%
\begin{pgfscope}%
\definecolor{textcolor}{rgb}{0.000000,0.000000,0.000000}%
\pgfsetstrokecolor{textcolor}%
\pgfsetfillcolor{textcolor}%
\pgftext[x=1.974363in,y=0.265432in,,top]{\color{textcolor}{\sffamily\fontsize{8.000000}{9.600000}\selectfont\catcode`\^=\active\def^{\ifmmode\sp\else\^{}\fi}\catcode`\%=\active\def
\end{pgfscope}%
\begin{pgfscope}%
\pgfsetbuttcap%
\pgfsetroundjoin%
\definecolor{currentfill}{rgb}{0.000000,0.000000,0.000000}%
\pgfsetfillcolor{currentfill}%
\pgfsetlinewidth{0.803000pt}%
\definecolor{currentstroke}{rgb}{0.000000,0.000000,0.000000}%
\pgfsetstrokecolor{currentstroke}%
\pgfsetdash{}{0pt}%
\pgfsys@defobject{currentmarker}{\pgfqpoint{0.000000in}{-0.048611in}}{\pgfqpoint{0.000000in}{0.000000in}}{%
\pgfpathmoveto{\pgfqpoint{0.000000in}{0.000000in}}%
\pgfpathlineto{\pgfqpoint{0.000000in}{-0.048611in}}%
\pgfusepath{stroke,fill}%
}%
\begin{pgfscope}%
\pgfsys@transformshift{2.634942in}{0.362654in}%
\pgfsys@useobject{currentmarker}{}%
\end{pgfscope}%
\end{pgfscope}%
\begin{pgfscope}%
\definecolor{textcolor}{rgb}{0.000000,0.000000,0.000000}%
\pgfsetstrokecolor{textcolor}%
\pgfsetfillcolor{textcolor}%
\pgftext[x=2.634942in,y=0.265432in,,top]{\color{textcolor}{\sffamily\fontsize{8.000000}{9.600000}\selectfont\catcode`\^=\active\def^{\ifmmode\sp\else\^{}\fi}\catcode`\%=\active\def
\end{pgfscope}%
\begin{pgfscope}%
\pgfsetbuttcap%
\pgfsetroundjoin%
\definecolor{currentfill}{rgb}{0.000000,0.000000,0.000000}%
\pgfsetfillcolor{currentfill}%
\pgfsetlinewidth{0.803000pt}%
\definecolor{currentstroke}{rgb}{0.000000,0.000000,0.000000}%
\pgfsetstrokecolor{currentstroke}%
\pgfsetdash{}{0pt}%
\pgfsys@defobject{currentmarker}{\pgfqpoint{0.000000in}{-0.048611in}}{\pgfqpoint{0.000000in}{0.000000in}}{%
\pgfpathmoveto{\pgfqpoint{0.000000in}{0.000000in}}%
\pgfpathlineto{\pgfqpoint{0.000000in}{-0.048611in}}%
\pgfusepath{stroke,fill}%
}%
\begin{pgfscope}%
\pgfsys@transformshift{3.295520in}{0.362654in}%
\pgfsys@useobject{currentmarker}{}%
\end{pgfscope}%
\end{pgfscope}%
\begin{pgfscope}%
\definecolor{textcolor}{rgb}{0.000000,0.000000,0.000000}%
\pgfsetstrokecolor{textcolor}%
\pgfsetfillcolor{textcolor}%
\pgftext[x=3.295520in,y=0.265432in,,top]{\color{textcolor}{\sffamily\fontsize{8.000000}{9.600000}\selectfont\catcode`\^=\active\def^{\ifmmode\sp\else\^{}\fi}\catcode`\%=\active\def
\end{pgfscope}%
\begin{pgfscope}%
\definecolor{textcolor}{rgb}{0.000000,0.000000,0.000000}%
\pgfsetstrokecolor{textcolor}%
\pgfsetfillcolor{textcolor}%
\pgftext[x=1.809219in,y=0.111111in,,top]{\color{textcolor}{\sffamily\fontsize{8.000000}{9.600000}\selectfont\catcode`\^=\active\def^{\ifmmode\sp\else\^{}\fi}\catcode`\%=\active\def
\end{pgfscope}%
\begin{pgfscope}%
\pgfsetbuttcap%
\pgfsetroundjoin%
\definecolor{currentfill}{rgb}{0.000000,0.000000,0.000000}%
\pgfsetfillcolor{currentfill}%
\pgfsetlinewidth{0.803000pt}%
\definecolor{currentstroke}{rgb}{0.000000,0.000000,0.000000}%
\pgfsetstrokecolor{currentstroke}%
\pgfsetdash{}{0pt}%
\pgfsys@defobject{currentmarker}{\pgfqpoint{-0.048611in}{0.000000in}}{\pgfqpoint{-0.000000in}{0.000000in}}{%
\pgfpathmoveto{\pgfqpoint{-0.000000in}{0.000000in}}%
\pgfpathlineto{\pgfqpoint{-0.048611in}{0.000000in}}%
\pgfusepath{stroke,fill}%
}%
\begin{pgfscope}%
\pgfsys@transformshift{0.322917in}{0.526734in}%
\pgfsys@useobject{currentmarker}{}%
\end{pgfscope}%
\end{pgfscope}%
\begin{pgfscope}%
\definecolor{textcolor}{rgb}{0.000000,0.000000,0.000000}%
\pgfsetstrokecolor{textcolor}%
\pgfsetfillcolor{textcolor}%
\pgftext[x=0.166667in, y=0.488153in, left, base]{\color{textcolor}{\sffamily\fontsize{8.000000}{9.600000}\selectfont\catcode`\^=\active\def^{\ifmmode\sp\else\^{}\fi}\catcode`\%=\active\def
\end{pgfscope}%
\begin{pgfscope}%
\pgfsetbuttcap%
\pgfsetroundjoin%
\definecolor{currentfill}{rgb}{0.000000,0.000000,0.000000}%
\pgfsetfillcolor{currentfill}%
\pgfsetlinewidth{0.803000pt}%
\definecolor{currentstroke}{rgb}{0.000000,0.000000,0.000000}%
\pgfsetstrokecolor{currentstroke}%
\pgfsetdash{}{0pt}%
\pgfsys@defobject{currentmarker}{\pgfqpoint{-0.048611in}{0.000000in}}{\pgfqpoint{-0.000000in}{0.000000in}}{%
\pgfpathmoveto{\pgfqpoint{-0.000000in}{0.000000in}}%
\pgfpathlineto{\pgfqpoint{-0.048611in}{0.000000in}}%
\pgfusepath{stroke,fill}%
}%
\begin{pgfscope}%
\pgfsys@transformshift{0.322917in}{0.854892in}%
\pgfsys@useobject{currentmarker}{}%
\end{pgfscope}%
\end{pgfscope}%
\begin{pgfscope}%
\definecolor{textcolor}{rgb}{0.000000,0.000000,0.000000}%
\pgfsetstrokecolor{textcolor}%
\pgfsetfillcolor{textcolor}%
\pgftext[x=0.166667in, y=0.816312in, left, base]{\color{textcolor}{\sffamily\fontsize{8.000000}{9.600000}\selectfont\catcode`\^=\active\def^{\ifmmode\sp\else\^{}\fi}\catcode`\%=\active\def
\end{pgfscope}%
\begin{pgfscope}%
\pgfsetbuttcap%
\pgfsetroundjoin%
\definecolor{currentfill}{rgb}{0.000000,0.000000,0.000000}%
\pgfsetfillcolor{currentfill}%
\pgfsetlinewidth{0.803000pt}%
\definecolor{currentstroke}{rgb}{0.000000,0.000000,0.000000}%
\pgfsetstrokecolor{currentstroke}%
\pgfsetdash{}{0pt}%
\pgfsys@defobject{currentmarker}{\pgfqpoint{-0.048611in}{0.000000in}}{\pgfqpoint{-0.000000in}{0.000000in}}{%
\pgfpathmoveto{\pgfqpoint{-0.000000in}{0.000000in}}%
\pgfpathlineto{\pgfqpoint{-0.048611in}{0.000000in}}%
\pgfusepath{stroke,fill}%
}%
\begin{pgfscope}%
\pgfsys@transformshift{0.322917in}{1.183050in}%
\pgfsys@useobject{currentmarker}{}%
\end{pgfscope}%
\end{pgfscope}%
\begin{pgfscope}%
\definecolor{textcolor}{rgb}{0.000000,0.000000,0.000000}%
\pgfsetstrokecolor{textcolor}%
\pgfsetfillcolor{textcolor}%
\pgftext[x=0.166667in, y=1.144470in, left, base]{\color{textcolor}{\sffamily\fontsize{8.000000}{9.600000}\selectfont\catcode`\^=\active\def^{\ifmmode\sp\else\^{}\fi}\catcode`\%=\active\def
\end{pgfscope}%
\begin{pgfscope}%
\pgfsetbuttcap%
\pgfsetroundjoin%
\definecolor{currentfill}{rgb}{0.000000,0.000000,0.000000}%
\pgfsetfillcolor{currentfill}%
\pgfsetlinewidth{0.803000pt}%
\definecolor{currentstroke}{rgb}{0.000000,0.000000,0.000000}%
\pgfsetstrokecolor{currentstroke}%
\pgfsetdash{}{0pt}%
\pgfsys@defobject{currentmarker}{\pgfqpoint{-0.048611in}{0.000000in}}{\pgfqpoint{-0.000000in}{0.000000in}}{%
\pgfpathmoveto{\pgfqpoint{-0.000000in}{0.000000in}}%
\pgfpathlineto{\pgfqpoint{-0.048611in}{0.000000in}}%
\pgfusepath{stroke,fill}%
}%
\begin{pgfscope}%
\pgfsys@transformshift{0.322917in}{1.511208in}%
\pgfsys@useobject{currentmarker}{}%
\end{pgfscope}%
\end{pgfscope}%
\begin{pgfscope}%
\definecolor{textcolor}{rgb}{0.000000,0.000000,0.000000}%
\pgfsetstrokecolor{textcolor}%
\pgfsetfillcolor{textcolor}%
\pgftext[x=0.166667in, y=1.472628in, left, base]{\color{textcolor}{\sffamily\fontsize{8.000000}{9.600000}\selectfont\catcode`\^=\active\def^{\ifmmode\sp\else\^{}\fi}\catcode`\%=\active\def
\end{pgfscope}%
\begin{pgfscope}%
\pgfsetbuttcap%
\pgfsetroundjoin%
\definecolor{currentfill}{rgb}{0.000000,0.000000,0.000000}%
\pgfsetfillcolor{currentfill}%
\pgfsetlinewidth{0.803000pt}%
\definecolor{currentstroke}{rgb}{0.000000,0.000000,0.000000}%
\pgfsetstrokecolor{currentstroke}%
\pgfsetdash{}{0pt}%
\pgfsys@defobject{currentmarker}{\pgfqpoint{-0.048611in}{0.000000in}}{\pgfqpoint{-0.000000in}{0.000000in}}{%
\pgfpathmoveto{\pgfqpoint{-0.000000in}{0.000000in}}%
\pgfpathlineto{\pgfqpoint{-0.048611in}{0.000000in}}%
\pgfusepath{stroke,fill}%
}%
\begin{pgfscope}%
\pgfsys@transformshift{0.322917in}{1.839367in}%
\pgfsys@useobject{currentmarker}{}%
\end{pgfscope}%
\end{pgfscope}%
\begin{pgfscope}%
\definecolor{textcolor}{rgb}{0.000000,0.000000,0.000000}%
\pgfsetstrokecolor{textcolor}%
\pgfsetfillcolor{textcolor}%
\pgftext[x=0.166667in, y=1.800786in, left, base]{\color{textcolor}{\sffamily\fontsize{8.000000}{9.600000}\selectfont\catcode`\^=\active\def^{\ifmmode\sp\else\^{}\fi}\catcode`\%=\active\def
\end{pgfscope}%
\begin{pgfscope}%
\definecolor{textcolor}{rgb}{0.000000,0.000000,0.000000}%
\pgfsetstrokecolor{textcolor}%
\pgfsetfillcolor{textcolor}%
\pgftext[x=0.111111in,y=1.101011in,,bottom,rotate=90.000000]{\color{textcolor}{\sffamily\fontsize{8.000000}{9.600000}\selectfont\catcode`\^=\active\def^{\ifmmode\sp\else\^{}\fi}\catcode`\%=\active\def
\end{pgfscope}%
\begin{pgfscope}%
\pgfpathrectangle{\pgfqpoint{0.322917in}{0.362654in}}{\pgfqpoint{2.972603in}{1.476712in}}%
\pgfusepath{clip}%
\pgfsetbuttcap%
\pgfsetroundjoin%
\pgfsetlinewidth{0.501875pt}%
\definecolor{currentstroke}{rgb}{0.121569,0.466667,0.705882}%
\pgfsetstrokecolor{currentstroke}%
\pgfsetdash{{1.850000pt}{0.800000pt}}{0.000000pt}%
\pgfpathmoveto{\pgfqpoint{0.322917in}{0.526734in}}%
\pgfpathlineto{\pgfqpoint{0.433014in}{0.526734in}}%
\pgfpathlineto{\pgfqpoint{0.543110in}{0.526734in}}%
\pgfpathlineto{\pgfqpoint{0.653207in}{0.526734in}}%
\pgfpathlineto{\pgfqpoint{0.763303in}{0.526734in}}%
\pgfpathlineto{\pgfqpoint{0.873399in}{0.526734in}}%
\pgfpathlineto{\pgfqpoint{0.983496in}{0.526734in}}%
\pgfpathlineto{\pgfqpoint{1.093592in}{0.526734in}}%
\pgfpathlineto{\pgfqpoint{1.203689in}{0.526734in}}%
\pgfpathlineto{\pgfqpoint{1.313785in}{0.526734in}}%
\pgfpathlineto{\pgfqpoint{1.423881in}{0.526734in}}%
\pgfpathlineto{\pgfqpoint{1.533978in}{0.526734in}}%
\pgfpathlineto{\pgfqpoint{1.644074in}{0.526734in}}%
\pgfpathlineto{\pgfqpoint{1.754171in}{0.526734in}}%
\pgfpathlineto{\pgfqpoint{1.864267in}{0.526734in}}%
\pgfpathlineto{\pgfqpoint{1.974363in}{0.526734in}}%
\pgfpathlineto{\pgfqpoint{2.084460in}{0.526734in}}%
\pgfpathlineto{\pgfqpoint{2.194556in}{0.526734in}}%
\pgfpathlineto{\pgfqpoint{2.304653in}{0.526734in}}%
\pgfpathlineto{\pgfqpoint{2.414749in}{0.526734in}}%
\pgfpathlineto{\pgfqpoint{2.524845in}{0.526734in}}%
\pgfpathlineto{\pgfqpoint{2.634942in}{0.526734in}}%
\pgfpathlineto{\pgfqpoint{2.745038in}{0.526734in}}%
\pgfpathlineto{\pgfqpoint{2.855135in}{0.526734in}}%
\pgfpathlineto{\pgfqpoint{2.965231in}{0.526734in}}%
\pgfpathlineto{\pgfqpoint{3.075327in}{0.526734in}}%
\pgfpathlineto{\pgfqpoint{3.185424in}{0.526734in}}%
\pgfpathlineto{\pgfqpoint{3.295520in}{0.526734in}}%
\pgfpathlineto{\pgfqpoint{3.305520in}{0.526734in}}%
\pgfusepath{stroke}%
\end{pgfscope}%
\begin{pgfscope}%
\pgfpathrectangle{\pgfqpoint{0.322917in}{0.362654in}}{\pgfqpoint{2.972603in}{1.476712in}}%
\pgfusepath{clip}%
\pgfsetrectcap%
\pgfsetroundjoin%
\pgfsetlinewidth{0.501875pt}%
\definecolor{currentstroke}{rgb}{0.501961,0.501961,0.501961}%
\pgfsetstrokecolor{currentstroke}%
\pgfsetdash{}{0pt}%
\pgfpathmoveto{\pgfqpoint{1.379843in}{0.352654in}}%
\pgfpathlineto{\pgfqpoint{1.379843in}{1.101011in}}%
\pgfpathlineto{\pgfqpoint{1.247727in}{1.101011in}}%
\pgfpathlineto{\pgfqpoint{1.247727in}{0.352654in}}%
\pgfusepath{stroke}%
\end{pgfscope}%
\begin{pgfscope}%
\pgfpathrectangle{\pgfqpoint{0.322917in}{0.362654in}}{\pgfqpoint{2.972603in}{1.476712in}}%
\pgfusepath{clip}%
\pgfsetrectcap%
\pgfsetroundjoin%
\pgfsetlinewidth{0.501875pt}%
\definecolor{currentstroke}{rgb}{0.501961,0.501961,0.501961}%
\pgfsetstrokecolor{currentstroke}%
\pgfsetdash{}{0pt}%
\pgfpathmoveto{\pgfqpoint{2.040421in}{0.526734in}}%
\pgfpathlineto{\pgfqpoint{2.172537in}{0.526734in}}%
\pgfpathlineto{\pgfqpoint{2.172537in}{1.849367in}}%
\pgfpathmoveto{\pgfqpoint{2.040421in}{1.849367in}}%
\pgfpathlineto{\pgfqpoint{2.040421in}{0.526734in}}%
\pgfusepath{stroke}%
\end{pgfscope}%
\begin{pgfscope}%
\pgfpathrectangle{\pgfqpoint{0.322917in}{0.362654in}}{\pgfqpoint{2.972603in}{1.476712in}}%
\pgfusepath{clip}%
\pgfsetrectcap%
\pgfsetroundjoin%
\pgfsetlinewidth{0.501875pt}%
\definecolor{currentstroke}{rgb}{0.501961,0.501961,0.501961}%
\pgfsetstrokecolor{currentstroke}%
\pgfsetdash{}{0pt}%
\pgfpathmoveto{\pgfqpoint{3.097347in}{0.352654in}}%
\pgfpathlineto{\pgfqpoint{3.097347in}{1.101011in}}%
\pgfpathlineto{\pgfqpoint{2.965231in}{1.101011in}}%
\pgfpathlineto{\pgfqpoint{2.965231in}{0.352654in}}%
\pgfusepath{stroke}%
\end{pgfscope}%
\begin{pgfscope}%
\pgfpathrectangle{\pgfqpoint{0.322917in}{0.362654in}}{\pgfqpoint{2.972603in}{1.476712in}}%
\pgfusepath{clip}%
\pgfsetrectcap%
\pgfsetroundjoin%
\pgfsetlinewidth{0.752812pt}%
\definecolor{currentstroke}{rgb}{1.000000,0.647059,0.000000}%
\pgfsetstrokecolor{currentstroke}%
\pgfsetdash{}{0pt}%
\pgfpathmoveto{\pgfqpoint{0.322917in}{0.526734in}}%
\pgfpathlineto{\pgfqpoint{0.438099in}{0.535013in}}%
\pgfpathlineto{\pgfqpoint{0.552253in}{0.563611in}}%
\pgfpathlineto{\pgfqpoint{0.664968in}{0.616795in}}%
\pgfpathlineto{\pgfqpoint{0.775809in}{0.696567in}}%
\pgfpathlineto{\pgfqpoint{0.885261in}{0.793258in}}%
\pgfpathlineto{\pgfqpoint{0.994017in}{0.889706in}}%
\pgfpathlineto{\pgfqpoint{1.102316in}{0.967076in}}%
\pgfpathlineto{\pgfqpoint{1.210294in}{1.018567in}}%
\pgfpathlineto{\pgfqpoint{1.317520in}{1.041423in}}%
\pgfpathlineto{\pgfqpoint{1.423252in}{1.034725in}}%
\pgfpathlineto{\pgfqpoint{1.527103in}{0.998918in}}%
\pgfpathlineto{\pgfqpoint{1.628738in}{0.934566in}}%
\pgfpathlineto{\pgfqpoint{1.728475in}{0.856725in}}%
\pgfpathlineto{\pgfqpoint{1.826831in}{0.768807in}}%
\pgfpathlineto{\pgfqpoint{1.924528in}{0.681757in}}%
\pgfpathlineto{\pgfqpoint{2.021877in}{0.619090in}}%
\pgfpathlineto{\pgfqpoint{2.118509in}{0.584635in}}%
\pgfpathlineto{\pgfqpoint{2.213975in}{0.579414in}}%
\pgfpathlineto{\pgfqpoint{2.307852in}{0.603386in}}%
\pgfpathlineto{\pgfqpoint{2.399771in}{0.656105in}}%
\pgfpathlineto{\pgfqpoint{2.489414in}{0.736908in}}%
\pgfpathlineto{\pgfqpoint{2.577230in}{0.839718in}}%
\pgfpathlineto{\pgfqpoint{2.664144in}{0.946190in}}%
\pgfpathlineto{\pgfqpoint{2.750847in}{1.033892in}}%
\pgfpathlineto{\pgfqpoint{2.837243in}{1.094652in}}%
\pgfpathlineto{\pgfqpoint{2.922873in}{1.126190in}}%
\pgfpathlineto{\pgfqpoint{3.007248in}{1.128194in}}%
\pgfpathlineto{\pgfqpoint{3.089919in}{1.101011in}}%
\pgfpathlineto{\pgfqpoint{3.170492in}{1.045246in}}%
\pgfpathlineto{\pgfqpoint{3.248702in}{0.961583in}}%
\pgfusepath{stroke}%
\end{pgfscope}%
\begin{pgfscope}%
\pgfpathrectangle{\pgfqpoint{0.322917in}{0.362654in}}{\pgfqpoint{2.972603in}{1.476712in}}%
\pgfusepath{clip}%
\pgfsetbuttcap%
\pgfsetroundjoin%
\definecolor{currentfill}{rgb}{1.000000,0.647059,0.000000}%
\pgfsetfillcolor{currentfill}%
\pgfsetlinewidth{1.003750pt}%
\definecolor{currentstroke}{rgb}{1.000000,0.647059,0.000000}%
\pgfsetstrokecolor{currentstroke}%
\pgfsetdash{}{0pt}%
\pgfsys@defobject{currentmarker}{\pgfqpoint{-0.013889in}{-0.013889in}}{\pgfqpoint{0.013889in}{0.013889in}}{%
\pgfpathmoveto{\pgfqpoint{0.000000in}{-0.013889in}}%
\pgfpathcurveto{\pgfqpoint{0.003683in}{-0.013889in}}{\pgfqpoint{0.007216in}{-0.012425in}}{\pgfqpoint{0.009821in}{-0.009821in}}%
\pgfpathcurveto{\pgfqpoint{0.012425in}{-0.007216in}}{\pgfqpoint{0.013889in}{-0.003683in}}{\pgfqpoint{0.013889in}{0.000000in}}%
\pgfpathcurveto{\pgfqpoint{0.013889in}{0.003683in}}{\pgfqpoint{0.012425in}{0.007216in}}{\pgfqpoint{0.009821in}{0.009821in}}%
\pgfpathcurveto{\pgfqpoint{0.007216in}{0.012425in}}{\pgfqpoint{0.003683in}{0.013889in}}{\pgfqpoint{0.000000in}{0.013889in}}%
\pgfpathcurveto{\pgfqpoint{-0.003683in}{0.013889in}}{\pgfqpoint{-0.007216in}{0.012425in}}{\pgfqpoint{-0.009821in}{0.009821in}}%
\pgfpathcurveto{\pgfqpoint{-0.012425in}{0.007216in}}{\pgfqpoint{-0.013889in}{0.003683in}}{\pgfqpoint{-0.013889in}{0.000000in}}%
\pgfpathcurveto{\pgfqpoint{-0.013889in}{-0.003683in}}{\pgfqpoint{-0.012425in}{-0.007216in}}{\pgfqpoint{-0.009821in}{-0.009821in}}%
\pgfpathcurveto{\pgfqpoint{-0.007216in}{-0.012425in}}{\pgfqpoint{-0.003683in}{-0.013889in}}{\pgfqpoint{0.000000in}{-0.013889in}}%
\pgfpathlineto{\pgfqpoint{0.000000in}{-0.013889in}}%
\pgfpathclose%
\pgfusepath{stroke,fill}%
}%
\begin{pgfscope}%
\pgfsys@transformshift{0.322917in}{0.526734in}%
\pgfsys@useobject{currentmarker}{}%
\end{pgfscope}%
\begin{pgfscope}%
\pgfsys@transformshift{0.438099in}{0.535013in}%
\pgfsys@useobject{currentmarker}{}%
\end{pgfscope}%
\begin{pgfscope}%
\pgfsys@transformshift{0.552253in}{0.563611in}%
\pgfsys@useobject{currentmarker}{}%
\end{pgfscope}%
\begin{pgfscope}%
\pgfsys@transformshift{0.664968in}{0.616795in}%
\pgfsys@useobject{currentmarker}{}%
\end{pgfscope}%
\begin{pgfscope}%
\pgfsys@transformshift{0.775809in}{0.696567in}%
\pgfsys@useobject{currentmarker}{}%
\end{pgfscope}%
\begin{pgfscope}%
\pgfsys@transformshift{0.885261in}{0.793258in}%
\pgfsys@useobject{currentmarker}{}%
\end{pgfscope}%
\begin{pgfscope}%
\pgfsys@transformshift{0.994017in}{0.889706in}%
\pgfsys@useobject{currentmarker}{}%
\end{pgfscope}%
\begin{pgfscope}%
\pgfsys@transformshift{1.102316in}{0.967076in}%
\pgfsys@useobject{currentmarker}{}%
\end{pgfscope}%
\begin{pgfscope}%
\pgfsys@transformshift{1.210294in}{1.018567in}%
\pgfsys@useobject{currentmarker}{}%
\end{pgfscope}%
\begin{pgfscope}%
\pgfsys@transformshift{1.317520in}{1.041423in}%
\pgfsys@useobject{currentmarker}{}%
\end{pgfscope}%
\begin{pgfscope}%
\pgfsys@transformshift{1.423252in}{1.034725in}%
\pgfsys@useobject{currentmarker}{}%
\end{pgfscope}%
\begin{pgfscope}%
\pgfsys@transformshift{1.527103in}{0.998918in}%
\pgfsys@useobject{currentmarker}{}%
\end{pgfscope}%
\begin{pgfscope}%
\pgfsys@transformshift{1.628738in}{0.934566in}%
\pgfsys@useobject{currentmarker}{}%
\end{pgfscope}%
\begin{pgfscope}%
\pgfsys@transformshift{1.728475in}{0.856725in}%
\pgfsys@useobject{currentmarker}{}%
\end{pgfscope}%
\begin{pgfscope}%
\pgfsys@transformshift{1.826831in}{0.768807in}%
\pgfsys@useobject{currentmarker}{}%
\end{pgfscope}%
\begin{pgfscope}%
\pgfsys@transformshift{1.924528in}{0.681757in}%
\pgfsys@useobject{currentmarker}{}%
\end{pgfscope}%
\begin{pgfscope}%
\pgfsys@transformshift{2.021877in}{0.619090in}%
\pgfsys@useobject{currentmarker}{}%
\end{pgfscope}%
\begin{pgfscope}%
\pgfsys@transformshift{2.118509in}{0.584635in}%
\pgfsys@useobject{currentmarker}{}%
\end{pgfscope}%
\begin{pgfscope}%
\pgfsys@transformshift{2.213975in}{0.579414in}%
\pgfsys@useobject{currentmarker}{}%
\end{pgfscope}%
\begin{pgfscope}%
\pgfsys@transformshift{2.307852in}{0.603386in}%
\pgfsys@useobject{currentmarker}{}%
\end{pgfscope}%
\begin{pgfscope}%
\pgfsys@transformshift{2.399771in}{0.656105in}%
\pgfsys@useobject{currentmarker}{}%
\end{pgfscope}%
\begin{pgfscope}%
\pgfsys@transformshift{2.489414in}{0.736908in}%
\pgfsys@useobject{currentmarker}{}%
\end{pgfscope}%
\begin{pgfscope}%
\pgfsys@transformshift{2.577230in}{0.839718in}%
\pgfsys@useobject{currentmarker}{}%
\end{pgfscope}%
\begin{pgfscope}%
\pgfsys@transformshift{2.664144in}{0.946190in}%
\pgfsys@useobject{currentmarker}{}%
\end{pgfscope}%
\begin{pgfscope}%
\pgfsys@transformshift{2.750847in}{1.033892in}%
\pgfsys@useobject{currentmarker}{}%
\end{pgfscope}%
\begin{pgfscope}%
\pgfsys@transformshift{2.837243in}{1.094652in}%
\pgfsys@useobject{currentmarker}{}%
\end{pgfscope}%
\begin{pgfscope}%
\pgfsys@transformshift{2.922873in}{1.126190in}%
\pgfsys@useobject{currentmarker}{}%
\end{pgfscope}%
\begin{pgfscope}%
\pgfsys@transformshift{3.007248in}{1.128194in}%
\pgfsys@useobject{currentmarker}{}%
\end{pgfscope}%
\begin{pgfscope}%
\pgfsys@transformshift{3.089919in}{1.101011in}%
\pgfsys@useobject{currentmarker}{}%
\end{pgfscope}%
\begin{pgfscope}%
\pgfsys@transformshift{3.170492in}{1.045246in}%
\pgfsys@useobject{currentmarker}{}%
\end{pgfscope}%
\begin{pgfscope}%
\pgfsys@transformshift{3.248702in}{0.961583in}%
\pgfsys@useobject{currentmarker}{}%
\end{pgfscope}%
\end{pgfscope}%
\begin{pgfscope}%
\pgfpathrectangle{\pgfqpoint{0.322917in}{0.362654in}}{\pgfqpoint{2.972603in}{1.476712in}}%
\pgfusepath{clip}%
\pgfsetrectcap%
\pgfsetroundjoin%
\pgfsetlinewidth{0.501875pt}%
\definecolor{currentstroke}{rgb}{0.000000,0.000000,1.000000}%
\pgfsetstrokecolor{currentstroke}%
\pgfsetdash{}{0pt}%
\pgfpathmoveto{\pgfqpoint{0.322917in}{0.526734in}}%
\pgfpathlineto{\pgfqpoint{0.438099in}{0.535013in}}%
\pgfpathlineto{\pgfqpoint{0.552253in}{0.563611in}}%
\pgfpathlineto{\pgfqpoint{0.665360in}{0.614902in}}%
\pgfpathlineto{\pgfqpoint{0.777051in}{0.691040in}}%
\pgfpathlineto{\pgfqpoint{0.887312in}{0.788369in}}%
\pgfpathlineto{\pgfqpoint{0.996732in}{0.890720in}}%
\pgfpathlineto{\pgfqpoint{1.105708in}{0.976912in}}%
\pgfpathlineto{\pgfqpoint{1.214176in}{1.036875in}}%
\pgfpathlineto{\pgfqpoint{1.321695in}{1.067545in}}%
\pgfpathlineto{\pgfqpoint{1.427796in}{1.068591in}}%
\pgfpathlineto{\pgfqpoint{1.532542in}{1.044005in}}%
\pgfpathlineto{\pgfqpoint{1.635698in}{0.993189in}}%
\pgfpathlineto{\pgfqpoint{1.736675in}{0.913924in}}%
\pgfpathlineto{\pgfqpoint{1.835345in}{0.807326in}}%
\pgfpathlineto{\pgfqpoint{1.932481in}{0.690901in}}%
\pgfpathlineto{\pgfqpoint{2.029303in}{0.590524in}}%
\pgfpathlineto{\pgfqpoint{2.125855in}{0.516425in}}%
\pgfpathlineto{\pgfqpoint{2.221708in}{0.471673in}}%
\pgfpathlineto{\pgfqpoint{2.316375in}{0.456703in}}%
\pgfpathlineto{\pgfqpoint{2.409414in}{0.471166in}}%
\pgfpathlineto{\pgfqpoint{2.500432in}{0.514502in}}%
\pgfpathlineto{\pgfqpoint{2.589084in}{0.586076in}}%
\pgfpathlineto{\pgfqpoint{2.675072in}{0.685213in}}%
\pgfpathlineto{\pgfqpoint{2.759077in}{0.796546in}}%
\pgfpathlineto{\pgfqpoint{2.841960in}{0.913072in}}%
\pgfpathlineto{\pgfqpoint{2.924441in}{1.025786in}}%
\pgfpathlineto{\pgfqpoint{3.007060in}{1.115071in}}%
\pgfpathlineto{\pgfqpoint{3.089523in}{1.177564in}}%
\pgfpathlineto{\pgfqpoint{3.171428in}{1.211852in}}%
\pgfpathlineto{\pgfqpoint{3.252420in}{1.217509in}}%
\pgfusepath{stroke}%
\end{pgfscope}%
\begin{pgfscope}%
\pgfpathrectangle{\pgfqpoint{0.322917in}{0.362654in}}{\pgfqpoint{2.972603in}{1.476712in}}%
\pgfusepath{clip}%
\pgfsetbuttcap%
\pgfsetroundjoin%
\definecolor{currentfill}{rgb}{0.000000,0.000000,1.000000}%
\pgfsetfillcolor{currentfill}%
\pgfsetlinewidth{1.003750pt}%
\definecolor{currentstroke}{rgb}{0.000000,0.000000,1.000000}%
\pgfsetstrokecolor{currentstroke}%
\pgfsetdash{}{0pt}%
\pgfsys@defobject{currentmarker}{\pgfqpoint{-0.006944in}{-0.006944in}}{\pgfqpoint{0.006944in}{0.006944in}}{%
\pgfpathmoveto{\pgfqpoint{0.000000in}{-0.006944in}}%
\pgfpathcurveto{\pgfqpoint{0.001842in}{-0.006944in}}{\pgfqpoint{0.003608in}{-0.006213in}}{\pgfqpoint{0.004910in}{-0.004910in}}%
\pgfpathcurveto{\pgfqpoint{0.006213in}{-0.003608in}}{\pgfqpoint{0.006944in}{-0.001842in}}{\pgfqpoint{0.006944in}{0.000000in}}%
\pgfpathcurveto{\pgfqpoint{0.006944in}{0.001842in}}{\pgfqpoint{0.006213in}{0.003608in}}{\pgfqpoint{0.004910in}{0.004910in}}%
\pgfpathcurveto{\pgfqpoint{0.003608in}{0.006213in}}{\pgfqpoint{0.001842in}{0.006944in}}{\pgfqpoint{0.000000in}{0.006944in}}%
\pgfpathcurveto{\pgfqpoint{-0.001842in}{0.006944in}}{\pgfqpoint{-0.003608in}{0.006213in}}{\pgfqpoint{-0.004910in}{0.004910in}}%
\pgfpathcurveto{\pgfqpoint{-0.006213in}{0.003608in}}{\pgfqpoint{-0.006944in}{0.001842in}}{\pgfqpoint{-0.006944in}{0.000000in}}%
\pgfpathcurveto{\pgfqpoint{-0.006944in}{-0.001842in}}{\pgfqpoint{-0.006213in}{-0.003608in}}{\pgfqpoint{-0.004910in}{-0.004910in}}%
\pgfpathcurveto{\pgfqpoint{-0.003608in}{-0.006213in}}{\pgfqpoint{-0.001842in}{-0.006944in}}{\pgfqpoint{0.000000in}{-0.006944in}}%
\pgfpathlineto{\pgfqpoint{0.000000in}{-0.006944in}}%
\pgfpathclose%
\pgfusepath{stroke,fill}%
}%
\begin{pgfscope}%
\pgfsys@transformshift{0.322917in}{0.526734in}%
\pgfsys@useobject{currentmarker}{}%
\end{pgfscope}%
\begin{pgfscope}%
\pgfsys@transformshift{0.438099in}{0.535013in}%
\pgfsys@useobject{currentmarker}{}%
\end{pgfscope}%
\begin{pgfscope}%
\pgfsys@transformshift{0.552253in}{0.563611in}%
\pgfsys@useobject{currentmarker}{}%
\end{pgfscope}%
\begin{pgfscope}%
\pgfsys@transformshift{0.665360in}{0.614902in}%
\pgfsys@useobject{currentmarker}{}%
\end{pgfscope}%
\begin{pgfscope}%
\pgfsys@transformshift{0.777051in}{0.691040in}%
\pgfsys@useobject{currentmarker}{}%
\end{pgfscope}%
\begin{pgfscope}%
\pgfsys@transformshift{0.887312in}{0.788369in}%
\pgfsys@useobject{currentmarker}{}%
\end{pgfscope}%
\begin{pgfscope}%
\pgfsys@transformshift{0.996732in}{0.890720in}%
\pgfsys@useobject{currentmarker}{}%
\end{pgfscope}%
\begin{pgfscope}%
\pgfsys@transformshift{1.105708in}{0.976912in}%
\pgfsys@useobject{currentmarker}{}%
\end{pgfscope}%
\begin{pgfscope}%
\pgfsys@transformshift{1.214176in}{1.036875in}%
\pgfsys@useobject{currentmarker}{}%
\end{pgfscope}%
\begin{pgfscope}%
\pgfsys@transformshift{1.321695in}{1.067545in}%
\pgfsys@useobject{currentmarker}{}%
\end{pgfscope}%
\begin{pgfscope}%
\pgfsys@transformshift{1.427796in}{1.068591in}%
\pgfsys@useobject{currentmarker}{}%
\end{pgfscope}%
\begin{pgfscope}%
\pgfsys@transformshift{1.532542in}{1.044005in}%
\pgfsys@useobject{currentmarker}{}%
\end{pgfscope}%
\begin{pgfscope}%
\pgfsys@transformshift{1.635698in}{0.993189in}%
\pgfsys@useobject{currentmarker}{}%
\end{pgfscope}%
\begin{pgfscope}%
\pgfsys@transformshift{1.736675in}{0.913924in}%
\pgfsys@useobject{currentmarker}{}%
\end{pgfscope}%
\begin{pgfscope}%
\pgfsys@transformshift{1.835345in}{0.807326in}%
\pgfsys@useobject{currentmarker}{}%
\end{pgfscope}%
\begin{pgfscope}%
\pgfsys@transformshift{1.932481in}{0.690901in}%
\pgfsys@useobject{currentmarker}{}%
\end{pgfscope}%
\begin{pgfscope}%
\pgfsys@transformshift{2.029303in}{0.590524in}%
\pgfsys@useobject{currentmarker}{}%
\end{pgfscope}%
\begin{pgfscope}%
\pgfsys@transformshift{2.125855in}{0.516425in}%
\pgfsys@useobject{currentmarker}{}%
\end{pgfscope}%
\begin{pgfscope}%
\pgfsys@transformshift{2.221708in}{0.471673in}%
\pgfsys@useobject{currentmarker}{}%
\end{pgfscope}%
\begin{pgfscope}%
\pgfsys@transformshift{2.316375in}{0.456703in}%
\pgfsys@useobject{currentmarker}{}%
\end{pgfscope}%
\begin{pgfscope}%
\pgfsys@transformshift{2.409414in}{0.471166in}%
\pgfsys@useobject{currentmarker}{}%
\end{pgfscope}%
\begin{pgfscope}%
\pgfsys@transformshift{2.500432in}{0.514502in}%
\pgfsys@useobject{currentmarker}{}%
\end{pgfscope}%
\begin{pgfscope}%
\pgfsys@transformshift{2.589084in}{0.586076in}%
\pgfsys@useobject{currentmarker}{}%
\end{pgfscope}%
\begin{pgfscope}%
\pgfsys@transformshift{2.675072in}{0.685213in}%
\pgfsys@useobject{currentmarker}{}%
\end{pgfscope}%
\begin{pgfscope}%
\pgfsys@transformshift{2.759077in}{0.796546in}%
\pgfsys@useobject{currentmarker}{}%
\end{pgfscope}%
\begin{pgfscope}%
\pgfsys@transformshift{2.841960in}{0.913072in}%
\pgfsys@useobject{currentmarker}{}%
\end{pgfscope}%
\begin{pgfscope}%
\pgfsys@transformshift{2.924441in}{1.025786in}%
\pgfsys@useobject{currentmarker}{}%
\end{pgfscope}%
\begin{pgfscope}%
\pgfsys@transformshift{3.007060in}{1.115071in}%
\pgfsys@useobject{currentmarker}{}%
\end{pgfscope}%
\begin{pgfscope}%
\pgfsys@transformshift{3.089523in}{1.177564in}%
\pgfsys@useobject{currentmarker}{}%
\end{pgfscope}%
\begin{pgfscope}%
\pgfsys@transformshift{3.171428in}{1.211852in}%
\pgfsys@useobject{currentmarker}{}%
\end{pgfscope}%
\begin{pgfscope}%
\pgfsys@transformshift{3.252420in}{1.217509in}%
\pgfsys@useobject{currentmarker}{}%
\end{pgfscope}%
\end{pgfscope}%
\begin{pgfscope}%
\pgfsetrectcap%
\pgfsetmiterjoin%
\pgfsetlinewidth{0.803000pt}%
\definecolor{currentstroke}{rgb}{0.000000,0.000000,0.000000}%
\pgfsetstrokecolor{currentstroke}%
\pgfsetdash{}{0pt}%
\pgfpathmoveto{\pgfqpoint{0.322917in}{0.362654in}}%
\pgfpathlineto{\pgfqpoint{0.322917in}{1.839367in}}%
\pgfusepath{stroke}%
\end{pgfscope}%
\begin{pgfscope}%
\pgfsetrectcap%
\pgfsetmiterjoin%
\pgfsetlinewidth{0.803000pt}%
\definecolor{currentstroke}{rgb}{0.000000,0.000000,0.000000}%
\pgfsetstrokecolor{currentstroke}%
\pgfsetdash{}{0pt}%
\pgfpathmoveto{\pgfqpoint{3.295520in}{0.362654in}}%
\pgfpathlineto{\pgfqpoint{3.295520in}{1.839367in}}%
\pgfusepath{stroke}%
\end{pgfscope}%
\begin{pgfscope}%
\pgfsetrectcap%
\pgfsetmiterjoin%
\pgfsetlinewidth{0.803000pt}%
\definecolor{currentstroke}{rgb}{0.000000,0.000000,0.000000}%
\pgfsetstrokecolor{currentstroke}%
\pgfsetdash{}{0pt}%
\pgfpathmoveto{\pgfqpoint{0.322917in}{0.362654in}}%
\pgfpathlineto{\pgfqpoint{3.295520in}{0.362654in}}%
\pgfusepath{stroke}%
\end{pgfscope}%
\begin{pgfscope}%
\pgfsetrectcap%
\pgfsetmiterjoin%
\pgfsetlinewidth{0.803000pt}%
\definecolor{currentstroke}{rgb}{0.000000,0.000000,0.000000}%
\pgfsetstrokecolor{currentstroke}%
\pgfsetdash{}{0pt}%
\pgfpathmoveto{\pgfqpoint{0.322917in}{1.839367in}}%
\pgfpathlineto{\pgfqpoint{3.295520in}{1.839367in}}%
\pgfusepath{stroke}%
\end{pgfscope}%
\begin{pgfscope}%
\definecolor{textcolor}{rgb}{0.000000,0.000000,0.000000}%
\pgfsetstrokecolor{textcolor}%
\pgfsetfillcolor{textcolor}%
\pgftext[x=0.322917in,y=0.575345in,,base]{\color{textcolor}{\sffamily\fontsize{6.000000}{7.200000}\selectfont\catcode`\^=\active\def^{\ifmmode\sp\else\^{}\fi}\catcode`\%=\active\def
\end{pgfscope}%
\begin{pgfscope}%
\definecolor{textcolor}{rgb}{0.000000,0.000000,0.000000}%
\pgfsetstrokecolor{textcolor}%
\pgfsetfillcolor{textcolor}%
\pgftext[x=0.552253in,y=0.612222in,,base]{\color{textcolor}{\sffamily\fontsize{6.000000}{7.200000}\selectfont\catcode`\^=\active\def^{\ifmmode\sp\else\^{}\fi}\catcode`\%=\active\def
\end{pgfscope}%
\begin{pgfscope}%
\definecolor{textcolor}{rgb}{0.000000,0.000000,0.000000}%
\pgfsetstrokecolor{textcolor}%
\pgfsetfillcolor{textcolor}%
\pgftext[x=0.775809in,y=0.745179in,,base]{\color{textcolor}{\sffamily\fontsize{6.000000}{7.200000}\selectfont\catcode`\^=\active\def^{\ifmmode\sp\else\^{}\fi}\catcode`\%=\active\def
\end{pgfscope}%
\begin{pgfscope}%
\definecolor{textcolor}{rgb}{0.000000,0.000000,0.000000}%
\pgfsetstrokecolor{textcolor}%
\pgfsetfillcolor{textcolor}%
\pgftext[x=0.994017in,y=0.938317in,,base]{\color{textcolor}{\sffamily\fontsize{6.000000}{7.200000}\selectfont\catcode`\^=\active\def^{\ifmmode\sp\else\^{}\fi}\catcode`\%=\active\def
\end{pgfscope}%
\begin{pgfscope}%
\definecolor{textcolor}{rgb}{0.000000,0.000000,0.000000}%
\pgfsetstrokecolor{textcolor}%
\pgfsetfillcolor{textcolor}%
\pgftext[x=1.210294in,y=1.067179in,,base]{\color{textcolor}{\sffamily\fontsize{6.000000}{7.200000}\selectfont\catcode`\^=\active\def^{\ifmmode\sp\else\^{}\fi}\catcode`\%=\active\def
\end{pgfscope}%
\begin{pgfscope}%
\definecolor{textcolor}{rgb}{0.000000,0.000000,0.000000}%
\pgfsetstrokecolor{textcolor}%
\pgfsetfillcolor{textcolor}%
\pgftext[x=1.423252in,y=1.083336in,,base]{\color{textcolor}{\sffamily\fontsize{6.000000}{7.200000}\selectfont\catcode`\^=\active\def^{\ifmmode\sp\else\^{}\fi}\catcode`\%=\active\def
\end{pgfscope}%
\begin{pgfscope}%
\definecolor{textcolor}{rgb}{0.000000,0.000000,0.000000}%
\pgfsetstrokecolor{textcolor}%
\pgfsetfillcolor{textcolor}%
\pgftext[x=1.628738in,y=0.983177in,,base]{\color{textcolor}{\sffamily\fontsize{6.000000}{7.200000}\selectfont\catcode`\^=\active\def^{\ifmmode\sp\else\^{}\fi}\catcode`\%=\active\def
\end{pgfscope}%
\begin{pgfscope}%
\definecolor{textcolor}{rgb}{0.000000,0.000000,0.000000}%
\pgfsetstrokecolor{textcolor}%
\pgfsetfillcolor{textcolor}%
\pgftext[x=1.826831in,y=0.817418in,,base]{\color{textcolor}{\sffamily\fontsize{6.000000}{7.200000}\selectfont\catcode`\^=\active\def^{\ifmmode\sp\else\^{}\fi}\catcode`\%=\active\def
\end{pgfscope}%
\begin{pgfscope}%
\definecolor{textcolor}{rgb}{0.000000,0.000000,0.000000}%
\pgfsetstrokecolor{textcolor}%
\pgfsetfillcolor{textcolor}%
\pgftext[x=2.021877in,y=0.667702in,,base]{\color{textcolor}{\sffamily\fontsize{6.000000}{7.200000}\selectfont\catcode`\^=\active\def^{\ifmmode\sp\else\^{}\fi}\catcode`\%=\active\def
\end{pgfscope}%
\begin{pgfscope}%
\definecolor{textcolor}{rgb}{0.000000,0.000000,0.000000}%
\pgfsetstrokecolor{textcolor}%
\pgfsetfillcolor{textcolor}%
\pgftext[x=2.213975in,y=0.628025in,,base]{\color{textcolor}{\sffamily\fontsize{6.000000}{7.200000}\selectfont\catcode`\^=\active\def^{\ifmmode\sp\else\^{}\fi}\catcode`\%=\active\def
\end{pgfscope}%
\begin{pgfscope}%
\definecolor{textcolor}{rgb}{0.000000,0.000000,0.000000}%
\pgfsetstrokecolor{textcolor}%
\pgfsetfillcolor{textcolor}%
\pgftext[x=2.399771in,y=0.704716in,,base]{\color{textcolor}{\sffamily\fontsize{6.000000}{7.200000}\selectfont\catcode`\^=\active\def^{\ifmmode\sp\else\^{}\fi}\catcode`\%=\active\def
\end{pgfscope}%
\begin{pgfscope}%
\definecolor{textcolor}{rgb}{0.000000,0.000000,0.000000}%
\pgfsetstrokecolor{textcolor}%
\pgfsetfillcolor{textcolor}%
\pgftext[x=2.577230in,y=0.888329in,,base]{\color{textcolor}{\sffamily\fontsize{6.000000}{7.200000}\selectfont\catcode`\^=\active\def^{\ifmmode\sp\else\^{}\fi}\catcode`\%=\active\def
\end{pgfscope}%
\begin{pgfscope}%
\definecolor{textcolor}{rgb}{0.000000,0.000000,0.000000}%
\pgfsetstrokecolor{textcolor}%
\pgfsetfillcolor{textcolor}%
\pgftext[x=2.750847in,y=1.082504in,,base]{\color{textcolor}{\sffamily\fontsize{6.000000}{7.200000}\selectfont\catcode`\^=\active\def^{\ifmmode\sp\else\^{}\fi}\catcode`\%=\active\def
\end{pgfscope}%
\begin{pgfscope}%
\definecolor{textcolor}{rgb}{0.000000,0.000000,0.000000}%
\pgfsetstrokecolor{textcolor}%
\pgfsetfillcolor{textcolor}%
\pgftext[x=2.922873in,y=1.174801in,,base]{\color{textcolor}{\sffamily\fontsize{6.000000}{7.200000}\selectfont\catcode`\^=\active\def^{\ifmmode\sp\else\^{}\fi}\catcode`\%=\active\def
\end{pgfscope}%
\begin{pgfscope}%
\definecolor{textcolor}{rgb}{0.000000,0.000000,0.000000}%
\pgfsetstrokecolor{textcolor}%
\pgfsetfillcolor{textcolor}%
\pgftext[x=3.089919in,y=1.149622in,,base]{\color{textcolor}{\sffamily\fontsize{6.000000}{7.200000}\selectfont\catcode`\^=\active\def^{\ifmmode\sp\else\^{}\fi}\catcode`\%=\active\def
\end{pgfscope}%
\begin{pgfscope}%
\definecolor{textcolor}{rgb}{0.000000,0.000000,0.000000}%
\pgfsetstrokecolor{textcolor}%
\pgfsetfillcolor{textcolor}%
\pgftext[x=3.248702in,y=1.010194in,,base]{\color{textcolor}{\sffamily\fontsize{6.000000}{7.200000}\selectfont\catcode`\^=\active\def^{\ifmmode\sp\else\^{}\fi}\catcode`\%=\active\def
\end{pgfscope}%
\begin{pgfscope}%
\pgfsetbuttcap%
\pgfsetmiterjoin%
\definecolor{currentfill}{rgb}{1.000000,1.000000,1.000000}%
\pgfsetfillcolor{currentfill}%
\pgfsetfillopacity{0.800000}%
\pgfsetlinewidth{1.003750pt}%
\definecolor{currentstroke}{rgb}{0.800000,0.800000,0.800000}%
\pgfsetstrokecolor{currentstroke}%
\pgfsetstrokeopacity{0.800000}%
\pgfsetdash{}{0pt}%
\pgfpathmoveto{\pgfqpoint{0.390973in}{1.490447in}}%
\pgfpathlineto{\pgfqpoint{3.347644in}{1.490447in}}%
\pgfpathquadraticcurveto{\pgfqpoint{3.367088in}{1.490447in}}{\pgfqpoint{3.367088in}{1.509892in}}%
\pgfpathlineto{\pgfqpoint{3.367088in}{1.771311in}}%
\pgfpathquadraticcurveto{\pgfqpoint{3.367088in}{1.790756in}}{\pgfqpoint{3.347644in}{1.790756in}}%
\pgfpathlineto{\pgfqpoint{0.390973in}{1.790756in}}%
\pgfpathquadraticcurveto{\pgfqpoint{0.371529in}{1.790756in}}{\pgfqpoint{0.371529in}{1.771311in}}%
\pgfpathlineto{\pgfqpoint{0.371529in}{1.509892in}}%
\pgfpathquadraticcurveto{\pgfqpoint{0.371529in}{1.490447in}}{\pgfqpoint{0.390973in}{1.490447in}}%
\pgfpathlineto{\pgfqpoint{0.390973in}{1.490447in}}%
\pgfpathclose%
\pgfusepath{stroke,fill}%
\end{pgfscope}%
\begin{pgfscope}%
\pgfsetbuttcap%
\pgfsetroundjoin%
\pgfsetlinewidth{0.501875pt}%
\definecolor{currentstroke}{rgb}{0.121569,0.466667,0.705882}%
\pgfsetstrokecolor{currentstroke}%
\pgfsetdash{{1.850000pt}{0.800000pt}}{0.000000pt}%
\pgfpathmoveto{\pgfqpoint{0.410417in}{1.717839in}}%
\pgfpathlineto{\pgfqpoint{0.507640in}{1.717839in}}%
\pgfpathlineto{\pgfqpoint{0.604862in}{1.717839in}}%
\pgfusepath{stroke}%
\end{pgfscope}%
\begin{pgfscope}%
\definecolor{textcolor}{rgb}{0.000000,0.000000,0.000000}%
\pgfsetstrokecolor{textcolor}%
\pgfsetfillcolor{textcolor}%
\pgftext[x=0.682640in,y=1.683811in,left,base]{\color{textcolor}{\sffamily\fontsize{7.000000}{8.400000}\selectfont\catcode`\^=\active\def^{\ifmmode\sp\else\^{}\fi}\catcode`\%=\active\def
\end{pgfscope}%
\begin{pgfscope}%
\pgfsetrectcap%
\pgfsetroundjoin%
\pgfsetlinewidth{0.501875pt}%
\definecolor{currentstroke}{rgb}{0.501961,0.501961,0.501961}%
\pgfsetstrokecolor{currentstroke}%
\pgfsetdash{}{0pt}%
\pgfpathmoveto{\pgfqpoint{0.410417in}{1.582268in}}%
\pgfpathlineto{\pgfqpoint{0.507640in}{1.582268in}}%
\pgfpathlineto{\pgfqpoint{0.604862in}{1.582268in}}%
\pgfusepath{stroke}%
\end{pgfscope}%
\begin{pgfscope}%
\definecolor{textcolor}{rgb}{0.000000,0.000000,0.000000}%
\pgfsetstrokecolor{textcolor}%
\pgfsetfillcolor{textcolor}%
\pgftext[x=0.682640in,y=1.548240in,left,base]{\color{textcolor}{\sffamily\fontsize{7.000000}{8.400000}\selectfont\catcode`\^=\active\def^{\ifmmode\sp\else\^{}\fi}\catcode`\%=\active\def
\end{pgfscope}%
\begin{pgfscope}%
\pgfsetrectcap%
\pgfsetroundjoin%
\pgfsetlinewidth{0.752812pt}%
\definecolor{currentstroke}{rgb}{1.000000,0.647059,0.000000}%
\pgfsetstrokecolor{currentstroke}%
\pgfsetdash{}{0pt}%
\pgfpathmoveto{\pgfqpoint{1.879377in}{1.717839in}}%
\pgfpathlineto{\pgfqpoint{1.976599in}{1.717839in}}%
\pgfpathlineto{\pgfqpoint{2.073821in}{1.717839in}}%
\pgfusepath{stroke}%
\end{pgfscope}%
\begin{pgfscope}%
\pgfsetbuttcap%
\pgfsetroundjoin%
\definecolor{currentfill}{rgb}{1.000000,0.647059,0.000000}%
\pgfsetfillcolor{currentfill}%
\pgfsetlinewidth{1.003750pt}%
\definecolor{currentstroke}{rgb}{1.000000,0.647059,0.000000}%
\pgfsetstrokecolor{currentstroke}%
\pgfsetdash{}{0pt}%
\pgfsys@defobject{currentmarker}{\pgfqpoint{-0.013889in}{-0.013889in}}{\pgfqpoint{0.013889in}{0.013889in}}{%
\pgfpathmoveto{\pgfqpoint{0.000000in}{-0.013889in}}%
\pgfpathcurveto{\pgfqpoint{0.003683in}{-0.013889in}}{\pgfqpoint{0.007216in}{-0.012425in}}{\pgfqpoint{0.009821in}{-0.009821in}}%
\pgfpathcurveto{\pgfqpoint{0.012425in}{-0.007216in}}{\pgfqpoint{0.013889in}{-0.003683in}}{\pgfqpoint{0.013889in}{0.000000in}}%
\pgfpathcurveto{\pgfqpoint{0.013889in}{0.003683in}}{\pgfqpoint{0.012425in}{0.007216in}}{\pgfqpoint{0.009821in}{0.009821in}}%
\pgfpathcurveto{\pgfqpoint{0.007216in}{0.012425in}}{\pgfqpoint{0.003683in}{0.013889in}}{\pgfqpoint{0.000000in}{0.013889in}}%
\pgfpathcurveto{\pgfqpoint{-0.003683in}{0.013889in}}{\pgfqpoint{-0.007216in}{0.012425in}}{\pgfqpoint{-0.009821in}{0.009821in}}%
\pgfpathcurveto{\pgfqpoint{-0.012425in}{0.007216in}}{\pgfqpoint{-0.013889in}{0.003683in}}{\pgfqpoint{-0.013889in}{0.000000in}}%
\pgfpathcurveto{\pgfqpoint{-0.013889in}{-0.003683in}}{\pgfqpoint{-0.012425in}{-0.007216in}}{\pgfqpoint{-0.009821in}{-0.009821in}}%
\pgfpathcurveto{\pgfqpoint{-0.007216in}{-0.012425in}}{\pgfqpoint{-0.003683in}{-0.013889in}}{\pgfqpoint{0.000000in}{-0.013889in}}%
\pgfpathlineto{\pgfqpoint{0.000000in}{-0.013889in}}%
\pgfpathclose%
\pgfusepath{stroke,fill}%
}%
\begin{pgfscope}%
\pgfsys@transformshift{1.976599in}{1.717839in}%
\pgfsys@useobject{currentmarker}{}%
\end{pgfscope}%
\end{pgfscope}%
\begin{pgfscope}%
\definecolor{textcolor}{rgb}{0.000000,0.000000,0.000000}%
\pgfsetstrokecolor{textcolor}%
\pgfsetfillcolor{textcolor}%
\pgftext[x=2.151599in,y=1.683811in,left,base]{\color{textcolor}{\sffamily\fontsize{7.000000}{8.400000}\selectfont\catcode`\^=\active\def^{\ifmmode\sp\else\^{}\fi}\catcode`\%=\active\def
\end{pgfscope}%
\begin{pgfscope}%
\pgfsetrectcap%
\pgfsetroundjoin%
\pgfsetlinewidth{0.501875pt}%
\definecolor{currentstroke}{rgb}{0.000000,0.000000,1.000000}%
\pgfsetstrokecolor{currentstroke}%
\pgfsetdash{}{0pt}%
\pgfpathmoveto{\pgfqpoint{1.879377in}{1.582268in}}%
\pgfpathlineto{\pgfqpoint{1.976599in}{1.582268in}}%
\pgfpathlineto{\pgfqpoint{2.073821in}{1.582268in}}%
\pgfusepath{stroke}%
\end{pgfscope}%
\begin{pgfscope}%
\pgfsetbuttcap%
\pgfsetroundjoin%
\definecolor{currentfill}{rgb}{0.000000,0.000000,1.000000}%
\pgfsetfillcolor{currentfill}%
\pgfsetlinewidth{1.003750pt}%
\definecolor{currentstroke}{rgb}{0.000000,0.000000,1.000000}%
\pgfsetstrokecolor{currentstroke}%
\pgfsetdash{}{0pt}%
\pgfsys@defobject{currentmarker}{\pgfqpoint{-0.006944in}{-0.006944in}}{\pgfqpoint{0.006944in}{0.006944in}}{%
\pgfpathmoveto{\pgfqpoint{0.000000in}{-0.006944in}}%
\pgfpathcurveto{\pgfqpoint{0.001842in}{-0.006944in}}{\pgfqpoint{0.003608in}{-0.006213in}}{\pgfqpoint{0.004910in}{-0.004910in}}%
\pgfpathcurveto{\pgfqpoint{0.006213in}{-0.003608in}}{\pgfqpoint{0.006944in}{-0.001842in}}{\pgfqpoint{0.006944in}{0.000000in}}%
\pgfpathcurveto{\pgfqpoint{0.006944in}{0.001842in}}{\pgfqpoint{0.006213in}{0.003608in}}{\pgfqpoint{0.004910in}{0.004910in}}%
\pgfpathcurveto{\pgfqpoint{0.003608in}{0.006213in}}{\pgfqpoint{0.001842in}{0.006944in}}{\pgfqpoint{0.000000in}{0.006944in}}%
\pgfpathcurveto{\pgfqpoint{-0.001842in}{0.006944in}}{\pgfqpoint{-0.003608in}{0.006213in}}{\pgfqpoint{-0.004910in}{0.004910in}}%
\pgfpathcurveto{\pgfqpoint{-0.006213in}{0.003608in}}{\pgfqpoint{-0.006944in}{0.001842in}}{\pgfqpoint{-0.006944in}{0.000000in}}%
\pgfpathcurveto{\pgfqpoint{-0.006944in}{-0.001842in}}{\pgfqpoint{-0.006213in}{-0.003608in}}{\pgfqpoint{-0.004910in}{-0.004910in}}%
\pgfpathcurveto{\pgfqpoint{-0.003608in}{-0.006213in}}{\pgfqpoint{-0.001842in}{-0.006944in}}{\pgfqpoint{0.000000in}{-0.006944in}}%
\pgfpathlineto{\pgfqpoint{0.000000in}{-0.006944in}}%
\pgfpathclose%
\pgfusepath{stroke,fill}%
}%
\begin{pgfscope}%
\pgfsys@transformshift{1.976599in}{1.582268in}%
\pgfsys@useobject{currentmarker}{}%
\end{pgfscope}%
\end{pgfscope}%
\begin{pgfscope}%
\definecolor{textcolor}{rgb}{0.000000,0.000000,0.000000}%
\pgfsetstrokecolor{textcolor}%
\pgfsetfillcolor{textcolor}%
\pgftext[x=2.151599in,y=1.548240in,left,base]{\color{textcolor}{\sffamily\fontsize{7.000000}{8.400000}\selectfont\catcode`\^=\active\def^{\ifmmode\sp\else\^{}\fi}\catcode`\%=\active\def
\end{pgfscope}%
\end{pgfpicture}%
\makeatother%
\endgroup%

%% file: Intersection_Scenario6.pgf
\begingroup%
\makeatletter%
\begin{pgfpicture}%
\pgfpathrectangle{\pgfpointorigin}{\pgfqpoint{1.796494in}{1.020274in}}%
\pgfusepath{use as bounding box, clip}%
\begin{pgfscope}%
\pgfsetbuttcap%
\pgfsetmiterjoin%
\definecolor{currentfill}{rgb}{1.000000,1.000000,1.000000}%
\pgfsetfillcolor{currentfill}%
\pgfsetlinewidth{0.000000pt}%
\definecolor{currentstroke}{rgb}{1.000000,1.000000,1.000000}%
\pgfsetstrokecolor{currentstroke}%
\pgfsetdash{}{0pt}%
\pgfpathmoveto{\pgfqpoint{0.000000in}{0.000000in}}%
\pgfpathlineto{\pgfqpoint{1.796494in}{0.000000in}}%
\pgfpathlineto{\pgfqpoint{1.796494in}{1.020274in}}%
\pgfpathlineto{\pgfqpoint{0.000000in}{1.020274in}}%
\pgfpathlineto{\pgfqpoint{0.000000in}{0.000000in}}%
\pgfpathclose%
\pgfusepath{fill}%
\end{pgfscope}%
\begin{pgfscope}%
\pgfpathrectangle{\pgfqpoint{0.000000in}{0.000000in}}{\pgfqpoint{1.796494in}{1.020274in}}%
\pgfusepath{clip}%
\pgfsys@transformshift{0.000000in}{0.000000in}%
\pgftext[left,bottom]{\includegraphics[interpolate=true,width=1.800000in,height=1.030000in]{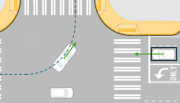}}%
\end{pgfscope}%
\begin{pgfscope}%
\pgfpathrectangle{\pgfqpoint{0.000000in}{0.000000in}}{\pgfqpoint{1.796494in}{1.020274in}}%
\pgfusepath{clip}%
\pgfsetbuttcap%
\pgfsetroundjoin%
\definecolor{currentfill}{rgb}{1.000000,0.549020,0.000000}%
\pgfsetfillcolor{currentfill}%
\pgfsetlinewidth{1.003750pt}%
\definecolor{currentstroke}{rgb}{1.000000,0.549020,0.000000}%
\pgfsetstrokecolor{currentstroke}%
\pgfsetdash{}{0pt}%
\pgfsys@defobject{currentmarker}{\pgfqpoint{-0.020833in}{-0.020833in}}{\pgfqpoint{0.020833in}{0.020833in}}{%
\pgfpathmoveto{\pgfqpoint{0.000000in}{-0.020833in}}%
\pgfpathcurveto{\pgfqpoint{0.005525in}{-0.020833in}}{\pgfqpoint{0.010825in}{-0.018638in}}{\pgfqpoint{0.014731in}{-0.014731in}}%
\pgfpathcurveto{\pgfqpoint{0.018638in}{-0.010825in}}{\pgfqpoint{0.020833in}{-0.005525in}}{\pgfqpoint{0.020833in}{0.000000in}}%
\pgfpathcurveto{\pgfqpoint{0.020833in}{0.005525in}}{\pgfqpoint{0.018638in}{0.010825in}}{\pgfqpoint{0.014731in}{0.014731in}}%
\pgfpathcurveto{\pgfqpoint{0.010825in}{0.018638in}}{\pgfqpoint{0.005525in}{0.020833in}}{\pgfqpoint{0.000000in}{0.020833in}}%
\pgfpathcurveto{\pgfqpoint{-0.005525in}{0.020833in}}{\pgfqpoint{-0.010825in}{0.018638in}}{\pgfqpoint{-0.014731in}{0.014731in}}%
\pgfpathcurveto{\pgfqpoint{-0.018638in}{0.010825in}}{\pgfqpoint{-0.020833in}{0.005525in}}{\pgfqpoint{-0.020833in}{0.000000in}}%
\pgfpathcurveto{\pgfqpoint{-0.020833in}{-0.005525in}}{\pgfqpoint{-0.018638in}{-0.010825in}}{\pgfqpoint{-0.014731in}{-0.014731in}}%
\pgfpathcurveto{\pgfqpoint{-0.010825in}{-0.018638in}}{\pgfqpoint{-0.005525in}{-0.020833in}}{\pgfqpoint{0.000000in}{-0.020833in}}%
\pgfpathlineto{\pgfqpoint{0.000000in}{-0.020833in}}%
\pgfpathclose%
\pgfusepath{stroke,fill}%
}%
\begin{pgfscope}%
\pgfsys@transformshift{0.268592in}{0.289294in}%
\pgfsys@useobject{currentmarker}{}%
\end{pgfscope}%
\end{pgfscope}%
\begin{pgfscope}%
\pgfpathrectangle{\pgfqpoint{0.000000in}{0.000000in}}{\pgfqpoint{1.796494in}{1.020274in}}%
\pgfusepath{clip}%
\pgfsetbuttcap%
\pgfsetroundjoin%
\definecolor{currentfill}{rgb}{1.000000,0.549020,0.000000}%
\pgfsetfillcolor{currentfill}%
\pgfsetlinewidth{1.003750pt}%
\definecolor{currentstroke}{rgb}{1.000000,0.549020,0.000000}%
\pgfsetstrokecolor{currentstroke}%
\pgfsetdash{}{0pt}%
\pgfsys@defobject{currentmarker}{\pgfqpoint{-0.020833in}{-0.020833in}}{\pgfqpoint{0.020833in}{0.020833in}}{%
\pgfpathmoveto{\pgfqpoint{0.000000in}{-0.020833in}}%
\pgfpathcurveto{\pgfqpoint{0.005525in}{-0.020833in}}{\pgfqpoint{0.010825in}{-0.018638in}}{\pgfqpoint{0.014731in}{-0.014731in}}%
\pgfpathcurveto{\pgfqpoint{0.018638in}{-0.010825in}}{\pgfqpoint{0.020833in}{-0.005525in}}{\pgfqpoint{0.020833in}{0.000000in}}%
\pgfpathcurveto{\pgfqpoint{0.020833in}{0.005525in}}{\pgfqpoint{0.018638in}{0.010825in}}{\pgfqpoint{0.014731in}{0.014731in}}%
\pgfpathcurveto{\pgfqpoint{0.010825in}{0.018638in}}{\pgfqpoint{0.005525in}{0.020833in}}{\pgfqpoint{0.000000in}{0.020833in}}%
\pgfpathcurveto{\pgfqpoint{-0.005525in}{0.020833in}}{\pgfqpoint{-0.010825in}{0.018638in}}{\pgfqpoint{-0.014731in}{0.014731in}}%
\pgfpathcurveto{\pgfqpoint{-0.018638in}{0.010825in}}{\pgfqpoint{-0.020833in}{0.005525in}}{\pgfqpoint{-0.020833in}{0.000000in}}%
\pgfpathcurveto{\pgfqpoint{-0.020833in}{-0.005525in}}{\pgfqpoint{-0.018638in}{-0.010825in}}{\pgfqpoint{-0.014731in}{-0.014731in}}%
\pgfpathcurveto{\pgfqpoint{-0.010825in}{-0.018638in}}{\pgfqpoint{-0.005525in}{-0.020833in}}{\pgfqpoint{0.000000in}{-0.020833in}}%
\pgfpathlineto{\pgfqpoint{0.000000in}{-0.020833in}}%
\pgfpathclose%
\pgfusepath{stroke,fill}%
}%
\begin{pgfscope}%
\pgfsys@transformshift{0.754532in}{0.784580in}%
\pgfsys@useobject{currentmarker}{}%
\end{pgfscope}%
\end{pgfscope}%
\begin{pgfscope}%
\definecolor{textcolor}{rgb}{0.000000,0.000000,0.000000}%
\pgfsetstrokecolor{textcolor}%
\pgfsetfillcolor{textcolor}%
\pgftext[x=0.310259in,y=0.145733in,,base]{\color{textcolor}{\sffamily\fontsize{8.000000}{9.600000}\selectfont\catcode`\^=\active\def^{\ifmmode\sp\else\^{}\fi}\catcode`\%=\active\def
\end{pgfscope}%
\begin{pgfscope}%
\definecolor{textcolor}{rgb}{0.000000,0.000000,0.000000}%
\pgfsetstrokecolor{textcolor}%
\pgfsetfillcolor{textcolor}%
\pgftext[x=0.793348in,y=0.793925in,left,base]{\color{textcolor}{\sffamily\fontsize{8.000000}{9.600000}\selectfont\catcode`\^=\active\def^{\ifmmode\sp\else\^{}\fi}\catcode`\%=\active\def
\end{pgfscope}%
\end{pgfpicture}%
\makeatother%
\endgroup%

%% file: MPC_Vx_R0.pgf
\begingroup%
\makeatletter%
\begin{pgfpicture}%
\pgfpathrectangle{\pgfpointorigin}{\pgfqpoint{3.478774in}{1.201158in}}%
\pgfusepath{use as bounding box, clip}%
\begin{pgfscope}%
\pgfsetbuttcap%
\pgfsetmiterjoin%
\definecolor{currentfill}{rgb}{1.000000,1.000000,1.000000}%
\pgfsetfillcolor{currentfill}%
\pgfsetlinewidth{0.000000pt}%
\definecolor{currentstroke}{rgb}{1.000000,1.000000,1.000000}%
\pgfsetstrokecolor{currentstroke}%
\pgfsetdash{}{0pt}%
\pgfpathmoveto{\pgfqpoint{0.000000in}{0.000000in}}%
\pgfpathlineto{\pgfqpoint{3.478774in}{0.000000in}}%
\pgfpathlineto{\pgfqpoint{3.478774in}{1.201158in}}%
\pgfpathlineto{\pgfqpoint{0.000000in}{1.201158in}}%
\pgfpathlineto{\pgfqpoint{0.000000in}{0.000000in}}%
\pgfpathclose%
\pgfusepath{fill}%
\end{pgfscope}%
\begin{pgfscope}%
\pgfsetbuttcap%
\pgfsetmiterjoin%
\definecolor{currentfill}{rgb}{1.000000,1.000000,1.000000}%
\pgfsetfillcolor{currentfill}%
\pgfsetlinewidth{0.000000pt}%
\definecolor{currentstroke}{rgb}{0.000000,0.000000,0.000000}%
\pgfsetstrokecolor{currentstroke}%
\pgfsetstrokeopacity{0.000000}%
\pgfsetdash{}{0pt}%
\pgfpathmoveto{\pgfqpoint{0.506171in}{0.388889in}}%
\pgfpathlineto{\pgfqpoint{3.478774in}{0.388889in}}%
\pgfpathlineto{\pgfqpoint{3.478774in}{1.194368in}}%
\pgfpathlineto{\pgfqpoint{0.506171in}{1.194368in}}%
\pgfpathlineto{\pgfqpoint{0.506171in}{0.388889in}}%
\pgfpathclose%
\pgfusepath{fill}%
\end{pgfscope}%
\begin{pgfscope}%
\pgfsetbuttcap%
\pgfsetroundjoin%
\definecolor{currentfill}{rgb}{0.000000,0.000000,0.000000}%
\pgfsetfillcolor{currentfill}%
\pgfsetlinewidth{0.803000pt}%
\definecolor{currentstroke}{rgb}{0.000000,0.000000,0.000000}%
\pgfsetstrokecolor{currentstroke}%
\pgfsetdash{}{0pt}%
\pgfsys@defobject{currentmarker}{\pgfqpoint{0.000000in}{-0.048611in}}{\pgfqpoint{0.000000in}{0.000000in}}{%
\pgfpathmoveto{\pgfqpoint{0.000000in}{0.000000in}}%
\pgfpathlineto{\pgfqpoint{0.000000in}{-0.048611in}}%
\pgfusepath{stroke,fill}%
}%
\begin{pgfscope}%
\pgfsys@transformshift{0.641290in}{0.388889in}%
\pgfsys@useobject{currentmarker}{}%
\end{pgfscope}%
\end{pgfscope}%
\begin{pgfscope}%
\definecolor{textcolor}{rgb}{0.000000,0.000000,0.000000}%
\pgfsetstrokecolor{textcolor}%
\pgfsetfillcolor{textcolor}%
\pgftext[x=0.641290in,y=0.291667in,,top]{\color{textcolor}{\sffamily\fontsize{9.000000}{10.800000}\selectfont\catcode`\^=\active\def^{\ifmmode\sp\else\^{}\fi}\catcode`\%=\active\def
\end{pgfscope}%
\begin{pgfscope}%
\pgfsetbuttcap%
\pgfsetroundjoin%
\definecolor{currentfill}{rgb}{0.000000,0.000000,0.000000}%
\pgfsetfillcolor{currentfill}%
\pgfsetlinewidth{0.803000pt}%
\definecolor{currentstroke}{rgb}{0.000000,0.000000,0.000000}%
\pgfsetstrokecolor{currentstroke}%
\pgfsetdash{}{0pt}%
\pgfsys@defobject{currentmarker}{\pgfqpoint{0.000000in}{-0.048611in}}{\pgfqpoint{0.000000in}{0.000000in}}{%
\pgfpathmoveto{\pgfqpoint{0.000000in}{0.000000in}}%
\pgfpathlineto{\pgfqpoint{0.000000in}{-0.048611in}}%
\pgfusepath{stroke,fill}%
}%
\begin{pgfscope}%
\pgfsys@transformshift{1.271842in}{0.388889in}%
\pgfsys@useobject{currentmarker}{}%
\end{pgfscope}%
\end{pgfscope}%
\begin{pgfscope}%
\definecolor{textcolor}{rgb}{0.000000,0.000000,0.000000}%
\pgfsetstrokecolor{textcolor}%
\pgfsetfillcolor{textcolor}%
\pgftext[x=1.271842in,y=0.291667in,,top]{\color{textcolor}{\sffamily\fontsize{9.000000}{10.800000}\selectfont\catcode`\^=\active\def^{\ifmmode\sp\else\^{}\fi}\catcode`\%=\active\def
\end{pgfscope}%
\begin{pgfscope}%
\pgfsetbuttcap%
\pgfsetroundjoin%
\definecolor{currentfill}{rgb}{0.000000,0.000000,0.000000}%
\pgfsetfillcolor{currentfill}%
\pgfsetlinewidth{0.803000pt}%
\definecolor{currentstroke}{rgb}{0.000000,0.000000,0.000000}%
\pgfsetstrokecolor{currentstroke}%
\pgfsetdash{}{0pt}%
\pgfsys@defobject{currentmarker}{\pgfqpoint{0.000000in}{-0.048611in}}{\pgfqpoint{0.000000in}{0.000000in}}{%
\pgfpathmoveto{\pgfqpoint{0.000000in}{0.000000in}}%
\pgfpathlineto{\pgfqpoint{0.000000in}{-0.048611in}}%
\pgfusepath{stroke,fill}%
}%
\begin{pgfscope}%
\pgfsys@transformshift{1.902394in}{0.388889in}%
\pgfsys@useobject{currentmarker}{}%
\end{pgfscope}%
\end{pgfscope}%
\begin{pgfscope}%
\definecolor{textcolor}{rgb}{0.000000,0.000000,0.000000}%
\pgfsetstrokecolor{textcolor}%
\pgfsetfillcolor{textcolor}%
\pgftext[x=1.902394in,y=0.291667in,,top]{\color{textcolor}{\sffamily\fontsize{9.000000}{10.800000}\selectfont\catcode`\^=\active\def^{\ifmmode\sp\else\^{}\fi}\catcode`\%=\active\def
\end{pgfscope}%
\begin{pgfscope}%
\pgfsetbuttcap%
\pgfsetroundjoin%
\definecolor{currentfill}{rgb}{0.000000,0.000000,0.000000}%
\pgfsetfillcolor{currentfill}%
\pgfsetlinewidth{0.803000pt}%
\definecolor{currentstroke}{rgb}{0.000000,0.000000,0.000000}%
\pgfsetstrokecolor{currentstroke}%
\pgfsetdash{}{0pt}%
\pgfsys@defobject{currentmarker}{\pgfqpoint{0.000000in}{-0.048611in}}{\pgfqpoint{0.000000in}{0.000000in}}{%
\pgfpathmoveto{\pgfqpoint{0.000000in}{0.000000in}}%
\pgfpathlineto{\pgfqpoint{0.000000in}{-0.048611in}}%
\pgfusepath{stroke,fill}%
}%
\begin{pgfscope}%
\pgfsys@transformshift{2.532946in}{0.388889in}%
\pgfsys@useobject{currentmarker}{}%
\end{pgfscope}%
\end{pgfscope}%
\begin{pgfscope}%
\definecolor{textcolor}{rgb}{0.000000,0.000000,0.000000}%
\pgfsetstrokecolor{textcolor}%
\pgfsetfillcolor{textcolor}%
\pgftext[x=2.532946in,y=0.291667in,,top]{\color{textcolor}{\sffamily\fontsize{9.000000}{10.800000}\selectfont\catcode`\^=\active\def^{\ifmmode\sp\else\^{}\fi}\catcode`\%=\active\def
\end{pgfscope}%
\begin{pgfscope}%
\pgfsetbuttcap%
\pgfsetroundjoin%
\definecolor{currentfill}{rgb}{0.000000,0.000000,0.000000}%
\pgfsetfillcolor{currentfill}%
\pgfsetlinewidth{0.803000pt}%
\definecolor{currentstroke}{rgb}{0.000000,0.000000,0.000000}%
\pgfsetstrokecolor{currentstroke}%
\pgfsetdash{}{0pt}%
\pgfsys@defobject{currentmarker}{\pgfqpoint{0.000000in}{-0.048611in}}{\pgfqpoint{0.000000in}{0.000000in}}{%
\pgfpathmoveto{\pgfqpoint{0.000000in}{0.000000in}}%
\pgfpathlineto{\pgfqpoint{0.000000in}{-0.048611in}}%
\pgfusepath{stroke,fill}%
}%
\begin{pgfscope}%
\pgfsys@transformshift{3.163498in}{0.388889in}%
\pgfsys@useobject{currentmarker}{}%
\end{pgfscope}%
\end{pgfscope}%
\begin{pgfscope}%
\definecolor{textcolor}{rgb}{0.000000,0.000000,0.000000}%
\pgfsetstrokecolor{textcolor}%
\pgfsetfillcolor{textcolor}%
\pgftext[x=3.163498in,y=0.291667in,,top]{\color{textcolor}{\sffamily\fontsize{9.000000}{10.800000}\selectfont\catcode`\^=\active\def^{\ifmmode\sp\else\^{}\fi}\catcode`\%=\active\def
\end{pgfscope}%
\begin{pgfscope}%
\definecolor{textcolor}{rgb}{0.000000,0.000000,0.000000}%
\pgfsetstrokecolor{textcolor}%
\pgfsetfillcolor{textcolor}%
\pgftext[x=1.992473in,y=0.125000in,,top]{\color{textcolor}{\sffamily\fontsize{9.000000}{10.800000}\selectfont\catcode`\^=\active\def^{\ifmmode\sp\else\^{}\fi}\catcode`\%=\active\def
\end{pgfscope}%
\begin{pgfscope}%
\pgfsetbuttcap%
\pgfsetroundjoin%
\definecolor{currentfill}{rgb}{0.000000,0.000000,0.000000}%
\pgfsetfillcolor{currentfill}%
\pgfsetlinewidth{0.803000pt}%
\definecolor{currentstroke}{rgb}{0.000000,0.000000,0.000000}%
\pgfsetstrokecolor{currentstroke}%
\pgfsetdash{}{0pt}%
\pgfsys@defobject{currentmarker}{\pgfqpoint{-0.048611in}{0.000000in}}{\pgfqpoint{-0.000000in}{0.000000in}}{%
\pgfpathmoveto{\pgfqpoint{-0.000000in}{0.000000in}}%
\pgfpathlineto{\pgfqpoint{-0.048611in}{0.000000in}}%
\pgfusepath{stroke,fill}%
}%
\begin{pgfscope}%
\pgfsys@transformshift{0.506171in}{0.609790in}%
\pgfsys@useobject{currentmarker}{}%
\end{pgfscope}%
\end{pgfscope}%
\begin{pgfscope}%
\definecolor{textcolor}{rgb}{0.000000,0.000000,0.000000}%
\pgfsetstrokecolor{textcolor}%
\pgfsetfillcolor{textcolor}%
\pgftext[x=0.244791in, y=0.566388in, left, base]{\color{textcolor}{\sffamily\fontsize{9.000000}{10.800000}\selectfont\catcode`\^=\active\def^{\ifmmode\sp\else\^{}\fi}\catcode`\%=\active\def
\end{pgfscope}%
\begin{pgfscope}%
\pgfsetbuttcap%
\pgfsetroundjoin%
\definecolor{currentfill}{rgb}{0.000000,0.000000,0.000000}%
\pgfsetfillcolor{currentfill}%
\pgfsetlinewidth{0.803000pt}%
\definecolor{currentstroke}{rgb}{0.000000,0.000000,0.000000}%
\pgfsetstrokecolor{currentstroke}%
\pgfsetdash{}{0pt}%
\pgfsys@defobject{currentmarker}{\pgfqpoint{-0.048611in}{0.000000in}}{\pgfqpoint{-0.000000in}{0.000000in}}{%
\pgfpathmoveto{\pgfqpoint{-0.000000in}{0.000000in}}%
\pgfpathlineto{\pgfqpoint{-0.048611in}{0.000000in}}%
\pgfusepath{stroke,fill}%
}%
\begin{pgfscope}%
\pgfsys@transformshift{0.506171in}{0.883773in}%
\pgfsys@useobject{currentmarker}{}%
\end{pgfscope}%
\end{pgfscope}%
\begin{pgfscope}%
\definecolor{textcolor}{rgb}{0.000000,0.000000,0.000000}%
\pgfsetstrokecolor{textcolor}%
\pgfsetfillcolor{textcolor}%
\pgftext[x=0.244791in, y=0.840370in, left, base]{\color{textcolor}{\sffamily\fontsize{9.000000}{10.800000}\selectfont\catcode`\^=\active\def^{\ifmmode\sp\else\^{}\fi}\catcode`\%=\active\def
\end{pgfscope}%
\begin{pgfscope}%
\pgfsetbuttcap%
\pgfsetroundjoin%
\definecolor{currentfill}{rgb}{0.000000,0.000000,0.000000}%
\pgfsetfillcolor{currentfill}%
\pgfsetlinewidth{0.803000pt}%
\definecolor{currentstroke}{rgb}{0.000000,0.000000,0.000000}%
\pgfsetstrokecolor{currentstroke}%
\pgfsetdash{}{0pt}%
\pgfsys@defobject{currentmarker}{\pgfqpoint{-0.048611in}{0.000000in}}{\pgfqpoint{-0.000000in}{0.000000in}}{%
\pgfpathmoveto{\pgfqpoint{-0.000000in}{0.000000in}}%
\pgfpathlineto{\pgfqpoint{-0.048611in}{0.000000in}}%
\pgfusepath{stroke,fill}%
}%
\begin{pgfscope}%
\pgfsys@transformshift{0.506171in}{1.157756in}%
\pgfsys@useobject{currentmarker}{}%
\end{pgfscope}%
\end{pgfscope}%
\begin{pgfscope}%
\definecolor{textcolor}{rgb}{0.000000,0.000000,0.000000}%
\pgfsetstrokecolor{textcolor}%
\pgfsetfillcolor{textcolor}%
\pgftext[x=0.180556in, y=1.114353in, left, base]{\color{textcolor}{\sffamily\fontsize{9.000000}{10.800000}\selectfont\catcode`\^=\active\def^{\ifmmode\sp\else\^{}\fi}\catcode`\%=\active\def
\end{pgfscope}%
\begin{pgfscope}%
\definecolor{textcolor}{rgb}{0.000000,0.000000,0.000000}%
\pgfsetstrokecolor{textcolor}%
\pgfsetfillcolor{textcolor}%
\pgftext[x=0.125000in,y=0.791629in,,bottom,rotate=90.000000]{\color{textcolor}{\sffamily\fontsize{9.000000}{10.800000}\selectfont\catcode`\^=\active\def^{\ifmmode\sp\else\^{}\fi}\catcode`\%=\active\def
\end{pgfscope}%
\begin{pgfscope}%
\pgfpathrectangle{\pgfqpoint{0.506171in}{0.388889in}}{\pgfqpoint{2.972603in}{0.805479in}}%
\pgfusepath{clip}%
\pgfsetrectcap%
\pgfsetroundjoin%
\pgfsetlinewidth{1.003750pt}%
\definecolor{currentstroke}{rgb}{0.121569,0.466667,0.705882}%
\pgfsetstrokecolor{currentstroke}%
\pgfsetdash{}{0pt}%
\pgfpathmoveto{\pgfqpoint{0.641290in}{1.157756in}}%
\pgfpathlineto{\pgfqpoint{0.686329in}{1.157756in}}%
\pgfpathlineto{\pgfqpoint{0.731369in}{1.157717in}}%
\pgfpathlineto{\pgfqpoint{0.776408in}{1.157698in}}%
\pgfpathlineto{\pgfqpoint{0.821447in}{1.157626in}}%
\pgfpathlineto{\pgfqpoint{0.866487in}{1.157577in}}%
\pgfpathlineto{\pgfqpoint{0.911526in}{1.157410in}}%
\pgfpathlineto{\pgfqpoint{0.956566in}{1.157289in}}%
\pgfpathlineto{\pgfqpoint{1.001605in}{1.157116in}}%
\pgfpathlineto{\pgfqpoint{1.046645in}{1.157063in}}%
\pgfpathlineto{\pgfqpoint{1.091684in}{1.156894in}}%
\pgfpathlineto{\pgfqpoint{1.136723in}{1.156874in}}%
\pgfpathlineto{\pgfqpoint{1.181763in}{1.156717in}}%
\pgfpathlineto{\pgfqpoint{1.226802in}{1.156150in}}%
\pgfpathlineto{\pgfqpoint{1.271842in}{1.155010in}}%
\pgfpathlineto{\pgfqpoint{1.316881in}{1.153695in}}%
\pgfpathlineto{\pgfqpoint{1.361921in}{1.146201in}}%
\pgfpathlineto{\pgfqpoint{1.406960in}{1.136551in}}%
\pgfpathlineto{\pgfqpoint{1.452000in}{1.129837in}}%
\pgfpathlineto{\pgfqpoint{1.497039in}{1.125594in}}%
\pgfpathlineto{\pgfqpoint{1.542078in}{1.122273in}}%
\pgfpathlineto{\pgfqpoint{1.587118in}{1.118587in}}%
\pgfpathlineto{\pgfqpoint{1.632157in}{1.115215in}}%
\pgfpathlineto{\pgfqpoint{1.677197in}{1.086348in}}%
\pgfpathlineto{\pgfqpoint{1.722236in}{1.052286in}}%
\pgfpathlineto{\pgfqpoint{1.767276in}{1.011628in}}%
\pgfpathlineto{\pgfqpoint{1.812315in}{0.969431in}}%
\pgfpathlineto{\pgfqpoint{1.857354in}{0.927140in}}%
\pgfpathlineto{\pgfqpoint{1.902394in}{0.885646in}}%
\pgfpathlineto{\pgfqpoint{1.947433in}{0.846056in}}%
\pgfpathlineto{\pgfqpoint{1.992473in}{0.813005in}}%
\pgfpathlineto{\pgfqpoint{2.037512in}{0.781578in}}%
\pgfpathlineto{\pgfqpoint{2.082552in}{0.751627in}}%
\pgfpathlineto{\pgfqpoint{2.127591in}{0.723003in}}%
\pgfpathlineto{\pgfqpoint{2.172631in}{0.695297in}}%
\pgfpathlineto{\pgfqpoint{2.217670in}{0.668402in}}%
\pgfpathlineto{\pgfqpoint{2.262709in}{0.643115in}}%
\pgfpathlineto{\pgfqpoint{2.307749in}{0.619493in}}%
\pgfpathlineto{\pgfqpoint{2.352788in}{0.597794in}}%
\pgfpathlineto{\pgfqpoint{2.397828in}{0.578497in}}%
\pgfpathlineto{\pgfqpoint{2.442867in}{0.562339in}}%
\pgfpathlineto{\pgfqpoint{2.487907in}{0.550931in}}%
\pgfpathlineto{\pgfqpoint{2.532946in}{0.548289in}}%
\pgfpathlineto{\pgfqpoint{2.577985in}{0.554032in}}%
\pgfpathlineto{\pgfqpoint{2.623025in}{0.563980in}}%
\pgfpathlineto{\pgfqpoint{2.668064in}{0.576038in}}%
\pgfpathlineto{\pgfqpoint{2.713104in}{0.589241in}}%
\pgfpathlineto{\pgfqpoint{2.758143in}{0.603006in}}%
\pgfpathlineto{\pgfqpoint{2.803183in}{0.617143in}}%
\pgfpathlineto{\pgfqpoint{2.848222in}{0.632022in}}%
\pgfpathlineto{\pgfqpoint{2.893261in}{0.647832in}}%
\pgfpathlineto{\pgfqpoint{2.938301in}{0.664729in}}%
\pgfpathlineto{\pgfqpoint{2.983340in}{0.682665in}}%
\pgfpathlineto{\pgfqpoint{3.028380in}{0.700319in}}%
\pgfpathlineto{\pgfqpoint{3.073419in}{0.683166in}}%
\pgfpathlineto{\pgfqpoint{3.118459in}{0.642226in}}%
\pgfpathlineto{\pgfqpoint{3.163498in}{0.601034in}}%
\pgfpathlineto{\pgfqpoint{3.208538in}{0.561929in}}%
\pgfpathlineto{\pgfqpoint{3.253577in}{0.526018in}}%
\pgfpathlineto{\pgfqpoint{3.298616in}{0.492501in}}%
\pgfpathlineto{\pgfqpoint{3.343656in}{0.460470in}}%
\pgfusepath{stroke}%
\end{pgfscope}%
\begin{pgfscope}%
\pgfpathrectangle{\pgfqpoint{0.506171in}{0.388889in}}{\pgfqpoint{2.972603in}{0.805479in}}%
\pgfusepath{clip}%
\pgfsetrectcap%
\pgfsetroundjoin%
\pgfsetlinewidth{1.003750pt}%
\definecolor{currentstroke}{rgb}{1.000000,0.498039,0.054902}%
\pgfsetstrokecolor{currentstroke}%
\pgfsetdash{}{0pt}%
\pgfpathmoveto{\pgfqpoint{0.641290in}{1.157756in}}%
\pgfpathlineto{\pgfqpoint{0.686329in}{1.157756in}}%
\pgfpathlineto{\pgfqpoint{0.731369in}{1.157734in}}%
\pgfpathlineto{\pgfqpoint{0.776408in}{1.157727in}}%
\pgfpathlineto{\pgfqpoint{0.821447in}{1.157720in}}%
\pgfpathlineto{\pgfqpoint{0.866487in}{1.157714in}}%
\pgfpathlineto{\pgfqpoint{0.911526in}{1.157708in}}%
\pgfpathlineto{\pgfqpoint{0.956566in}{1.157704in}}%
\pgfpathlineto{\pgfqpoint{1.001605in}{1.157701in}}%
\pgfpathlineto{\pgfqpoint{1.046645in}{1.157698in}}%
\pgfpathlineto{\pgfqpoint{1.091684in}{1.157698in}}%
\pgfpathlineto{\pgfqpoint{1.136723in}{1.157698in}}%
\pgfpathlineto{\pgfqpoint{1.181763in}{1.157679in}}%
\pgfpathlineto{\pgfqpoint{1.226802in}{1.157674in}}%
\pgfpathlineto{\pgfqpoint{1.271842in}{1.156327in}}%
\pgfpathlineto{\pgfqpoint{1.316881in}{1.152059in}}%
\pgfpathlineto{\pgfqpoint{1.361921in}{1.129945in}}%
\pgfpathlineto{\pgfqpoint{1.406960in}{1.105485in}}%
\pgfpathlineto{\pgfqpoint{1.452000in}{1.082702in}}%
\pgfpathlineto{\pgfqpoint{1.497039in}{1.061298in}}%
\pgfpathlineto{\pgfqpoint{1.542078in}{1.040919in}}%
\pgfpathlineto{\pgfqpoint{1.587118in}{1.021237in}}%
\pgfpathlineto{\pgfqpoint{1.632157in}{1.001962in}}%
\pgfpathlineto{\pgfqpoint{1.677197in}{0.982800in}}%
\pgfpathlineto{\pgfqpoint{1.722236in}{0.936148in}}%
\pgfpathlineto{\pgfqpoint{1.767276in}{0.887186in}}%
\pgfpathlineto{\pgfqpoint{1.812315in}{0.837168in}}%
\pgfpathlineto{\pgfqpoint{1.857354in}{0.786453in}}%
\pgfpathlineto{\pgfqpoint{1.902394in}{0.734483in}}%
\pgfpathlineto{\pgfqpoint{1.947433in}{0.691816in}}%
\pgfpathlineto{\pgfqpoint{1.992473in}{0.654858in}}%
\pgfpathlineto{\pgfqpoint{2.037512in}{0.623027in}}%
\pgfpathlineto{\pgfqpoint{2.082552in}{0.593789in}}%
\pgfpathlineto{\pgfqpoint{2.127591in}{0.564020in}}%
\pgfpathlineto{\pgfqpoint{2.172631in}{0.535439in}}%
\pgfpathlineto{\pgfqpoint{2.217670in}{0.508464in}}%
\pgfpathlineto{\pgfqpoint{2.262709in}{0.486477in}}%
\pgfpathlineto{\pgfqpoint{2.307749in}{0.481526in}}%
\pgfpathlineto{\pgfqpoint{2.352788in}{0.483206in}}%
\pgfpathlineto{\pgfqpoint{2.397828in}{0.490283in}}%
\pgfpathlineto{\pgfqpoint{2.442867in}{0.500932in}}%
\pgfpathlineto{\pgfqpoint{2.487907in}{0.513552in}}%
\pgfpathlineto{\pgfqpoint{2.532946in}{0.527140in}}%
\pgfpathlineto{\pgfqpoint{2.577985in}{0.541122in}}%
\pgfpathlineto{\pgfqpoint{2.623025in}{0.555133in}}%
\pgfpathlineto{\pgfqpoint{2.668064in}{0.568933in}}%
\pgfpathlineto{\pgfqpoint{2.713104in}{0.582425in}}%
\pgfpathlineto{\pgfqpoint{2.758143in}{0.595749in}}%
\pgfpathlineto{\pgfqpoint{2.803183in}{0.612412in}}%
\pgfpathlineto{\pgfqpoint{2.848222in}{0.616692in}}%
\pgfpathlineto{\pgfqpoint{2.893261in}{0.586463in}}%
\pgfpathlineto{\pgfqpoint{2.938301in}{0.546836in}}%
\pgfpathlineto{\pgfqpoint{2.983340in}{0.508715in}}%
\pgfpathlineto{\pgfqpoint{3.028380in}{0.477641in}}%
\pgfpathlineto{\pgfqpoint{3.073419in}{0.456095in}}%
\pgfpathlineto{\pgfqpoint{3.118459in}{0.438532in}}%
\pgfpathlineto{\pgfqpoint{3.163498in}{0.429825in}}%
\pgfpathlineto{\pgfqpoint{3.208538in}{0.425502in}}%
\pgfpathlineto{\pgfqpoint{3.253577in}{0.427953in}}%
\pgfpathlineto{\pgfqpoint{3.298616in}{0.438260in}}%
\pgfpathlineto{\pgfqpoint{3.343656in}{0.447401in}}%
\pgfusepath{stroke}%
\end{pgfscope}%
\begin{pgfscope}%
\pgfsetrectcap%
\pgfsetmiterjoin%
\pgfsetlinewidth{0.803000pt}%
\definecolor{currentstroke}{rgb}{0.000000,0.000000,0.000000}%
\pgfsetstrokecolor{currentstroke}%
\pgfsetdash{}{0pt}%
\pgfpathmoveto{\pgfqpoint{0.506171in}{0.388889in}}%
\pgfpathlineto{\pgfqpoint{0.506171in}{1.194368in}}%
\pgfusepath{stroke}%
\end{pgfscope}%
\begin{pgfscope}%
\pgfsetrectcap%
\pgfsetmiterjoin%
\pgfsetlinewidth{0.803000pt}%
\definecolor{currentstroke}{rgb}{0.000000,0.000000,0.000000}%
\pgfsetstrokecolor{currentstroke}%
\pgfsetdash{}{0pt}%
\pgfpathmoveto{\pgfqpoint{3.478774in}{0.388889in}}%
\pgfpathlineto{\pgfqpoint{3.478774in}{1.194368in}}%
\pgfusepath{stroke}%
\end{pgfscope}%
\begin{pgfscope}%
\pgfsetrectcap%
\pgfsetmiterjoin%
\pgfsetlinewidth{0.803000pt}%
\definecolor{currentstroke}{rgb}{0.000000,0.000000,0.000000}%
\pgfsetstrokecolor{currentstroke}%
\pgfsetdash{}{0pt}%
\pgfpathmoveto{\pgfqpoint{0.506171in}{0.388889in}}%
\pgfpathlineto{\pgfqpoint{3.478774in}{0.388889in}}%
\pgfusepath{stroke}%
\end{pgfscope}%
\begin{pgfscope}%
\pgfsetrectcap%
\pgfsetmiterjoin%
\pgfsetlinewidth{0.803000pt}%
\definecolor{currentstroke}{rgb}{0.000000,0.000000,0.000000}%
\pgfsetstrokecolor{currentstroke}%
\pgfsetdash{}{0pt}%
\pgfpathmoveto{\pgfqpoint{0.506171in}{1.194368in}}%
\pgfpathlineto{\pgfqpoint{3.478774in}{1.194368in}}%
\pgfusepath{stroke}%
\end{pgfscope}%
\begin{pgfscope}%
\pgfsetbuttcap%
\pgfsetmiterjoin%
\definecolor{currentfill}{rgb}{1.000000,1.000000,1.000000}%
\pgfsetfillcolor{currentfill}%
\pgfsetfillopacity{0.800000}%
\pgfsetlinewidth{1.003750pt}%
\definecolor{currentstroke}{rgb}{0.800000,0.800000,0.800000}%
\pgfsetstrokecolor{currentstroke}%
\pgfsetstrokeopacity{0.800000}%
\pgfsetdash{}{0pt}%
\pgfpathmoveto{\pgfqpoint{2.699641in}{0.795603in}}%
\pgfpathlineto{\pgfqpoint{3.400996in}{0.795603in}}%
\pgfpathquadraticcurveto{\pgfqpoint{3.423219in}{0.795603in}}{\pgfqpoint{3.423219in}{0.817825in}}%
\pgfpathlineto{\pgfqpoint{3.423219in}{1.116591in}}%
\pgfpathquadraticcurveto{\pgfqpoint{3.423219in}{1.138813in}}{\pgfqpoint{3.400996in}{1.138813in}}%
\pgfpathlineto{\pgfqpoint{2.699641in}{1.138813in}}%
\pgfpathquadraticcurveto{\pgfqpoint{2.677419in}{1.138813in}}{\pgfqpoint{2.677419in}{1.116591in}}%
\pgfpathlineto{\pgfqpoint{2.677419in}{0.817825in}}%
\pgfpathquadraticcurveto{\pgfqpoint{2.677419in}{0.795603in}}{\pgfqpoint{2.699641in}{0.795603in}}%
\pgfpathlineto{\pgfqpoint{2.699641in}{0.795603in}}%
\pgfpathclose%
\pgfusepath{stroke,fill}%
\end{pgfscope}%
\begin{pgfscope}%
\pgfsetrectcap%
\pgfsetroundjoin%
\pgfsetlinewidth{1.003750pt}%
\definecolor{currentstroke}{rgb}{0.121569,0.466667,0.705882}%
\pgfsetstrokecolor{currentstroke}%
\pgfsetdash{}{0pt}%
\pgfpathmoveto{\pgfqpoint{2.721863in}{1.055479in}}%
\pgfpathlineto{\pgfqpoint{2.832974in}{1.055479in}}%
\pgfpathlineto{\pgfqpoint{2.944086in}{1.055479in}}%
\pgfusepath{stroke}%
\end{pgfscope}%
\begin{pgfscope}%
\definecolor{textcolor}{rgb}{0.000000,0.000000,0.000000}%
\pgfsetstrokecolor{textcolor}%
\pgfsetfillcolor{textcolor}%
\pgftext[x=3.032974in,y=1.016591in,left,base]{\color{textcolor}{\sffamily\fontsize{8.000000}{9.600000}\selectfont\catcode`\^=\active\def^{\ifmmode\sp\else\^{}\fi}\catcode`\%=\active\def
\end{pgfscope}%
\begin{pgfscope}%
\pgfsetrectcap%
\pgfsetroundjoin%
\pgfsetlinewidth{1.003750pt}%
\definecolor{currentstroke}{rgb}{1.000000,0.498039,0.054902}%
\pgfsetstrokecolor{currentstroke}%
\pgfsetdash{}{0pt}%
\pgfpathmoveto{\pgfqpoint{2.721863in}{0.900541in}}%
\pgfpathlineto{\pgfqpoint{2.832974in}{0.900541in}}%
\pgfpathlineto{\pgfqpoint{2.944086in}{0.900541in}}%
\pgfusepath{stroke}%
\end{pgfscope}%
\begin{pgfscope}%
\definecolor{textcolor}{rgb}{0.000000,0.000000,0.000000}%
\pgfsetstrokecolor{textcolor}%
\pgfsetfillcolor{textcolor}%
\pgftext[x=3.032974in,y=0.861652in,left,base]{\color{textcolor}{\sffamily\fontsize{8.000000}{9.600000}\selectfont\catcode`\^=\active\def^{\ifmmode\sp\else\^{}\fi}\catcode`\%=\active\def
\end{pgfscope}%
\end{pgfpicture}%
\makeatother%
\endgroup%